\documentclass[iop]{emulateapj}%
\usepackage{amsmath}%
\usepackage{amsfonts}%
\usepackage{amssymb}%
\usepackage{graphicx}%
\usepackage{epsfig}%
\usepackage[font=small, labelsep=space]{caption}
\usepackage{natbib}

\usepackage{color}
\definecolor{dkgreen}{RGB}{70,128,30}
\definecolor{orange}{RGB}{210,65,28}

\def\JHU{1}
\def\STSCI{2}
\def\UTA{3}
\def\MSU{4}
\def\Siena{5}

\begin{document}

\slugcomment{\em Accepted for publication in the Astrophysical Journal}

\title{The Relationship Between Brightest Cluster Galaxy Star Formation and the Intracluster Medium in CLASH}
\author{Kevin Fogarty\altaffilmark{\JHU}, Marc Postman\altaffilmark{\STSCI}, Rebecca Larson\altaffilmark{\UTA}, Megan Donahue\altaffilmark{\MSU}, John Moustakas\altaffilmark{\Siena}}

\altaffiltext{\JHU}{Department of Physics and Astronomy, Johns Hopkins University Baltimore, MD, USA}
\altaffiltext{\STSCI}{Space Telescope Science Institute Baltimore, MD, USA}
\altaffiltext{\UTA}{University of Texas Austin, TX, USA}
\altaffiltext{\MSU}{Michigan State University East Lansing, MI, USA}
\altaffiltext{\Siena}{Siena College Loudonville, NY, USA}

\begin{abstract}  
We study the nature of feedback mechanisms in the 11 CLASH brightest cluster galaxies (BCGs) that exhibit extended ultraviolet and nebular line emission features. We estimate star formation rates (SFRs), dust masses, and starburst durations using a Bayesian photometry-fitting technique that accounts for both stellar and dust emission from the UV through far IR. By comparing these quantities to intracluster medium (ICM) cooling times and free-fall times derived from X-ray observations and lensing estimates of the cluster mass distribution, we discover a tight relationship between the BCG SFR and the ICM cooling time to free-fall time ratio, $t_{cool}/t_{ff}$, with an upper limit on the intrinsic scatter of 0.15 dex. Furthermore, starburst durations may correlate with ICM cooling times at a radius of $0.025R_{500}$, and the two quantities converge upon reaching the Gyr regime. Our results provide a direct observational link between the thermodynamical state of the ICM and the intensity and duration of BCG star formation activity, and appear consistent with a scenario where active galactic nuclei (AGN) induce condensation of thermally unstable ICM overdensities that fuel long-duration ($>$ 1 Gyr) BCG starbursts. This scenario can explain (a) how gas with a low cooling time is depleted without causing a cooling flow and (b) the scaling relationship between SFR and $t_{cool}/t_{ff}$.  We also find that the scaling relation between SFR and dust mass in BCGs with SFRs $<100$ M$_{\odot}$ yr$^{-1}$ is similar to star-forming field galaxies; BCGs with large ($>100$ M$_{\odot}$ yr$^{-1}$) SFRs have dust masses comparable to extreme starbursts.
\end{abstract}


\section{Introduction}\label{sec-intro}

\renewcommand*{\thefootnote}{\arabic{footnote}}
\setcounter{footnote}{0}

The brightest cluster galaxy (BCG) in the center of the Perseus cluster, NGC1275, was long known to be an emission-line system with ionized hydrogen gas spanning $\sim 100$ kpc \citep{1970Lynds_NGC1275}. While BCGs are typically quiescent systems, the correlation of emission-line BCGs inside so-called ``cooling flow'' clusters was noted by the early studies of X-ray clusters and the optical spectra of their BCGs \citep[e.g.][]{1981Heckman_BCGLines, 1983Cowie_CoreSpecPhot, 1985Hu_CoreSpec, 1987Johnstone_BCGSpectra}. Such clusters, more recently termed ``cool core'' clusters since their central gas X-ray temperatures are somewhat cooler than their outskirts, have highly peaked central X-ray surface brightness profiles arising from relatively high central gas densities. Initially, it was thought that since the cooling time for this gas (the ratio of the thermal energy content to the radiative loss rate) was short compared to the Hubble time, that such gas would cool, lose pressure, and gradually allow more gas to settle into the center of such a cluster: a ``cooling flow''. The rates inferred were 100-1000 M$_{\odot}$ yr$^{-1}$ in some cases (see, e.g. \cite{1994Fabian_CoolingFlow} for a review). Subsequent high-resolution X-ray spectroscopy from XMM-Newton convincingly showed that this simple model was incorrect \citep[e.g.][]{2003Peterson_CoolingFlow, 2006Peterson_CoolingFlow}. Nevertheless, observations of the BCGs in samples of X-ray luminous galaxy clusters have revealed that in up to 70$\%$ of cool-core clusters, the otherwise quiescent elliptical BCG shows signs of ultraviolet and nebular line emission. This emission, in some cases, is consistent with star formation rates (SFRs) on the order of 100 M$_{\odot}$ yr$^{-1}$ \citep[e.g.][]{1989McNamara_CCSFR, 1999Crawford_BCS, 2007Edwards_BCGLines, 2010Donahue_REXCESS}. This activity is observed to be related to the presence of low entropy intracluster gas (the intracluster medium, or ICM) in the cluster core. Since low-entropy ICM gas has a short cooling time, this correlation suggests that the observed star formation is fueled by cold gas that has condensed from a hot gas reservoir, however at a far gentler rate than the simple cooling flow model predicted \citep{2006Rafferty_Feedback, 2008Rafferty_BCGSF, 2008Cavagnolo_Entropy, 2009Cavagnolo_ACCEPT, 2010McDonald_Ha, 2011Hoffer_SFREntropy}. One viable source of heat replacing energy lost via radiative cooling is energy injected by active galactic nuclei (AGN) into the ICM.  The mechanical work indicated by the size and ICM gas pressure surrounding X-ray cavities filled by radio-emitting plasma demonstrates that there is sufficient energy to counter radiative cooling \citep{2006Rafferty_Feedback, 2007McNamara_AGNFeedback, 2012Fabian_AGNFeedback, 2012McNamara_MechanicalFeedback}.

There is a growing body of evidence supporting AGN feedback-driven precipitation and condensation as being responsible for balancing heating and cooling in cool-core galaxy clusters. AGN jet feedback is thought to inject energy into the ICM, thus offsetting cooling while also triggering condensation of thermodynamically unstable volumes of ICM plasma, which then precipitate onto the BCG and fuel star formation and further AGN activity  \citep{2014Voit_Feedback, 2015Voit_FeedbackACCEPT, 2014Li_ColdClumps, 2015Li_SFAGN, 2012Gaspari_Feedback, 2013Gaspari_CCA, 2015Gaspari_AGNFeedback, 2015Fogarty_CLASH, 2016Gaspari_Condensation, 2015Prasad_Feedback, 2016Yang_Feedback, 2016Meece_AGNFeedback}. Recently, the precipitation aspect of this model was shown to be consistent with observations of molecular gas accretion onto the AGN in the Abell 2597 BCG \citep{2016Tremblay_Accretion}. The morphologies of UV and nebular emission structures in observed BCGs in cool core clusters \citep[e.g.][]{2015Donahue_CLASH, 2015Tremblay_BCGUV} are reproduced in simulations of AGN feedback-regulated condensation and precipitation in low-entropy ICM gas \citep{2015Li_SFAGN, 2017Gaspari_Condensation}. Furthermore, submillimeter observations of cool core clusters reveal reservoirs of as much as $\sim 10^{11}$ M$_{\odot}$ of molecular gas \citep{2002Edge_Molecular, 2008ODea_BCGSF, 2016Russell_Phoenix}. The picture that is emerging is one of a complex feedback-driven interaction between the BCG and ICM in the cluster core which produces substantial gas condensation but suppresses runaway ICM cooling. 

The condensation model follows thermal instabilities that are both triggered and regulated by AGN feedback. In this model, the hot ICM gas is near hydrostatic equilibrium with the gravitational potential, and the gas entropy at large cluster-centric distances is governed by cosmological processes.  At very small radii close to the central AGN, the gas can develop a nearly flat entropy profile that allows thermal instabilities to rapidly grow. However, \cite{2016Meece_AGNFeedback} showed that even a modest fraction of AGN feedback in the form of mechanical jets can transport energy beyond the region local to the AGN. Rather than resulting in unphysical catastrophic cooling, such systems can self-regulate. The resulting entropy profile within 5-10 kpc of the simulated AGN with jet feedback is somewhat shallower than in the outer parts of the cluster, but condensation of low-entropy inhomogeneities which have been uplifted to greater altitudes can produce multi-phase structure at radii larger than 10 kpc \citep{2016Meece_AGNFeedback, 2016Voit_Model}. Key to this model is the prediction that ICM plasma with a cooling-to-freefall time ratio ($t_{cool}/t_{ff}$) $\lesssim 10$ is sufficiently unstable that jets trigger condensation. Supporting this prediction are observations indicating that $t_{cool}/t_{ff} \leq 20$ in the center of a galaxy cluster is a good predictor of BCG activity \citep{2015Voit_FeedbackACCEPT, 2016Voit_Model}. In simulations, material will condense out of the ambient ICM when $t_{cool}/t_{ff} \lesssim 10$ locally \citep{2014Li_ColdClumps, 2015Li_SFAGN}.

Alternatively, \cite{2016McNamara_Feedback} posit that most cold molecular gas condenses in the vicinity of BCGs when AGN jets uplift low entropy plasma from within the BCG to high enough radii for the plasma to condense. In this scenario, jet uplifting can cause core plasma to condense if drag prevents it from sinking back to its original altitude in a time shorter than the cooling time of the gas. Observations of massive (up to $10^{10}$ M$_{\odot}$) flows of molecular gas with velocities of several hundred km s$^{-1}$ in cool core clusters such as Abell 1835 and Abell 1664 suggest that this mode of condensation and precipitation plays an important role in AGN-regulated feedback  \citep{2014McNamara_A1835, 2014Russell_A1664}. Both condensation due to uplifting and condensation in the ambient ICM in a cluster core are studied in models of cluster-scale feedback \citep{2015Li_SFAGN, 2016Gaspari_Condensation, 2016Voit_Model}. 
 
\cite{2016Molendi_Cooling} propose a possible scenario where star formation in BCGs lags behind cooling by up to a Gyr, owing to molecular gas in the environs of the BCG forming stars at only a few percent efficiency, as one possible explanation for the presence of star forming BCGs in the absence of cooling flows. This scenario can explain relationships between BCG SFRs and the thermodynamical state of the ICM, but would imply that the density and temperature of the hot ICM in the vicinity of the BCG is related to the amount of gas that had cooled in a previous cooling phase. 

In this paper, we analyze the spectral energy distributions of the UV-luminous BCGs in the full X-ray selected cluster sample from the CLASH\footnote{Cluster Lensing And Supernova survey with \textit{Hubble}} program. Our aim is to explore physical connections between BCG star formation activity and the properties of the surrounding ICM and constrain models for possible mechanisms of condensation in the ICM. We take full advantage deep, multi-band imaging of the relatively uniform and massive sample of galaxy clusters observed by CLASH and we are able to reveal relationships that may have previously been confounded by analyses of less deeply observed and/or less uniform samples. Specifically, by studying the BCGs of cool-core clusters in CLASH, we explore how cooling and thermal instability timescales ($t_{cool}$ and $t_{cool}/t_{ff}$) in the ICM may relate to BCG star formation.

Our paper is structured as follows: in Section~\ref{sec-obs} we describe the observational data set used to construct the spectral energy distributions (SEDs) for the 11 active CLASH BCGs studied in this sample. In Section~\ref{sec-methods}, we describe constructing and fitting UV-through-FIR SEDs, as well as the X-ray derived parameters we use to study the ICM. In Section~\ref{sec-results} we present our results, which we discuss in Section~\ref{sec-discuss}. We summarize our conclusions in Section~\ref{sec-conclude}. Throughout our analysis we adopt a $\Lambda$CDM cosmology with $\Omega_{m} = 0.3$, $\Omega_{\Lambda} = 0.7$, $H_{0}$ = 70 km/s/Mpc, and $h = 0.7$.

\section{Observations}\label{sec-obs}
 
The near UV through far IR SEDs in this paper are based on photometry from the CLASH \textit{HST} data set in combination with mid- and far-IR data from \textit{Spitzer} and \textit{Herschel}. X-ray data used for measuring the temperature, density and metallicity profiles of the ICM were taken from the \textit{Chandra} archive.

\subsection{HST: UV through near IR photometry}\label{sec-hstphot}
 
A detailed summary of the science-level data products for CLASH may be found in \cite{2012Postman_CLASH}. For the 11 X-ray selected clusters with evidence of BCG star formation activity, we used the CLASH photometric data set covering 16 filters spanning from $\sim$2000-17000 $\textrm{\AA}$ in the observer frame. We used drizzled mosaics with a 65 milliarcsecond pixel scale, the same image data used in \cite{2015Fogarty_CLASH}. These data are publicly available via MAST HLSP \footnote{https://archive.stsci.edu/prepds/clash/}. Drizzled mosaics were constructed using the {\tt MosaicDrizzle} pipeline \citep{2011Koekemoer_Drizzle}. In keeping with our previous work, we calculated a single Milky Way foreground reddening correction for each BCG in each filter using the \cite{1998Schlegel_Dust} dust maps. Our drizzled images are background corrected using an iterative 3-sigma clipping technique. Owing to additional uncertainty in the flat-fielding of WFC3/UVIS photometry over large spatial scales, discussed in \cite{2015Fogarty_CLASH}, the median flux measured in an annulus around each BCG was subtracted from each of the UV filters as well.

\subsection{Spitzer: mid-IR photometry}\label{sec-mirphot}
 
\textit{Spitzer}/IRAC 3.6 $\mu$m and 4.5 $\mu$m mosaicked observations and catalogs are available for all of the BCGs studied in this paper, and 5.7 $\mu$m and 7.9 $\mu$m observations are available for Abell 383, MACS1423.8+2404, and RXJ1347.5-1145. Fluxes for \textit{Spitzer} IR sources in the CLASH fields were taken from the publicly available CLASH/\textit{Spitzer} catalog \footnote{http://irsa.ipac.caltech.edu/data/SPITZER/CLASH/}. We describe aperture selection and our method for correcting for crowded fields in the \textit{Spitzer} photometry in Section~\ref{sec-bcgphot}.
 
Photometry for each channel was measured on mosaic images generated using the {\tt MOPEX} software package. The default {\tt MOPEX} settings for the catalog use the 3.6 $\mu$m channel, and use the fiducial image frame for this channel to generate the mosaics for longer wavelength IRAC channels for each CLASH cluster. The \textit{Spitzer} mosaic images have a pixel scale of 0.6 arcseconds. Flux values were obtained with {\tt Source Extractor} in double-image mode using {\tt MOPEX} weights \citep{1996Bertin_SE}.  The {\tt Source Extractor}-generated catalogs consist of aperture photometry for sources in fixed-diameter apertures ranging from 2 to 40 pixels. Full details about the parameters used with {\tt MOPEX} and {\tt Source Extractor} to generate the CLASH catalog are given in the online documentation\footnote{http://irsa.ipac.caltech.edu/data/SPITZER/CLASH/docs/ \\ README.CLASHSpitzer}.

\subsection{Herschel: far-IR photometry}\label{sec-firphot}
 
We used archival data from \textit{Herschel}/PACS (\textit{100 $\mu$m, 160 $\mu$m}) and \textit{Herschel}/SPIRE (250 $\mu$m, 350 $\mu$m, and 500 $\mu$m) to extend BCG photometry into the far-infrared.  Table \ref{table:Herschel} details the observations used and their exposure times. We obtained level 2 archival data using the \textit{Herschel} Science Archive reduction pipeline for SPIRE observations and the HSA MADMAP reduction pipeline for PACS observations. Observations were co-added and photometric parameters were measured using the {\tt HIPE} software package. We describe  \textit{Herschel} aperture selection and background subtraction in Section~\ref{sec-bcgphot}.  

\begin{table}[]  
\footnotesize  
\caption{\textit{Herschel} Observations}
\label{table:Herschel}  
\vspace{1mm}  
\centering  
{  
\begin{tabular}{llcc} 
\hline
\hline 
 & Instrument & Observation ID & Exposure Time \\
BCG & & & (seconds) \\   
\hline  
Abell 383 & PACS & 1342189151 & 7704 \\
& & 1342189152 & 7704 \\
& & 1342189153 & 7185 \\
& & 1342189154 & 7185 \\
& SPIRE & 1342189503 & 5803 \\
& & 1342201147 & 3172 \\
\hline
MACS0329.7$-$0211  & PACS & 1342249280 & 7704 \\
& & 1342249281 & 7704 \\
& SPIRE &  1342214564 & 169 \\
& & 1342239844 & 1411 \\
\hline 
MACS0429.6$-$0253 & PACS & 1342250641 & 7704 \\
& & 1342250836 & 7704 \\
& SPIRE & 1342239932 & 169 \\
& & 1342241124 & 1411 \\
\hline
MACS1115.9+0219	 & PACS & 1342247672 & 7704 \\
& & 1342247691 & 7704 \\
& SPIRE & 1342223226 & 169 \\
& & 1342256866 & 1411 \\
\hline
MACS1423.8+2404	 & PACS & 1342188215 & 9850 \\
& & 1342188216 & 9850 \\ 
& SPIRE & 1342188159 & 6636 \\
\hline
MACS1720.3+3536	 & PACS & 1342243800 & 7704 \\
& & 1342243801 & 7704 \\
& SPIRE & 1342229601 & 169 \\
& & 1342239976 & 1411 \\
\hline
MACS1931.8$-$2653  & PACS & 1342241619 & 7704 \\
& & 1342241681 & 7704 \\
& SPIRE & 1342215993 & 169 \\
& & 1342254639 & 1411 \\
\hline
MS2137$-$2353  & PACS & 1342187803 & 9850 \\
& & 1342187804 & 9850 \\
& SPIRE & 1342195938 & 5786 \\
\hline
RXJ1347.5$-$1145  & PACS & 1342213836 & 9420 \\
& & 1342213837 & 9420 \\
& SPIRE & 1342201256 & 859 \\
& & 1342201257 & 859 \\
& & 1342201258 & 859 \\
& & 1342201259 & 859 \\
& & 1342201260 & 859 \\
& & 1342201261 & 859 \\
& & 1342201262 & 859 \\
& & 1342201263 & 859 \\
& & 1342247859 & 1952 \\
& & 1342247860 & 1952 \\
& & 1342247861 & 1952 \\
\hline
RXJ1532.9+3021  & PACS & 1342258435 & 7704 \\
& & 1342258435 & 7704 \\
& SPIRE & 1342234776 & 169 \\
& & 1342261681 & 1411 \\
\hline
RXJ2129.7+0005  & PACS & 1342187256 & 9569 \\
& & 1342187257 & 9569 \\
& & 1342196791 & 7185 \\
& & 1342196792 & 7185 \\
& SPIRE & 1342188167 & 6636 \\
& & 1342209312 & 3172 \\
\hline
\end{tabular}  
\begin{flushleft}
\end{flushleft}  
}  
\end{table}

\subsection{Chandra: X-ray Observations}\label{sec-xray}
 
\textit{Chandra} data exists for all CLASH clusters, with exposure times for individual observations ranging from 19.3 ks (for Abell 383) to 115.1 ks (for MACS1423.8+2404). These data were assembled in \cite{2014Donahue_CLASHX} to construct the parameter profiles that we use to analyze the thermodynamics of the ICM in the environs of each BCG. We also examine the archival data for evidence of X-ray loud AGN, in order to determine whether it is important to consider AGN emission models in our SED fits. We found evidence for only one X-ray loud AGN, in the cluster MACS1931.8-2635. The AGN classification is based on the presence of a \textit{Chandra} point source, with a spectrum showing evidence for hard X-ray ($>$5 keV) emission over that expected from hot gas with $kT\sim 5-7$ keV.

\section{Methods}\label{sec-methods}

In \cite{2015Fogarty_CLASH}, we identified 11 BCGs in the CLASH X-ray selected sample with evidence of ongoing star formation. In each of these BCGs, we detected extended UV emission in WFC3/UVIS photometry and extended H$\alpha$ + [\ion{N}{2}] emission by differencing ACS images. The extended UV and line emission features were shown in \cite{2015Donahue_CLASH} and \cite{2015Fogarty_CLASH} to be the site of nebular emission and star formation. These 11 BCGs form the star-forming BCG sample we study in this paper. 

\subsection{BCG Photometry}\label{sec-bcgphot}
 
We constructed multi-instrument UV-through-IR SEDs for the UV-bright filamentary features in each UV-luminous CLASH BCG. Since we are interested in measuring the star formation properties of the BCGs, for the \textit{Hubble} UV through near-IR photometry we extracted flux from within the apertures described in \cite{2015Fogarty_CLASH} (apertures for each BCG are shown in Figure 3 in that paper). These apertures were selected to encompass the region in each BCG exhibiting a UV luminosity of $7.14\times 10^{24}$ erg s$^{-1}$ Hz$^{-1}$ pix$^{-2}$, corresponding to a star formation surface density of $\Sigma_{\rm SFR}$ $\geq 0.001$ M$_{\odot}$ yr$^{-1}$ pix$^{-2}$, after accounting for dust reddening. Measuring fluxes in these apertures maximizes the contribution made by the recently formed stellar population to the SED, while it minimizes the contribution made by the passive stellar population in the bulk of the BCGs. Our procedure for measuring fluxes minimizes the risk of underestimating dust attenuation in the star forming regions of the BCGs since we  are not averaging dust attenuation over the dusty star-forming and relatively dust-free quiescent parts of the galaxy. We do not match the \textit{Hubble} point spread functions (PSF) since the photometric aperture sizes we use are much larger than the sizes of the PSFs for the \textit{HST} passbands.

\begin{figure}
\begin{center}
\begin{tabular}{c}
\includegraphics[height=9cm]{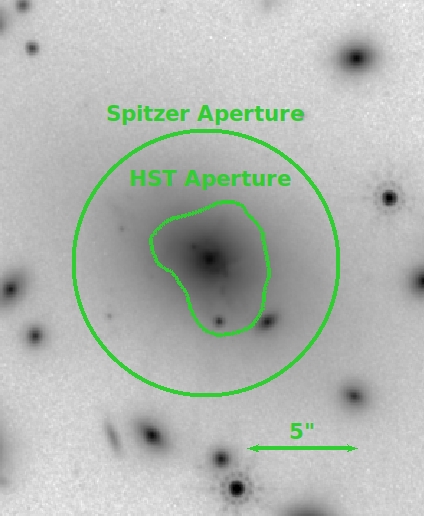}
\end{tabular}
\end{center}
\caption[]
{ \label{fig:Spitzer_Ap} The UV-bright \textit{Hubble} aperture and \textit{Spitzer} photometric aperture for RXJ1532.9+3021 are shown on logarithmically scaled F160W band photometry of the cluster. The \textit{Spitzer} aperture encompasses the \textit{Hubble} aperture, and includes flux from the old stellar population in the BCG as well as contaminating light from nearby early-type satellite galaxies that needs to be modeled and subtracted to estimate the mid-IR flux within the \textit{Hubble} aperture.}
\end{figure}

Estimating mid-IR fluxes using \textit{Spitzer} was more complicated, since star-forming features are not well resolved spatially in \textit{Spitzer} photometry. We measured IR fluxes from \textit{Spitzer} in apertures in the CLASH/\textit{Spitzer} catalogs that were selected to encompass the BCG while minimizing inclusion of satellite galaxies. Both the old stellar population and dust emission (primarily in the form of PAH features) contribute significantly to the flux in these filters. Since the angular resolution of \textit{Spitzer} ($>$ 1.45 arcsec) is insufficient to resolve the star forming features we wish to study, we needed to subtract the contribution made by old stellar light from outside our \textit{Hubble} apertures to the \textit{Spitzer} IR fluxes. We accomplished this by first measuring the {\it HST}/WFC3 F160W flux in the region covered by the \textit{Spitzer} apertures but not the \textit{Hubble} apertures. We assumed this flux is due entirely to old stellar light, and estimated (IRAC-F160W) color by modeling old stellar populations experiencing dust attenuation $\leq 0.5$ A$_{\textrm{V}}$ using the Bayesian SED fitting algorithm {\tt iSEDfit} (see Section~\ref{sec-sedfit} for details) for each BCG. We used these colors to scale the F160W fluxes corresponding to the old stellar population outside the {\it Hubble} apertures to {\it Spitzer} IRAC fluxes and subtracted these scaled fluxes from our \textit{Spitzer} photometry. The resulting \textit{Spitzer} IRAC photometry corresponds to the fluxes in the \textit{Hubble} apertures. An example pair of {\it Spitzer} and {\it Hubble}  apertures is shown in Figure \ref{fig:Spitzer_Ap}.

We note that H$_{2}$ vibrational modes and other near-IR emission lines have been detected between 5-25$\mu$m in star-forming BCGs \citep{2011Donahue_H2}. However, contamination from emission lines similar to those observed in \cite{2011Donahue_H2} would only contribute a few percent to the mid-IR fluxes derived for the stellar and dust emission in star forming regions, while the uncertainties on these fluxes after subtracting the excess old stellar light component is $\gtrsim 20\%$. Therefore, we do not attempt to estimate the contribution of these lines to the \textit{Spitzer} mid-IR photometry.
 
We extracted \textit{Herschel} PACS and SPIRE photometry from archival data using the {\tt HIPE} software package. Since \textit{Herschel} bands are dominated by dust re-emission of UV flux, we assumed that all of the \textit{Herschel} flux in each BCG comes from the star forming features. We measured photometry in circular apertures large enough to cover the PSF of each \textit{Herschel} filter. Aperture radii were chosen to be 8$"$, 12$"$ , 18$"$ , 24$"$ , and 36$"$ for the 100 $\mu$m, 160 $\mu$m, 250 $\mu$m, 350 $\mu$m, and 500 $\mu$m filters respectively. To obtain fluxes we took  the mean of pairs of cross scans, while we estimated uncertainties using the difference between scans. For each \textit{Herschel} band, we found typical uncertainties on the order of $\lesssim 2\%$ for the PACS filters and $\sim 10\%$ for the SPIRE filters. Since the BCGs occupy crowded fields, we measured crowded source backgrounds in annuli centered on the BCG with inner radii of 16$"$, 24$"$, 36$"$, 48$"$, 72$"$ for the respective PACS and SPIRE filters and outer radii of 180$"$.
 
 \subsubsection{Photometric Errors}\label{sec-photerr}
 
Error budgets for the fluxes in our SEDs include uncertainty from counting statistics, systematic uncertainties from weight maps (produced as output from Drizzling in the case of the \textit{Hubble} data and provided by the level 2 data pipeline in the case of the \textit{Herschel} PACS and SPIRE data), and the absolute calibration uncertainty for each instrument. In the case of the \textit{Spitzer} IRAC data, the total counting and formal systematic uncertainties are available in the CLASH/\textit{Spitzer} catalogs from the fluxes and {\tt MOPEX} weight maps (see Section~\ref{sec-mirphot}). Calculating uncertainties for the mid-IR fluxes obtained using {\tt Spitzer} IRAC photometry required propagating the additional uncertainty involved in estimating and subtracting the excess old stellar light component from the mid-IR photometry. 
 
Source confusion noise is an important component of the error budget for SPIRE detections \citep{2010Nguyen_Confusion}. We incorporate confusion noise terms of 5.8 $\mu$Jy, 6.3 $\mu$Jy, and 6.8 $\mu$Jy from \cite{2010Nguyen_Confusion} into our estimates of the uncertainty on SPIRE 250 $\mu$m, 350 $\mu$m, and 500 $\mu$m fluxes respectively. In bands where we do not detect significant flux from the BCG, we use the confusion and instrument noise estimates computed in \cite{2010Nguyen_Confusion} to estimate 3$\sigma$ upper limits.
 
For \textit{HST} ACS and WFC3 data, we included a 5$\%$ total absolute and relative calibration uncertainty in our error budget. For the ACS and WFC3/IR filters, the absolute calibration is the dominant term in the overall error budget. Since several of the early CLASH WFC3/UVIS observations were affected by non-uniform flat-fielding at large (hundreds to thousands of pixels) scales, we calculated the scatter in identical apertures placed in empty patches of sky for each UV observation. This extra uncertainty component in the WFC3/UVIS error budget is in most cases similar to the absolute calibration uncertainty, although for several filters with only faintly detected BCG UV emission it is the main source of uncertainty. 
 
We adopted an additional 3$\%$ uncertainty for the \textit{Spitzer} IRAC fluxes, and 10$\%$ uncertainty for \textit{Herschel} PACS and SPIRE fluxes. The 3$\%$ figure accounts for the absolute flux calibration uncertainty in \textit{Spitzer} \citep{2005Reach_IRAC}. For \textit{Herschel},  we incorporate the absolute calibration uncertainty ($\sim 5\%$), the relative calibration uncertainty ($\sim 2\%$), and allow for an additional factor of $\sim 2$ applied to the systematic uncertainty to account for the fact that we are measuring the fluxes of marginally extended sources and that our method of estimating the uncertainty in the \textit{Herschel} images may underestimate the uncertainty \citep{2013Griffin_SPIRE, 2014Balog_PACS}.  

For most of the BCGs, the error budget for the \textit{Hubble} photometry is dominated by the absolute calibration uncertainty, the \textit{Spitzer} error budget is dominated by the uncertainty from modeling and subtracting the old stellar excess, the \textit{Herschel} PACS uncertainties are dominated either by counting statistics or the calibration, and the \textit{Herschel} SPIRE uncertainties are dominated by the calibration uncertainty or the confusion noise.

\subsection{SED Fitting}\label{sec-sedfit}
 
We fit SEDs for the CLASH BCGs using the Bayesian fitting code {\tt iSEDfit} \citep{2013Moustakas_iSEDFIT}. In order to take full advantage of the CLASH UV-FIR observations, we incorporated modifications to {\tt iSEDfit} to account for emission from dust. The {\tt iSEDfit} package fits SEDs by first generating a grid of parameters that describe models of dust--attenuated stellar emission by a composite stellar population. Grids are produced by randomly sampling a bounded section of parameter space. The sampling is weighted with either a uniform prior or a logarithmic scale prior. With these parameters, {\tt iSEDfit} constructs a grid of model spectra given a choice of synthetic stellar population (SSP), initial mass function (IMF), and dust attenuation model. The package then computes synthetic photometry by convolving these model spectra with the filter responses of the CLASH SEDs, and uses this grid of synthetic SEDs to sample the posterior probability distribution that the model stellar population and dust parameters produce the observed SED.  

The {\tt iSEDfit} package uses a Bayesian Monte Carlo approach to estimate probability distributions for model parameters. For each BCG, we constructed a model grid consisting of 20000 models. The likelihood and mass scaling $\mathcal{A}$ of each model given the observed SED was calculated by minimizing $\chi^{2}$ for each model,
\begin{equation}
\chi^{2} = \sum_{i=1}^{N}{\frac{\left(F_{i} - \mathcal{A}M_{i}\right)^{2}}{\sigma_{i}^{2}}},
\end{equation}
where $F_{i}$ and $\sigma_{i}$ are the SED fluxes and uncertainties, and $M_{i}$ are the model fluxes. The minimum $\chi^{2}$ and $\mathcal{A}$ are found by solving for $\partial \chi^{2} / \partial \mathcal{A} = 0$. We obtained a posterior probability distribution for the models using a weighted random sampling of the model grid, where model weights were determined by the likelihoods. Posterior probability distributions for individual physical parameters were obtained by taking the distribution of parameters of the sampled models. For the SFR, the posterior consists of the distribution of the instantaneous normalized SFR determined by the parameterized star formation history, multiplied by $\mathcal{A}$. For a detailed discussion of {\tt iSEDfit}, see \cite{2013Moustakas_iSEDFIT}.

\begin{figure}
\begin{center}
\begin{tabular}{c}
\includegraphics[height=6.5cm]{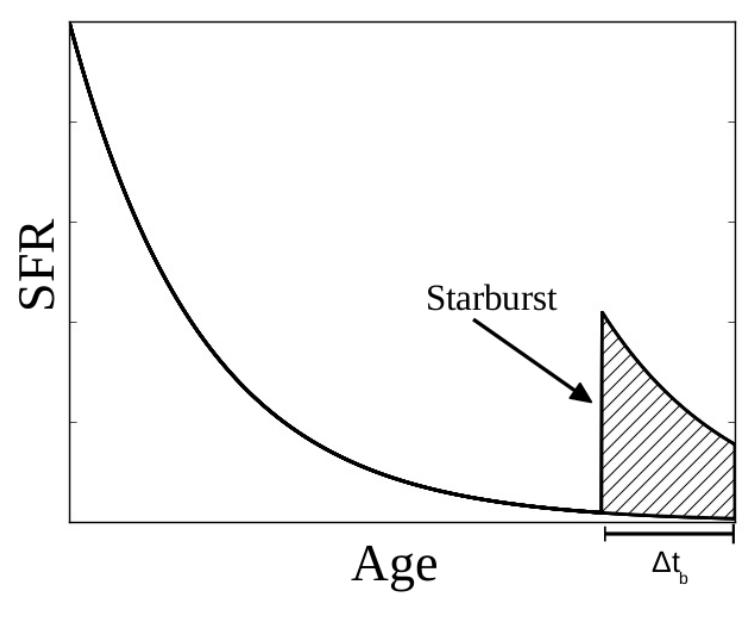}
\end{tabular}
\end{center}
\caption[]
{ \label{fig:SFH_Example} A schematic of the double-exponential star formation history adopted in this paper. The solid line shows the SFR as a function of time. The horizontal line under the lower-right part of the x-axis depicts the starburst duration $\Delta t_{b}$, which is the amount of time since the onset of the BCG starburst. The area under the curve is equal to the total mass of stars formed by the BCG, and the portion of the area highlighted with hatching is the contribution of the starburst to the BCG stellar mass.}
\end{figure}

We model the star formation history of each BCG with an exponentially decaying curve (the early-type population) and a super-imposed exponentially decaying burst at recent times. The initial exponentially decaying curve is parameterized by the age of the BCG $t$ and the decay rate of the curve $\tau$, while the exponentially decaying burst is parameterized by the duration of the starburst $\Delta t_{b}$, the decay rate of the burst, and the mass of the burst relative to the old stellar population (see Figure \ref{fig:SFH_Example}). The bounds on parameter space for the entire model and assumptions we made for the BCG stellar populations are given in Table \ref{table:SED_Params}.

\begin{figure}
\begin{center}
\begin{tabular}{c}
\includegraphics[height=9cm]{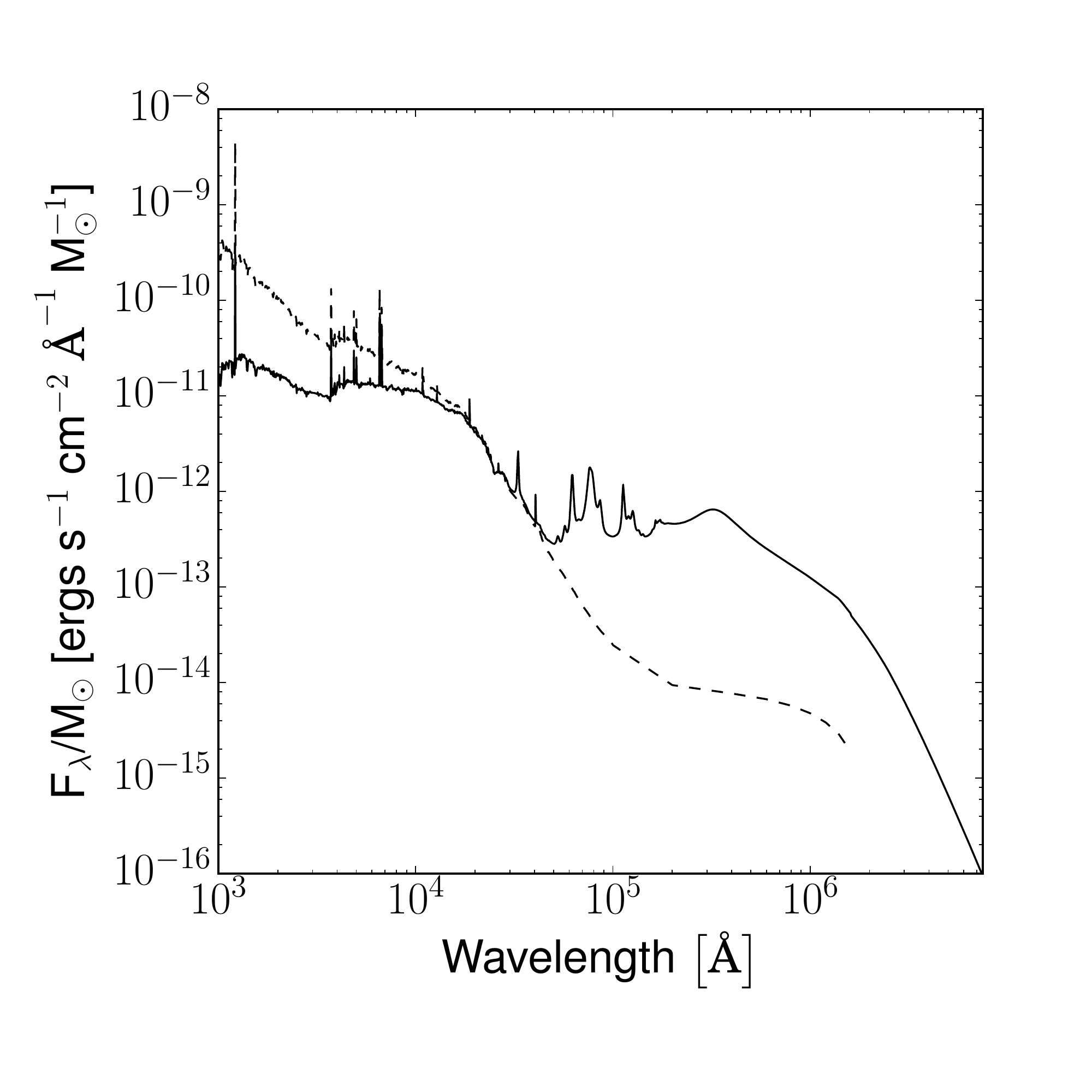}
\end{tabular}
\end{center}
\caption[]
{ \label{fig:Synthetic_Spectrum} An example synthetic spectrum produced by our modified version of {\tt iSEDfit}. The dashed line shows the model stellar emission spectrum in the absence of dust. The solid line shows the model spectrum after the stellar emission has been absorbed by dust obeying a modified \cite{2000Calzetti_Extinction} attenuation law and re-emitted in the far-IR.}
\end{figure}

We modified the SED--fitting routine to incorporate dust emission models from \cite{2007Draine_Dust} into the parameter grid and synthetic spectra used in {\tt iSEDfit}. The dust parameter space was sampled using bounds and priors given in Table \ref{table:SED_Params}. For each synthetic spectrum, the total luminosity of the dust spectrum was normalized by the difference between the un-attenuated and attenuated stellar spectrum. Figure \ref{fig:Synthetic_Spectrum} shows an example of the synthetic attenuated stellar plus dust spectrum, and the Appendix shows best-fit synthetic spectrum for each BCG.

With our modified version of {\tt iSEDfit}, we were able to test multiple stellar population and dust attenuation models in order to examine the dependence of our results on the choice of SSP, IMF, and attenuation law. We adopt a \cite{2003Bruzual_BC03} SSP, the \cite{1955Salpeter_IMF} IMF defined over the interval 0.1-100 M$_{\odot}$, and the \cite{2000Calzetti_Extinction} attenuation law. We chose to adopt a \cite{1955Salpeter_IMF} IMF in order to produce SFRs consistent with the SFRs estimated in our previous paper using the \cite{1998Kennicutt_SFR} relationship. In order to examine the effect of an $A_{\textrm{V}}$-dependent attenuation curve, we also performed fits using a modified Calzetti attenuation law, where we multiplied the Calzetti curve with an attenuation-dependent slope that matches the attenuation dependence of the curves published in \cite{2000Witt_Extinction}. To test if our results have any dependence on our choice to adopt a Calzetti attenuation law, we also ran {\tt iSEDfit} assuming clumpy SMC-like dust in a shell geometry. We found our results to be largely consistent with results using the regular Calzetti curve, as well as the attenuation-dependent version of the Calzetti curve. This is because, in the typical CLASH BCG, the attenuation is $A_{\textrm{V}} \lesssim 1.0$, where overall attenuation dependence on the shape of the curve has a only a modest effect. As a further test of the model dependence of our results, we performed SED fits assuming a \cite{2003Chabrier_IMF} IMF. Altering the IMF shifts the SFRs downwards by 0.25 dex, but otherwise does not significantly alter the stellar and dust parameters we seek to measure ($\Delta t_{b}$, M$_{d}$, A$_{\textrm{V}}$).

\begin{table*}[]
\footnotesize
\caption{SED Fitting Parameters}
\label{table:SED_Params}
\vspace{1mm}
\centering
{
\begin{tabular}{lrrr}
\hline
\hline
\textit{Stellar Population Model} \\
\hline
Synthetic Stellar Population & \cite{2003Bruzual_BC03} \\
Initial Mass Function & \cite{1955Salpeter_IMF} \\
Attenuation Law & \cite{2000Calzetti_Extinction} \\
Dust Emission & \cite{2007Draine_Dust} \\
\hline
\hline
                                                                       & \textit{Minimum} & \textit{Maximum} & \textit{Sampling}\\
\textit{Model Parameter Space Constraints} & \textit{Value} & \textit{Value} & \textit{Interval}\\
\hline
\textit{Old Stellar Population} \\
Age, $t$ & 6 Gyr & $z_{Age}^{a}$ & Linear \\
Decay Rate, $\tau$ & $0.05t$ & $0.2t$ & Linear \\
Metallicity & $3 \times 10^{-2} Z_{\odot}$ & $1.5 Z_{\odot}$ & Linear \\
 & & & \\
\textit{Burst Population} \\
Burst Duration, $\Delta t_{b}$ & $10^{-2}$ Gyr & 5.0 Gyr & Logarithmic$^{b}$ \\
Burst Decay Percentage & 0.01 & 0.99 & Linear \\
Relative Burst Mass, {\sc fburst}$^{c}$ & 0.0016 & 6.4 & Logarithmic \\
 & & & \\
\textit{Dust Parameters} \\
Attenuation A$_{V}$ & 0 & 2 & Linear \\
PAH Abundance Index $q_{PAH}$ & 0.10 & 4.58 & Linear$^{d}$ \\
$\gamma^{e}$ & 0.0 & 1.0 & Linear \\
$U_{min}^{e}$ & 0.10 & 25.0 & Logarithmic \\
$U_{max}^{e}$ & $10^{3}$ & $10^{7}$ & Logarithmic \\ 
 & & & \\
\hline
\end{tabular}  
\begin{flushleft}
$^{a}$ $Z_{Age}$ is the age of the universe at redshift the BCG redshift $Z$. \\
$^{b}$ Burst parameters were sampled logarithmically, since their qualitative effect on the model SED of the galaxy occurs on order-of-magnitude scales. The exception to this is the burst decay percentage, which is one minus the amplitude of current star formation activity relative to the amplitude of the burst $\Delta t_{b}$ yr ago. \\
$^{c}$ Mass of stars created by the starburst at $t$ relative to the mass of stars created by the exponentially decaying old stellar population at $t$. The burst mass percentage is calculated by {\sc fburst}/(1+{\sc fburst}). \\ 
$^{d}$ \cite{2007Draine_Dust} model parameters sampling intervals were chosen based on the model parameter distributions of the template spectra. \\ 
$^{e}$ The \cite{2007Draine_Dust} treats dust in a galaxy as consisting of two components. The first component consists of a fraction $\gamma$ of the dust is exposed to a power law distribution of starlight intensity, ranging from $U_{min}$ to $U_{max}$, while the second component consists of the remainder of the dust, and is only exposed to a starlight intensity $U_{min}$. $U$ is defined to be the intensity of starlight relative to the local radiation field, and $U_{min}$ and $U_{max}$ are bounds on the distribution of $U$.\\ 
\end{flushleft}  
}  
\end{table*}

\subsection{Cooling and Freefall Time Profiles}\label{sec-tprof}
 
We calculated radial profiles of the cooling time, defined to be the ratio of the thermal energy density to the rate of radiative energy density loss for an optically thin plasma undergoing Bremsstrahlung emission, for each cluster using
\begin{equation}\label{eq:tcool}
\begin{split}
t_{cool}\left(r\right) &= \frac{3}{2}\frac{n\left(r\right)k_{B}T\left(r\right)}{n_{e}\left(r\right)n_{H}\left(r\right)\Lambda\left[T\left(r\right), Z\left(r\right)\right]} \\
&= \frac{6.9}{2}\frac{k_{B}T\left(r\right)}{n_{e}\left(r\right)\Lambda\left[T\left(r\right), Z\left(r\right)\right]},
\end{split}
\end{equation}
where $k_{B}$ is the Boltzmann constant, $n\left(r\right)$ is the total number density profile of the plasma, $n_{H}\left(r\right)$ is the H number density, $n_{e}\left(r\right)$ is the electron number density, $T\left(r\right)$ is the temperature profile, $Z\left(r\right)$ the metallicity profile, and $\Lambda\left(T, Z\right)$ is the cooling function. We obtained values of the cooling function for specific temperatures and metallicities by interpolating the cooling function values in \cite{1993Sutherland_Cooling}, and assumed $n \approx 2.3n_{H}$ \citep{2009Cavagnolo_ACCEPT}. The ICM density, temperature, and metallicity profiles used in this study are available in \cite{2014Donahue_CLASHX}. We used the non-parametric Joint Analysis of Cluster Observations (JACO)  profiles (see \cite{2007Mahdavi_JACO} and \cite{2013Mahdavi_JACO} for a description of the JACO algorithm), which are reported in concentric shells spaced so that each annulus contains at least 1500 X-ray counts.

We measured $t_{cool}$ at specific radii by interpolating on $n_{e}\left(r\right)$, $T\left(r\right)$, and $Z\left(r\right)$ at the desired radius and solving Equation \ref{eq:tcool}. In order to determine the uncertainty on $t_{cool}$ we produced an ensemble of 1000 $n_{e}$, $T$, and $Z$ profiles, where the values in each profile were obtained by drawing from normal distributions defined by the observed values and uncertainties in each profile. By calculating the distribution of $t_{cool}$ given the interpolated values of $n_{e}$, $T$, and $Z$ for the ensemble of profiles, we sampled the probability distribution of $t_{cool}$ and thus estimated uncertainties. We constrained the ensemble of metallicity profiles to only allow values of $Z$ in the range $0.15-1.5 Z_{\odot}$, since ICM metallicities are typically $0.3 Z_{\odot}$ and tend to be $\gtrsim 0.6-0.8 Z_{\odot}$ in the centers of cool core clusters \citep{2004DeGrandi_Metallicity}. 
 
We calculated freefall times by assuming cluster masses obey an NFW profile \citep{1997Navarro_NFW}. NFW mass concentration parameters and values of M$_{200}$ were obtained from \cite{2015Merten_ConcMass}. We computed the enclosed mass for each cluster as a function of radius,
\begin{equation}
M_{enc}\left(r\right) = 4\pi\rho_{0}r_{s}^{3}\left[\ln\left(\frac{r_{s}+r}{r_{s}}\right) - \frac{r}{r_{s}+r}\right],
\end{equation}
where $\rho_{0}$ and $r_{s}$ are NFW scale factors determined by the mass and concentration parameter of the galaxy cluster, as well as the critical density at the cluster redshift. We then calculated freefall time profiles, 
\begin{equation}
t_{ff}\left(r\right) = \sqrt{\frac{2r^{3}}{GM_{enc}\left(r\right)}}.
\end{equation}
We estimated the contribution of BCG stellar mass to the free-fall time using the \cite{2016Cooke_CLASH} stellar mass estimates of CLASH BCGs. We did not estimate BCG stellar masses with our SED fits since our photometry does not cover the entire extent of the BCG. Stellar mass density profiles were estimated from the stellar mass of each BCG assuming a \cite{1990Hernquist_Profile} profile and the \cite{2003Shen_SizeMass} mass-size relationship \citep{2009Ruszkowski_BCGSim, 2013Laporte_BCGSim}. The inclusion of stellar mass in our estimates of $M_{enc}\left(r\right)$ does not significantly affect the freefall time profiles at radii $\gtrsim 25$ kpc (or similarly $\gtrsim 0.025 R_{500}$), which is not surprising given that BCG stellar mass dominates the cluster density profile only within the central $\sim 10$ kpc of a cluster \citep{2013Newman_Mass, 2017Monna_MassProfile, 2017Caminha_MassProfile}. Still, we include the BCG stellar mass in our estimation of the free-fall time profiles and find that the BCG stellar masses alter the free fall times by $\lesssim 4\%$ at $\sim 10$ kpc and $\lesssim 1\%$ at $\sim 20$ kpc. 

Comparing the characteristics of BCG starbursts to ICM thermodynamic parameters requires choosing a radius in the ICM profile at which to measure these parameters. We measure ICM parameters at $0.025R_{500}$, which is the smallest fraction of $R_{500}$ that does not require extrapolation of the X-ray profiles for the 11 CLASH clusters studied. We chose this radius since we expect feedback effects to be strongest near the centers of cool cores, and negligible outside the core. At radii outside the cool core the thermodynamical state of the ICM is only weakly tied to feedback and cooling, and therefore should be weakly related to the BCG \citep{1983Cowie_CoreSpecPhot, 1985Hu_CoreSpec}.
  
We also base our expectation of the radial dependence between the properties of the ICM and feedback on BCG simulations in \cite{2015Li_SFAGN}. At large radii ($\gtrsim 100$ kpc), the variation in $t_{cool}/t_{ff}$ with time in the simulation (and with the level of star formation and AGN-driven feedback) is weak, so a relationship between SFR and cluster dynamical state would be difficult to detect (see their Figure 9). On the other hand, $t_{cool}/t_{ff}$ varies wildly at small radii ($\lesssim$ 10 kpc), so in this case too, relationships involving $t_{cool}/t_{ff}$ and other parameters related to cooling and feedback in the cluster would be difficult to interpret (although the variation in simulations may be due to the idealized AGN entering periods of complete quiescence as the fuel supply reaches zero). However, since $0.025R_{500}$ is greater than 10 kpc for all the CLASH clusters, even if this effect is physically realistic we would not expect to observe it.

\begin{figure*}[t]
\begin{center}
\begin{tabular}{c}
\includegraphics[height=14cm]{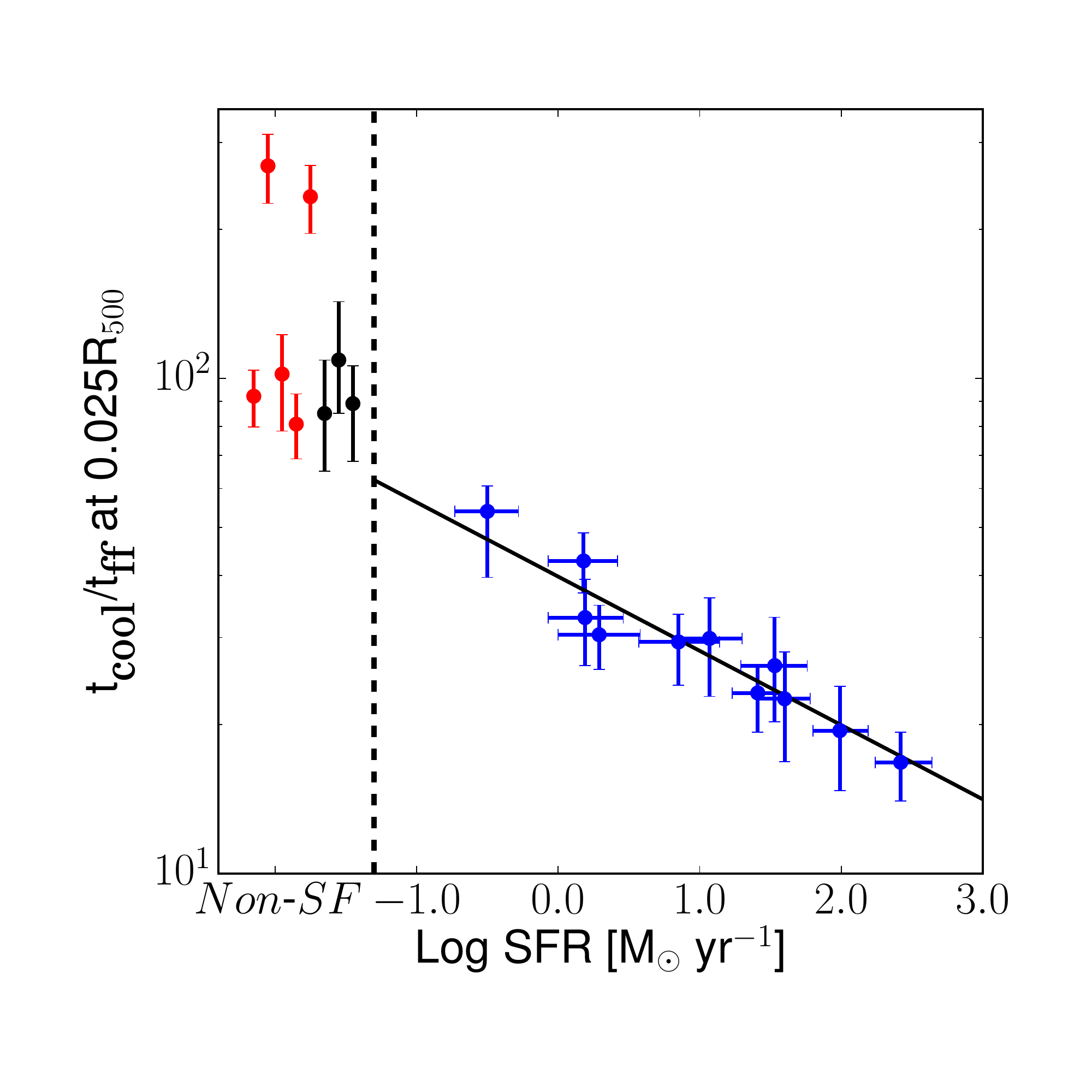}
\end{tabular}
\end{center}
\caption[]
{ \label{fig:SFR_Ratio} The $t_{cool}/t_{ff}$ ratio is shown as a function of Log$_{10}$ SFR. Values of the ratio $t_{cool}/t_{ff}$ were measured at a radius of $0.025 R_{500}$ for each cluster. 
The blue data points are the measured values from the active BCG sample.
The solid black line depicts the best-fit straight line to the data in log-log space. Values of $t_{cool}/t_{ff}$ at $0.025 R_{500}$ for CLASH clusters with non-starforming BCGs are shown to the left of the vertical dashed line. The black points show clusters where $t_{cool}/t_{ff}$ values were obtained using X-ray profiles from \cite{2014Donahue_CLASHX}, while the red points were obtained using temperature and density profiles from \cite{2009Cavagnolo_ACCEPT} and assuming a metallicity of 0.3 Z$_{\odot}$. Uncertainties for the SFR is taken to be the 68.3\% confidence interval for the marginal posterior probability distribution, which for a Gaussian distribution is equal to the 1$\sigma$ uncertainty. For $t_{cool}/t_{ff}$, the 1$\sigma$ uncertainties are shown. }
\end{figure*}

\section{Results}\label{sec-results}
 
Table \ref{table:SED_Dust_Params} lists our estimates for the best-fit SFR, the duration of the exponentially decaying starburst $\Delta t_{b}$, and dust mass $M_{d}$ for each active BCG, expressed as the mean of the marginalized posterior probability distribution with uncertainties given as the 68.3\% confidence interval. We also list the best-fit intrinsic optical attenuation, $A_{\textrm{V}}$. The SFRs in the CLASH BCGs span the range from $\sim 0.5$ M$_{\odot}$ yr$^{-1}$ to $\sim 250$ M$_{\odot}$ yr$^{-1}$. The BCGs have dust masses ranging from $10^{7}$ M$_{\odot}$ in the case of RXJ2129.7+0005 to $10^{9}$ $M_{\odot}$ in the case of MACS1931.8-2653. Our multiband photometry and the corresponding best fit SEDs for each star forming CLASH BCG are shown in the appendix in Figure \ref{fig:Best_Fits}.
 
Burst durations ($\Delta t_{b}$) range from $\lesssim 100$ Myr to several Gyr with very large uncertainties for any given galaxy. Estimating the posterior probability distribution of $\Delta t_{b}$ is more complicated than estimating the distribution of the SFR or $M_{d}$, owing to the strong dependence of $\Delta t_{b}$ on {\sc fburst}, the ratio of the stellar mass of the burst population to the stellar mass of the old population in the apertures we use. We find the mean Spearman correlation coefficient between $\Delta t_{b}$ and {\sc fburst} is $R_{S}=0.72$ (see Figure \ref{fig:DTB_FB} in the Appendix). 
 
Since the uncertainties on $\Delta t_{b}$ are large and the probability distribution of $\Delta t_{b}$ is highly correlated with {\sc fburst}, we can increase the precision of our $\Delta t_{b}$ estimates by imposing physically motivated constraints on {\sc fburst}. We do this by using literature estimates of the fraction of stellar mass contributed to BCGs by star formation at a redshift $< 1$ as the range of plausible burst mass percentages (referring back to Table \ref{table:SED_Params}, the percentage of stellar mass due to the starburst in our models is {\sc fburst}/(1 + {\sc fburst})). This range is on the order of $1\%-10\%$ \citep[e.g.][]{2008McIntosh_Growth, 2015Inagaki_Growth, 2016Cooke_IP}. By restricting the burst mass percentage (and therefore {\sc fburst}) to a single order of magnitude,  we reduce the uncertainty in our estimates of $\Delta t_{b}$  by slightly more than a factor of 1.5, and can provide estimates of the best-fit $\Delta t_{b}$ in some cases where we otherwise report lower limits.

Restricting {\sc fburst} for each SED is slightly complicated by the fact that we measure our SEDs in apertures that do not cover the entire surface area of the BCG. While the range of burst mass percentages we cite above correspond to $M_{\textrm{burst}}/M_{\textrm{BCG}}$, the stellar mass contributed by the starburst over the stellar mass of the BCG, the burst mass percentage in our SED models is $M_{\textrm{burst}}/M_{\textrm{aperture}}$, the stellar mass contributed by the starburst over the stellar mass in the aperture. However,
\begin{equation}
\frac{M_{\textrm{burst}}}{M_{\textrm{aperture}}} = \left(\frac{M_{\textrm{burst}}}{M_{\textrm{BCG}}}\right)\times\left(\frac{L_{F160,\textrm{BCG}}}{L_{F160, \textrm{aperture}}}\right),
\end{equation}
where we assume that the F160W near-IR luminosity, $L_{F160}$, is a reasonable proxy for stellar mass. We measured $L_{F160,\textrm{BCG}}/L_{F160,\textrm{aperture}}$ using the 2D image model of the BCG from the isophotal fitting procedure described in $\S$ 2.3.2 in \cite{2014Lauer_CLASH}. The 2D image models are free of contamination by light from surrounding galaxies. By multiplying the $1\%-10\%$ burst mass percentage range by $L_{F160,\textrm{BCG}}/L_{F160,\textrm{aperture}}$ for each BCG, we were able to find appropriate ranges of $M_{\textrm{burst}}/M_{\textrm{aperture}}$, and therefore {\sc fburst}, for each SED. For example, $L_{F160,\textrm{BCG}}/L_{F160,\textrm{aperture}}$ is $\sim 2.5 - 2.7$ in the case of RXJ1532.9+3021, implying that $2.5\% \lesssim M_{\textrm{burst}}/M_{\textrm{aperture}} \lesssim 25\%$. This implies a range of {\sc fburst} of 0.025-0.4. Since MACS1931.8-2635 does not have a 2D image model available, we used the same range for {\sc fburst}, but otherwise calculated the restricted range of {\sc fburst} for each BCG individually.

The best fit values for $\Delta t_{b}$ assuming restricted ranges of {\sc fburst} are reported in Table \ref{table:SED_Dust_Params}. By restricting the range of {\sc fburst} to values that correspond to modest contributions to the BCG stellar mass from recent star formation, we produce a more precise set of $\Delta t_{b}$ results compared to fits with no restrictions on {\sc fburst}. We adopt these values in the rest of our analysis.
 
The trends we report below between quantities measured with SED fits are not caused by correlations in the posterior probability distributions of individual fits. The posterior probability distributions of SFR, $\Delta t_{b}$, and $M_{d}$ for individual BCGs are either uncorrelated or only weakly  correlated. We report Spearman correlation coefficients in Table \ref{table:SED_Dust_Params} and two-dimensional posterior probability distributions in Figure \ref{fig:Posteriors} in the Appendix. Across the sample of CLASH BCGs, the posterior probability distributions of SFR and $M_{d}$ are the most correlated, but only with a mean Spearman correlation coefficient of 0.19, while $M_{d}$ and $\Delta t_{b}$ are the least correlated, with a mean Spearman correlation coefficient of 0.01. 
 
Our modified version of {\tt iSEDfit} produces fits with a mean $\chi_{\nu}^{2}$ of 0.93, and a range between $\chi_{\nu}^{2} = 0.57$ and $\chi_{\nu}^{2} = 1.63$. The $\chi_{\nu}^{2}$ statistic for the best fitting model in the grid for each BCG is listed in Table \ref{table:BCG_Statistical_Params}. Table \ref{table:BCG_Statistical_Params} lists $\chi_{\nu}^{2}$ values we obtain for each dust attenuation law we try using, coupled with a \cite{1955Salpeter_IMF} IMF, and the values we obtain using a \cite{2000Calzetti_Extinction} attenuation law coupled with a \cite{2003Chabrier_IMF} IMF. The choice of the adopted attenuation law and IMF can affect the quality of the fits to each individual SED but does not affect the average $\chi_{\nu}^{2}$ for the full sample nor does it affect our qualitative conclusions. When the \cite{2000Calzetti_Extinction} attenuation law is replaced by \cite{2000Witt_Extinction} clumpy SMC attenuation, the mean $\chi_{\nu}^{2}$ increases by 16\% to 1.08; when we use a modified \cite{2000Calzetti_Extinction} law instead the mean is 1.10.  

The best-fit stellar and dust parameters are similarly minimally affected by choice of attenuation law.  Adopting a \cite{2003Chabrier_IMF} IMF systematically shifts the SFRs $\sim$ 0.25 dex downwards but does not otherwise significantly alter the probability distributions or correlations between parameters, and does not strongly affect values of $\chi_{\nu}^{2}$. The fits using the Chabrier IMF have a mean $\chi_{\nu}^{2}$ of 0.98, and any offsets in SED fit parameters are comparable to, or smaller than, our reported uncertainties. We report the parameters obtained for each model used in the appendix, but in our analyses of the results we only discuss the SED fits obtained using the \cite{2000Calzetti_Extinction} law and \cite{1955Salpeter_IMF} IMF, unless otherwise noted. 

\begin{table*}[]  
\small  
\caption{BCG Stellar and Dust Parameters}
\label{table:SED_Dust_Params}  
\vspace{1mm}  
\centering  
{  
\begin{tabular}{lcccccccc}  
\hline
\hline
   & log$_{10}$ SFR & log$_{10}$ $\Delta t_{b}$ & log$_{10}$ $\Delta t_{b}$  & log$_{10}$ $M_{d}$ & A$_{\textrm{V}}$ & $R_{S}$$^{a}$ & $R_{S}$ & $R_{S}$ \\ 
 &                    &      & restr. {\sc fburst}$^{b}$  & & & SFR$\times \Delta t_{b}$ & SFR$\times$ $M_{d}$ & M$_{\textrm{d}}\times\Delta t_{b}$ \\
BCG  & (M$_{\odot}$ yr$^{-1}$) & (Gyr) & (Gyr) & (M$_{\odot}$) & (mag) & & & \\ 
\hline  
\\  
Abell 383 & $0.18^{+0.24}_{-0.25}$ & $> -0.49^{d}$ & $0.19^{+0.35}_{-0.34}$ & $7.47^{+0.84}_{-0.87}$ & $0.48^{+0.22}_{-0.24}$ & {-0.16} & {0.36} & {-0.03} \\ \\
 
MACS0329.7-0211 & $1.6^{+0.18}_{-0.2}$ & $ > -0.73$ & $0.01^{+0.37}_{-0.40}$ & $8.4^{+0.71}_{-0.73}$ & $0.56^{+0.18}_{-0.19}$ & {-0.06} & {0.07} & {0.07} \\ \\ 
 
MACS0429.6-0253 & $1.53^{+0.23}_{-0.24}$ & {$> -0.36$} & $0.09^{+0.36}_{-0.37}$ & $8.49^{+0.7}_{-0.68}$ & $0.75^{+0.23}_{-0.24}$ & {0.03} & {0.20} & {0.07} \\ \\
 
MACS1115.9+0219 & $0.85^{+0.29}_{-0.28}$ & {$> -0.40$} & $0.19^{+0.35}_{-0.36}$ & $7.6^{+0.85}_{-0.81}$ & $0.44^{+0.3}_{-0.3}$ & {-0.03} & {0.44} & {-0.01} \\ \\
 
MACS1423.8+2404 & $1.41^{+0.18}_{-0.18}$ & $> -0.65$ & $> -0.03$ & $8.47^{+0.69}_{-0.72}$ & $0.42^{+0.18}_{-0.18}$ & {-0.11} & {0.25} & {0.02} \\ \\ 
 
MACS1720.3+3536 & $0.19^{+0.27}_{-0.26}$ & $> -0.34$ & $> -0.01$ & $7.6^{+0.67}_{-0.71}$ & $0.6^{+0.26}_{-0.26}$ & {-0.10} & {0.10} & {-0.0} \\ \\
 
MACS1931.8-2653 & $2.42^{+0.22}_{-0.18}$ & $-0.97^{+0.53}_{-0.56}$ & $-1.01^{+0.34}_{-0.35}$ & $8.88^{+0.39}_{-0.4}$ & $0.87^{+0.21}_{-0.21}$ & {-0.52} & {-0.23} & {0.19} \\ \\ 
 
MS2137-2353 & $0.25^{+0.29}_{-0.29}$ & $> -0.43$ & $> 0.08$ & $7.41^{+0.86}_{-0.87}$ & $0.43^{+0.31}_{-0.31}$ & {-0.23} & {0.47} & {-0.21} \\ \\
 
RXJ1347.5-1145 & $1.07^{+0.23}_{-0.23}$ & {$> -0.39$} & $0.13^{+0.36}_{-0.35}$ & $7.76^{+0.96}_{-0.9}$ & $0.31^{+0.22}_{-0.23}$ & {0.03} & {0.47} & {-0.10} \\ \\ 
 
RXJ1532.9+3021 & $1.99^{+0.2}_{-0.19}$ & $-0.29^{+0.54}_{-0.59}$ & $-0.39^{+0.38}_{-0.41}$ & $8.77^{+0.47}_{-0.51}$ & $0.9^{+0.19}_{-0.19}$ & {-0.40} & {-0.16} & {0.11} \\ \\
 
RXJ2129.7+0005 & $-0.5^{+0.22}_{-0.23}$ & $0.29^{+0.28}_{-0.31}$ & $> 0.25$ & $6.81^{+0.87}_{-0.71}$ & $0.28^{+0.14}_{-0.17}$ & {-0.19} & {0.12} & {0.01} \\ \\
\hline
\end{tabular}  
\begin{flushleft}
$^{a}$ {Spearman} correlation coefficients for the sample of pairs of parameters obtained by sampling the posterior probability distribution of models. {The Spearman correlation coefficient measures the rank correlation of two datasets, and is between -1 (a perfect negative correlation) and 1 (a perfect positive correlation).}\\
$^{b}$ Results obtained for $\Delta t_{b}$ when the fractional burst strength, {\sc fburst}, is restricted to the range [0.025, 0.4] and the SED fit is re-run. \\
$^{c}$ Uncertainties denote the 1$\sigma$ credible intervals for each value. \\
$^{d}$ For log$_{10}$ $\Delta t_{b}$ posterior probability histograms that peak near the upper bound of the parameter space, we report the 1$\sigma$ confidence interval as a lower limit on log$_{10}$ $\Delta t_{b}$. \\  
\end{flushleft}  
}  
\end{table*}

\begin{table*}[]  
\footnotesize  
\caption{Best Fit $\chi^{2}$ Values}
\label{table:BCG_Statistical_Params}  
\vspace{1mm}  
\centering  
{  
\begin{tabular}{lcccc }
\hline 
\hline  
 & Calzetti & Modified Calzetti & Witt Clumpy SMC & Chabrier IMF \\ 
BCG & $\chi_{\nu}^{2}$ & $\chi_{\nu}^{2}$ & $\chi_{\nu}^{2}$ & $\chi_{\nu}^{2}$ \\   
\hline  
Abell 383 & 0.99 & 1.33 & 1.54 & 1.41 \\
MACS0329.7$-$0211  & 0.57 & 0.76 & 0.60 & 0.64 \\
MACS0429.6$-$0253 & 1.53 & 1.53 & 2.25 & 1.39 \\
MACS1115.9+0219	 & 0.59 & 0.53 & 0.46 & 0.52 \\
MACS1423.8+2404	 & 0.80 & 1.44 & 1.24 & 0.80 \\
MACS1720.3+3536	 & 1.41 & 1.48 & 1.02 & 1.31 \\
MACS1931.8$-$2653  & 0.66 & 1.39 & 1.22 & 1.17 \\
MS2137$-$2353  & 0.64 & 0.70 & 0.68 & 0.67 \\
RXJ1347.5$-$1145  & 0.72 & 0.72 & 0.66 & 0.78 \\
RXJ1532.9+3021  & 1.63 & 1.51 & 1.28 & 1.40 \\
RXJ2129.7+0005  & 0.70 & 0.73 & 0.79 & 0.73 \\
\hline
\end{tabular}  
\begin{flushleft}

\end{flushleft}  
}  
\end{table*}

\subsection{The Starburst - ICM Connection}\label{sec-sfricm}
 
For BCGs with detectable star-formation activity, we find a tight correlation between the BCG SFR and the ratio $t_{cool}/t_{ff}$ measured at $0.025 R_{500}$. The trend and the data are shown in Figure \ref{fig:SFR_Ratio}. The best fit line between these two parameters is 
\begin{equation}\label{eq:SFR_Ratio}
\log_{10} \frac{t_{cool}}{t_{ff}} = \left(1.6\pm0.4\right) - \left(0.15\pm0.03\right)\log_{10} \frac{\textrm{SFR}}{\textrm{M}_{\odot}\textrm{ yr}^{-1}},
\end{equation}
and is shown as the black curve in Figure \ref{fig:SFR_Ratio}. The data are fit by this trend with no detectable intrinsic scatter ($\sigma_{i} < 0.15$ dex at 3$\sigma$). We estimated the best fit lines and intrinsic scatters using the least squares method in \cite{2010Hogg_Methods} for fitting data with uncertainties in two dimensions and intrinsic scatter. The data have a Spearman correlation coefficient $R_{S} = -0.98$ and a Pearson correlation coefficient $R = -0.95$.
 
We estimated both the strength of the relationship we observe, and the strength of the claim that the relationship has little intrinsic scatter. We first investigated the possibility that there is no relationship between SFR and $t_{cool}/t_{ff}$, by calculating the marginal probability distribution of the slope using Equation 35 of \cite{2010Hogg_Methods}, and find that the slope of the relationship is less than 0 at 99.993\% confidence.

\begin{figure}
\begin{center}
\begin{tabular}{c}
\includegraphics[height=9cm]{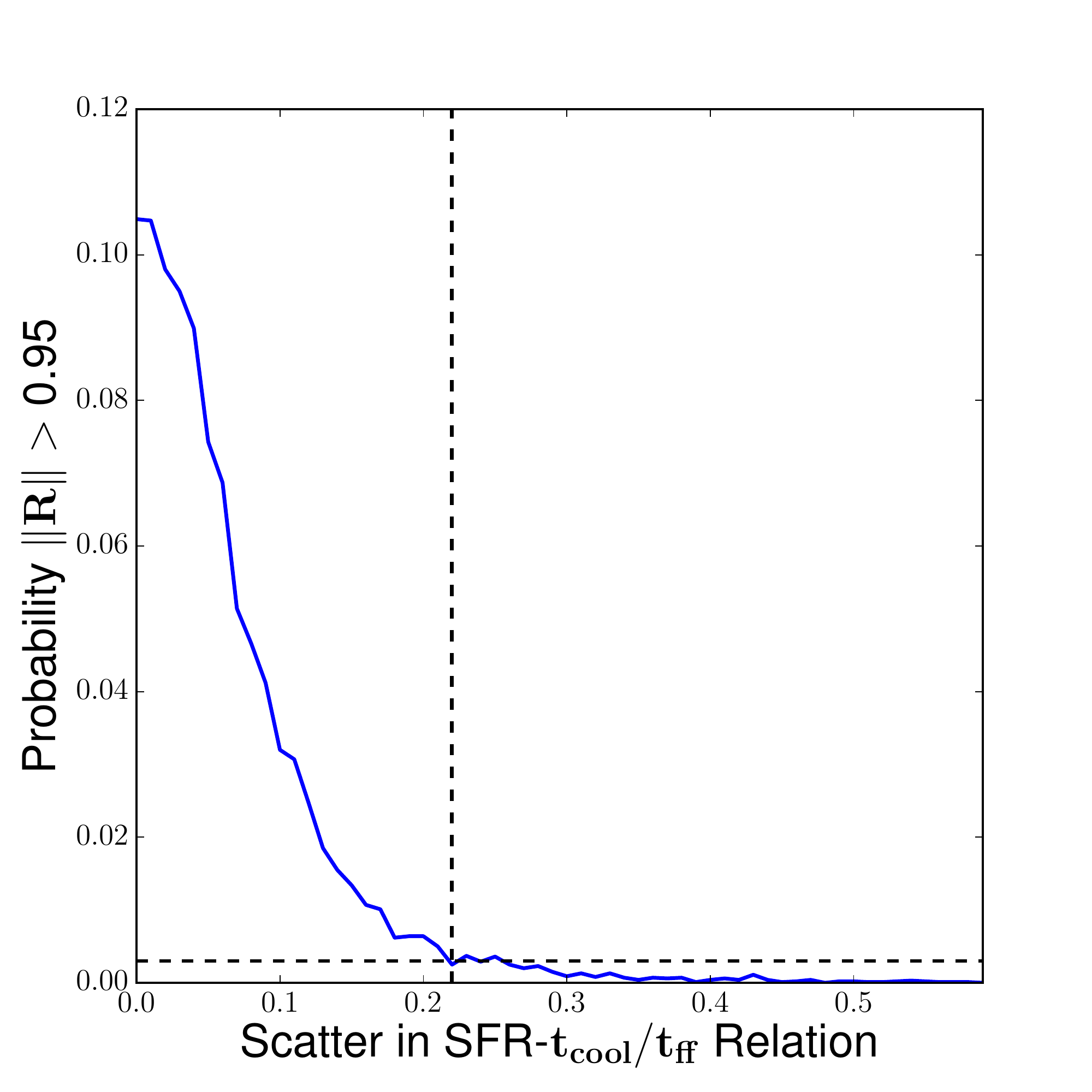}
\end{tabular}
\end{center}
\caption[]
{ \label{fig:Corr_Prob} The probability of a synthetic dataset of SFR and $t_{cool}/t_{ff}$ observations with an underlying relationship corresponding to the best-fit relationship measured for CLASH active BCGs and measured uncertainties comparable to the CLASH data set having a Pearson correlation $\| R\| > 0.95$ is shown as a function of intrinsic scatter. The horizontal dashed line denotes where the probability falls below 0.3\% (corresponding to a 3$\sigma$ outlier for a Gaussian process). The vertical dashed line denotes an intrinsic scatter of 0.22 dex, which is approximately where the probability curve dips below 0.3\%. }
\end{figure}

We then examined the possibility that a relationship obeying the trend in Equation \ref{eq:SFR_Ratio} but with a large intrinsic scatter produced the results we observe. We created an ensemble with $10^{4}$ synthetic datasets, consisting of 11 SFR values randomly sampled in logarithmic units from 0.1 M$_{\odot}$ yr$^{-1}$ to 1000 M$_{\odot}$ yr$^{-1}$, and calculated the corresponding predicted values of log$_{10}$ $t_{cool}/t_{ff}$. Assuming a given intrinsic scatter, $\sigma_{i}$, we generated an offset from the trend line for each of the 11 points by sampling a normal distribution of standard deviation $\sigma_{i}$ to obtain the amplitude of the offset and determined a direction of the offset by sampling a uniform distribution between 0 and $2\pi$. We added an additional offset for each point in the x-direction and in the y-direction by sampling normal distributions with standard deviations equal to the mean uncertainty of the observed SFRs and the observed $t_{cool}/t_{ff}$ values, respectively. The probability that data more tightly correlated than our dataset ($\|R\| = 0.95$) would be produced given $\sigma_{i}$ was calculated for the ensemble of datasets. Since we used log-normally distributed variables to generate our synthetic data, we used the Pearson coefficient $R$ to measure the strength of the correlation. We repeated this procedure for $\sigma_{i}$ ranging from 0.0 to 0.6, and found that for $\sigma_{i} \gtrsim 0.22$ dex, the probability of drawing a dataset with a trend tighter than the one we observe is $< 0.3\%$. The results of the above correlation test are presented in Figure \ref{fig:Corr_Prob}.

The ratio $t_{cool}/t_{ff}$ is thought to be a proxy for the thermal instability of ICM gas -- when the gas can cool quickly relative to the time it takes for it to infall, it can more readily become thermally unstable and collapse \citep{2012Gaspari_Feedback, 2014Li_ColdClumps}. Assuming ICM instability increases with decreasing $t_{cool}/t_{ff}$, our results suggest that BCG SFRs will increase with the thermal instability of the surrounding ICM.

\begin{figure}
\begin{center}
\begin{tabular}{c}
\includegraphics[height=9cm]{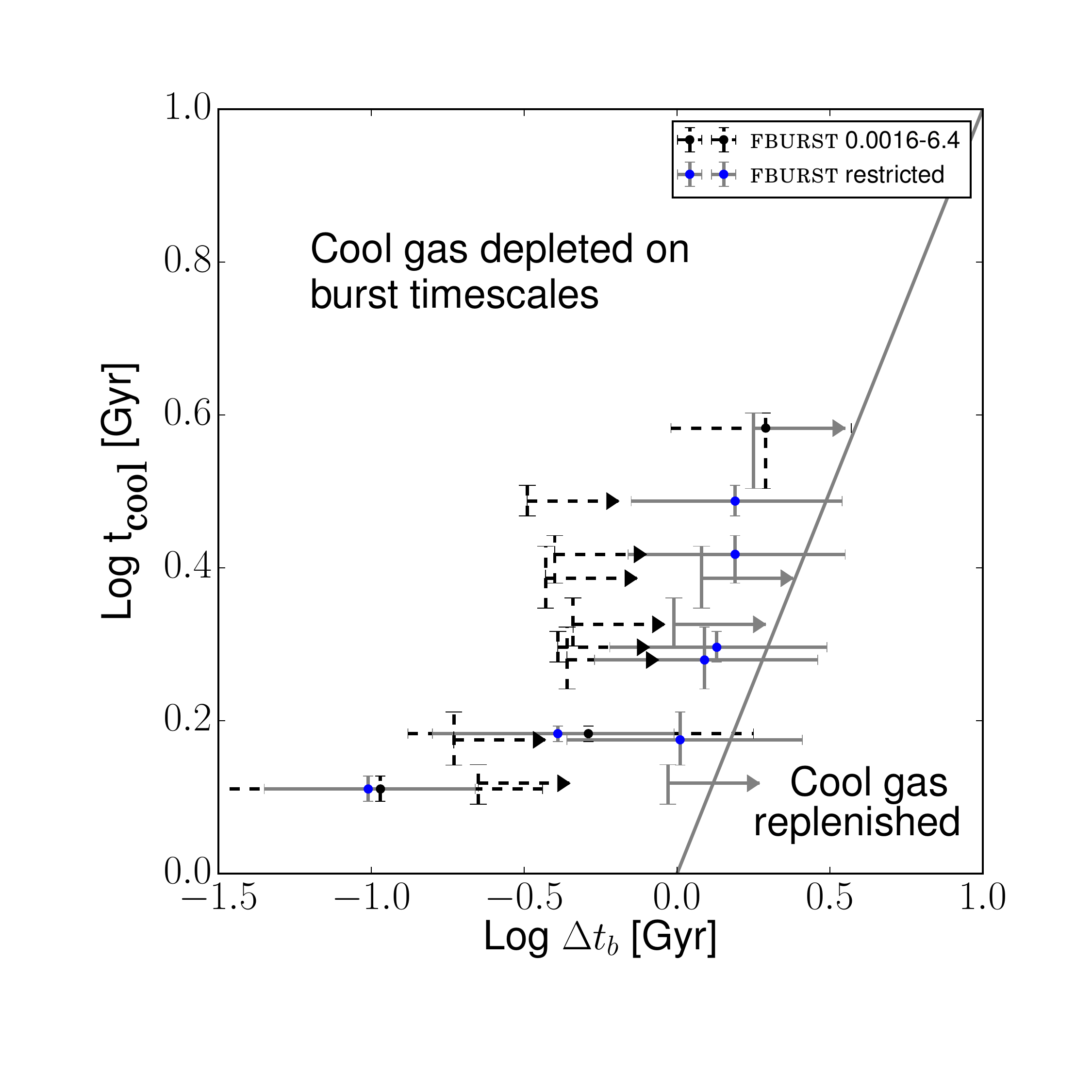}
\end{tabular}
\end{center}
\caption[]
{ \label{fig:DTB_TC} Log$_{10}$ $\Delta$ t$_{b}$ vs. Log$_{10}$ t$_{\textrm{cool}}$ is shown. Cooling times were measured at a radius of $0.025 R_{500}$ in each cluster. The solid grey line shows where $\Delta$ t$_{b}$ = t$_{\textrm{cool}}$. Blue points with grey errorbars show values of $\Delta t_{b}$ obtained when limiting the range of {\sc fburst} according to the procedure discussed in $\S$ 4. For reference, black points with dashed black errorbars show $\Delta t_{b}$ measured assuming $0.0016 <$ {\sc fburst} $< 6.4$. For points in the region to the left of the line, labelled â`cool gas depleted on burst timescales', the cooling time of gas at $0.025 R_{500}$ exceeds the duration of the starburst. For points in the region to the right of the line, labelled `cool gas replenished',  the duration of the starburst exceeds the cooling time of the gas at $0.025 R_{500}$. Uncertainties for both parameters are defined analogous to Figure~\ref{fig:SFR_Ratio} .}
\end{figure}

The relationship between the star formation duration $\Delta t_{b}$ and cooling time at $0.025R_{500}$, $t_{\textrm{cool}}$, is shown in Figure \ref{fig:DTB_TC}. The relationship between these two parameters is difficult to quantify owing to the large uncertainties on $\Delta t_{b}$. In order to estimate the strength of this correlation, we calculated the Spearman coefficients both including and excluding MACS1931.8-2653. In performing these computations, we also need to address the fact that some of the $\Delta t_{b}$ values are only lower limits. If we use the lower limit values to compute the $\Delta t_{b}$ ranking, then $\Delta t_{b}$ and $t_{cool}$ are strongly correlated, with a Spearman coefficient of 0.86 including MACS1931.8-2653, and a coefficient of 0.81 excluding it. If we assume that the lower limits in $\Delta t_{b}$ are tied for the highest rank, then the Spearman coefficient including MACS1931.8-2653 is 0.57, implying a weak correlation ($p < 0.065$), while if MACS1931.8-2653 is excluded, the data are not significantly correlated ($R_{S} = 0.427$, $p < 0.22$). In the limit of the worst possible $\Delta t_{b}$ ranking for the lower limit data (e.g., the data point with the highest ranked $t_{cool}$ amongst the `lower limit' points has the lowest relative $\Delta t_{b}$ rank), the two datasets are not correlated ($R_{S} = 0.433$, $p<0.18$ with MACS1931.8-2653, $R_{S}=0.24$, $p<0.50$ without). Therefore, cooling times and burst durations are likely weakly positively correlated (although this observation may be driven by MACS1931.8-2653), and $\Delta t_{b}$ and $t_{\textrm{cool}}$ become comparable in amplitude when $\Delta t_{b}$ reaches Gyr timescales.

All the BCGs either lie to the left of the line where $\Delta t_{b} = t_{\textrm{cool}}$ or are consistent with lying to the left of this line. This is the region where starbursts occur on timescales shorter than the timescale for the ICM to cool. In clusters where $\Delta t_{b}$ and $t_{cool}$ fall to the left of this line, star formation has been ongoing for a shorter duration than the time it would take for the ICM at $0.025R_{500}$ to radiatively cool. Therefore, assuming $t_{cool}$ profiles typically increase monotonically with radius, if the cold gas reservoir fueling star formation is the result of the ICM radiatively cooling, then the body of low-cooling time gas inside $0.025R_{500}$ that was present at the onset of star formation will be depleted. Alternatively, in clusters where $\Delta t_{b}$ and $t_{cool}$ fall to the right of this line, either higher cooling time gas will have radiatively cooled and replenished the gas inside $0.025R_{500}$ in order to continue forming the cold gas fueling star formation, or radiative cooling has been arrested and $t_{cool}$ is locally static.
 
\subsection{Star Formation and Dust Parameters}\label{sec-sfrdust}
 
\begin{figure}
\begin{center}
\begin{tabular}{c}
\includegraphics[height=9cm]{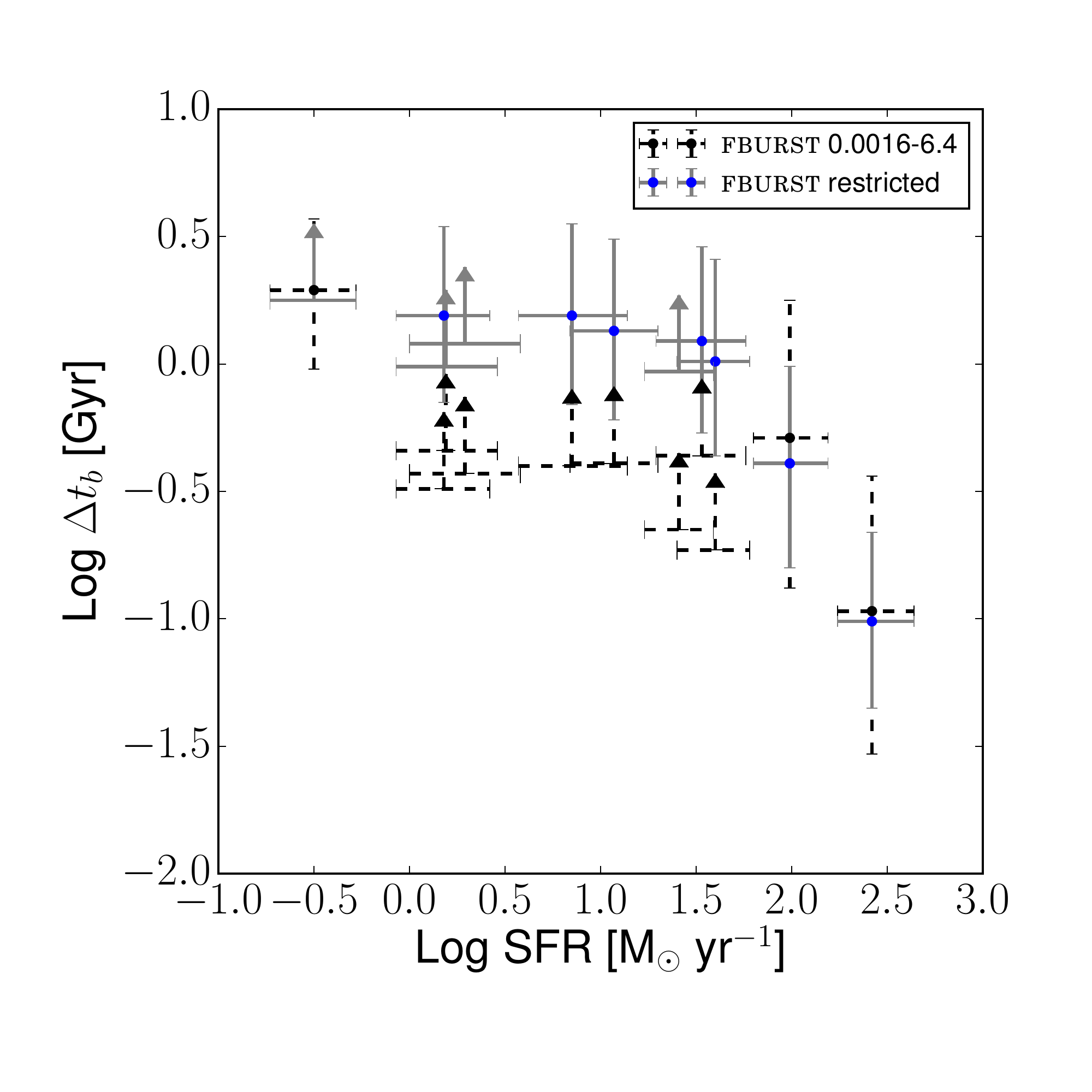}
\end{tabular}
\end{center}
\caption[]
{ \label{fig:SFR_DTB} Log$_{10}$ SFR vs. Log$_{10}$ $\Delta t_{b}$ is shown. Blue points with grey errorbars show values of $\Delta t_{b}$ obtained when limiting the range of {\sc fburst} according to the procedure discussed in $\S$ 4. For reference, black points with dashed black errorbars show $\Delta t_{b}$ measured assuming $0.0016 <$ {\sc fburst} $< 6.4$.  Uncertainties for both parameters are taken to be the 68.3$\%$ confidence interval for their respective marginal posterior probability distributions. }
\end{figure}

We find that both the dust mass and burst duration of CLASH BCGs are correlated with their SFRs. The relationship between $\Delta t_{b}$ and SFR is shown in Figure \ref{fig:SFR_DTB}.  Large SFRs are consistent with star formation episodes that have recently begun, and as the bursts persist to $\sim$ Gyr timescales, the SFRs diminish by several orders of magnitude.  

\begin{figure}
\begin{center}
\begin{tabular}{c}
\includegraphics[height=9cm]{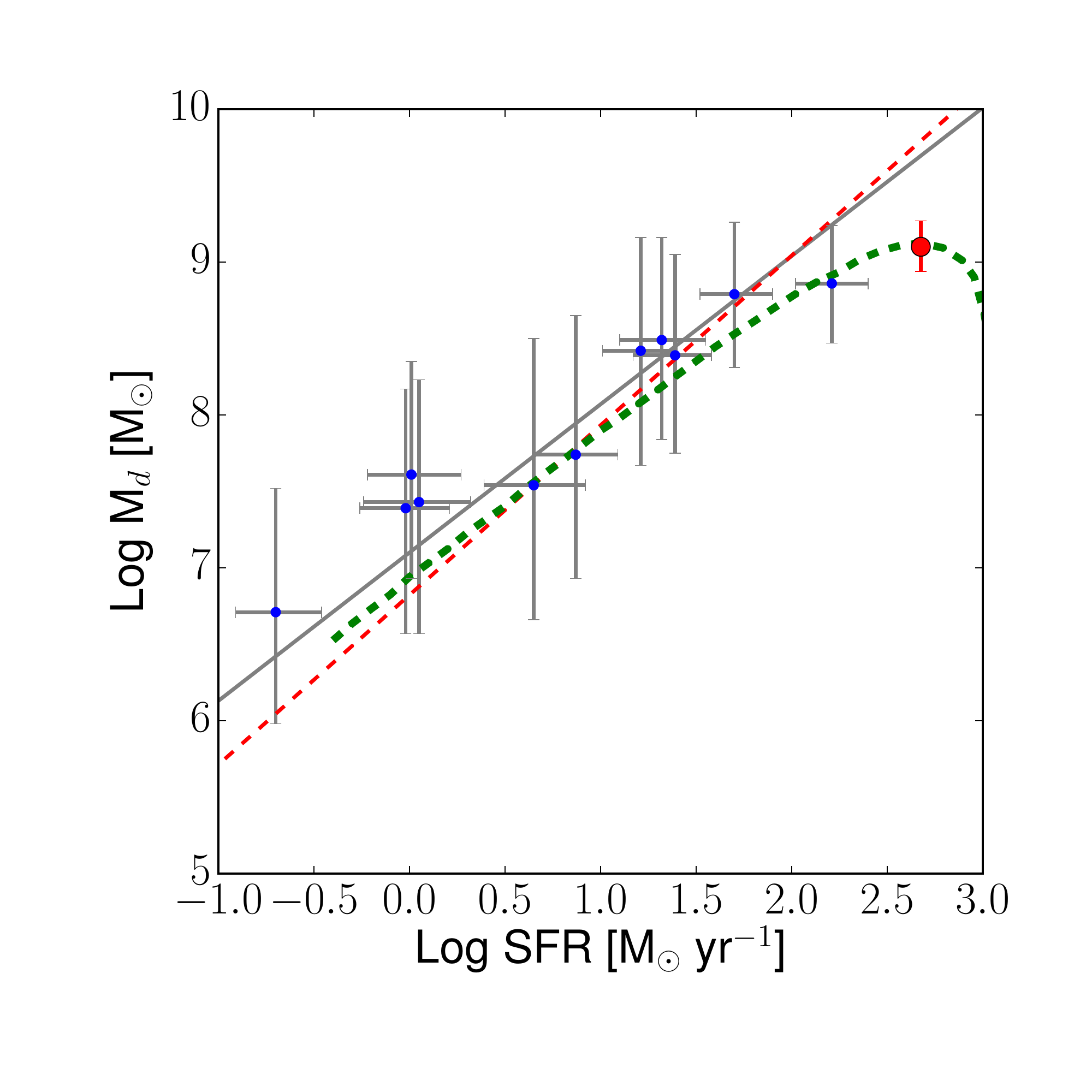}
\end{tabular}
\end{center}
\caption[]
{ \label{fig:SFR_MD} Log$_{10}$ SFR vs. Log$_{10}$ M$_{d}$ is shown. The solid grey line depicts the best fit trend line for these parameters for the CLASH active BCG sample. The dashed red line shows the SFR-M$_{d}$ relationship of \cite{2010daCunha_DustMass}, which was obtained using SDSS field galaxies. The dashed green line shows the \cite{2014Hjorth_SFRDust} relationship. The red data point is the BCG of the Phoenix cluster. Uncertainties for both parameters are taken to be the 68.3$\%$ confidence interval. \vspace{2.0mm}}
\end{figure}

Figure \ref{fig:SFR_MD} shows the relationship between $M_{d}$ and SFR. In order to be consistent with studies of SFR and M$_{d}$ conducted by \cite{2010daCunha_DustMass} and \cite{2014Hjorth_SFRDust}, when comparing these two quantities we analyze SFRs and dust masses obtained with a \cite{2003Chabrier_IMF} IMF. We continue to use \cite{1955Salpeter_IMF} when discussing results in the rest of our paper.

Dust masses and SFRs are correlated, with a best-fit trendline of 
\begin{equation}
\log_{10} \frac{M_{d}}{\textrm{M}_{\odot}} = \left(7.1^{+0.4}_{-0.3}\right) +\left(0.97^{+0.34}_{0.24}\right)\log_{10} \frac{\textrm{SFR}}{\textrm{M}_{\odot}\textrm{ yr}^{-1}}
\end{equation}
with an intrinsic scatter of $< 0.83$ dex ($3\sigma$ limit). The analysis of \cite{2010daCunha_DustMass} derived dust masses and SFRs for field galaxies using a UV-IR SED fitting technique that was similar to ours. We overlay their best-fit trendline, which has a best-fit slope of $1.11\pm0.01$ and intercept of $7.1\pm0.01$ in Figure \ref{fig:SFR_MD}. In order to constrain the behavior of the trend at the large SFR end, we include the dust mass and SFR of the Phoenix Cluster BCG \citep[e.g.][]{2013McDonald_Phoenix}. We adopt the \cite{2017Mittal_Phoenix} estimate of 454-494 M$_{\odot}$ yr$^{-1}$ for the Phoenix SFR and fit the far-IR SED of Phoenix to estimate $M_{d}$. This extended dataset shows a flattening slope at the high-SFR end of the relationship. When we overlay a trend drawn from \cite{2014Hjorth_SFRDust} that takes into account the evolution of star formation and dust for starbursting galaxies with large ($\gtrsim 1000$ M$_{\odot}$ yr$^{-1}$) SFRs, we find excellent correspondence to our data across the range of SFRs studied. While our confidence that we observe a change in the SFR-M$_{d}$ trend at the high SFR-end is limited by the size of our sample, our results are fully consistent with starforming BCGs producing dust reservoirs like starbursts in the field.

\subsection{Testing Impact of AGN Emission on SED Results on MACS1931.8-2653}\label{sec-agn}

We wish to understand the possible impact of AGN emission on our SED fitting results. Therefore, we analyzed the impact of AGN emission on the SED fit to MACS1931.8-2653, since this BCG shows the strongest evidence for AGN emission. If the AGN component has little impact on the SED fit to MACS1931.8-2653, we do not believe there will be substantial AGN contamination of our fits to the other starbursting BCGs in CLASH, which host weaker or quiescent AGN. The X-ray point source in this BCG is well-fit by an AGN power spectrum with a 2-10 keV luminosity of $5.32^{+0.40}_{-0.37}\times 10^{43}$ ergs s$^{-1}$ cm$^{-2}$ \citep{2013HL_AGN}.  Moreover, the IR template-fitting results of \cite{2015Santos_AGN} imply a substantial AGN contribution may exist in the case of MACS1931.8-2653, so it is important to investigate the effect of adding an AGN component to the model used to fit the UV-IR SED of this X-ray loud BCG.

As \cite{2015Santos_AGN} 
postulate a potentially large AGN contribution to the UV-IR SED of MACS1931.8-2653, we ran a separate SED fitting analysis for this BCG wherein we include the IR AGN model of \cite{2015Siebenmorgen_AGN}. We allowed {\tt iSEDfit} to sample the full range of parameters that describe the \cite{2015Siebenmorgen_AGN} AGN emission model library, which for our purposes are nuisance parameters. The contribution of the AGN to the total UV-IR luminosity of the galaxy was allowed to range between $0.001\times$ and $10\times$ the stellar contribution.
 
We find that the effect of AGN emission in MACS1931.8-2653 is marginal. The best fit $\chi^{2}$ degrades slightly, to 1.09, most likely owing to the addition of AGN model nuisance parameters. After incorporating AGN emission, we find the log$_{10}$ SFR = $2.36\pm0.19$ M$_{\odot}$ yr$^{-1}$, log$_{10}$ $\Delta t_{b}$ = $-0.69^{+0.46}_{-0.80}$ Gyr, and that log$_{10}$ M$_{d}$ = $8.86^{+0.34}_{-0.38}$ M$_{\odot}$. Compared to our AGN-free model, we find the SFR changes by $\sim 0.3\sigma$, $\Delta t_{b}$ by $\sim 0.5\sigma$, and M$_{d}$ by $\sim 0.05\sigma$. This is not surprising, since the AGN contribution to the UV-FIR luminosity relative to the stellar contribution is log$_{10}$ f$_{AGN}$ = $-1.65\pm0.87$.
 
One possible explanation for the discrepancy between these findings and the results published in \cite{2015Santos_AGN} for MACS1931.8-2653 is our decision not to incorporate WISE photometry in our SED fitting. WISE W3 and W4 filters cover the region of the MIR spectrum ($\sim 10$ $\mu$m) sensitive to the contribution of AGN flux. Since for our purposes, the AGN flux is a contaminant, it does not make sense to include these filters in our SED. The difference may also be explained, in part, by the fitting technique-- the theoretical modelling we employ to fit the data allows us greater flexibility to fit the data than the empirical templates used in \cite{2015Santos_AGN}.

\section{Discussion}\label{sec-discuss}
 
CLASH galaxy clusters provide a data-rich sample for the study of feedback in the environments of BCGs. Just over 50$\%$ of the CLASH X-ray selected clusters host BCGs that exhibit signs of vigorous feedback and their relatively low redshifts allow us to study their properties in detail. We find a strong observational relationship between BCG star formation and the ratio $t_{cool}/t_{ff}$, which is a proxy for thermal instability in the ICM. As we will discuss below, this relationship appears to strongly support the AGN-regulated cooling mechanism in the condensation and precipitation model advocated in \cite{2014Voit_Feedback} and \cite{2016Voit_Model}. However, several of the implications of our findings have not been anticipated by models of cooling and feedback in the ICM, and raise interesting questions about the dynamics of condensation and heating in the BCG and its environs.

\subsection{$t_{cool}/t_{ff}$ As A Proxy For Thermal Instability and ICM Condensation}\label{sec-tcool}
 
Models of AGN-driven condensation and precipitation that involve $t_{cool}/t_{ff}$ as a proxy for thermal instability in the ICM provide a natural foundation for interpreting our observed correlation between between $t_{cool}/t_{ff}$ and the SFR. If $t_{cool}/t_{ff}$ is related to the rate of molecular gas production around an active BCG, it should be correlated with the SFR. Specifically, recent simulations show that $t_{cool}/t_{ff}$ determines the critical overdensity at which ICM density perturbations can condense \citep{2015Singh_ICMPerturb}. We hypothesize this critical overdensity, in turn, determines the mass deposition efficiency $\epsilon_{MD}$, defined to be the mass fraction in a region of the ICM that cools into molecular gas. Assuming that density perturbations of the ICM follow a log-normal distribution in the core of the cluster, $\epsilon_{MD}$ at a radius $r$ ought to be
\begin{equation}
\begin{split}
\epsilon_{MD} &= \int_{\delta_{c}\left(\frac{t_{cool}}{t_{ff}}\right)}^{\infty}{\frac{1}{\sqrt{2\pi}\sigma\rho}e^{-\frac{\left(\ln\rho\right)^{2}}{2\sigma^{2}}}d\rho} \\
&= \frac{1}{2}Erfc\left(\frac{\ln\delta_{c}\left(\frac{t_{cool}}{t_{ff}}\right)}{\sqrt{2}\sigma}\right),
\end{split}
\end{equation}
where $\delta_{c}\left(\frac{t_{cool}}{t_{ff}}\right)$ is the critical overdensity for ICM condensation as a function of $t_{cool}/t_{ff}$ and $\sigma$ is the width of the ICM density perturbation distribution. The observed relation between SFR and $t_{cool}/t_{ff}$ would therefore imply that SFR scales with $\epsilon_{MD}$.
 
We can use the $t_{cool}/t_{ff}$-SFR relationship to infer properties of the relationship between $t_{cool}/t_{ff}$ and $\epsilon_{MD}$, and therefore constrain models of feedback-regulated cooling. As a first-order approximation, we assume that a condition close to equilibrium exists between cooling and star-formation for most of the duration of an episode of feedback-regulated cooling, so that SFR $\sim$ $\dot{M}_{g, real}$, where $\dot{M}_{g, real}$ is the actual ICM cooling rate. By measuring $\dot{M}_{g} \equiv M_{g}/t_{cool}$ within $0.025R_{500}$ with the X-ray parameter profiles used in this paper, we find that
\begin{equation}\label{eq-emd}
\log_{10} \epsilon_{MD} \sim \log_{10} \textrm{SFR} - \log_{10} \dot{M}_{g} = \frac{t_{cool}/t_{ff}-10}{-17.1^{+1.9}_{-2.3}}, 
\end{equation}
where for the purposes of this simple model we have set $\log_{10} \epsilon_{MD} = 0.0$ when $t_{cool}/t_{ff} = 10.0$, which is a `critical' value for mechanical feedback-triggered condensation \citep{2015Voit_FeedbackACCEPT}. In a relatively uniform sample of clusters like CLASH, where core gas masses occupy a narrow range ($\sim$0.4-1.4$\times 10^{13}$ M$_{\odot}$), $\dot{M}_{g}$ does plays a relatively minor role in Equation \ref{eq-emd}, resulting in a tight relationship between SFR and $t_{cool}/t_{ff}$. While it is important to bear in mind that we have made a simple estimate of $\epsilon_{MD}$ which may not be the actual mass deposition efficiency, our analysis demonstrates how the relationship between $t_{cool}/t_{ff}$ and star formation may be used to constrain the processes governing the cooling and condensation of gas in the ICM.
 
In \cite{2015Fogarty_CLASH}, we attempted to estimate the cooling rate of the ICM by measuring $\dot{M}_{g}$ for gas that was at a radius $< 35$ kpc or for gas that had an average $t_{cool}/t_{ff}$ ratio below 70~\footnote{The cooling times quoted in \cite{2015Fogarty_CLASH} were taken from the ACCEPT website (http://www.pa.msu.edu/astro/MC2/accept/). ACCEPT website cooling times are incorrect and need to be multiplied by a factor of 6.9/2. Values of $t_{cool}$ were calculated using ACCEPT profiles and assuming fixed metallicity.}. We found that while reddening-corrected UV photometric SFRs scale with $\dot{M}_{g}$ in both cases, star formation appears to be increasingly `inefficient' as the SFR decreases. The lowest SFRs we observed were $\sim 0.1-1\%$ of $\dot{M}_{g}$, while at the other extreme the two quantities were comparable. 
Such a trend would be expected if, as Equation ~\ref{eq-emd} indicates, the $\epsilon_{MD}$ scales with the SFR.

\subsubsection{The Role of $t_{ff}$ in the SFR-$t_{cool}/t_{ff}$ relationship}\label{sec-tff}
 
Recent work examining the critical condition for the onset of BCG activity in a cool core cluster suggests that this activity is driven by $t_{cool}$, not $t_{cool}/t_{ff}$ \citep{2017Hogan_Condensation}. These results may imply that we ought to observe a relationship between SFR-$t_{cool}$, without any significant contribution from $t_{ff}$. An SFR-$t_{cool}$ relationship may be observed if the dominant driver of condensation is the AGN jet propelling low cooling time gas from lower to higher altitudes, where it can condense.  Since the cooling time of the uplifted gas determines the radius where the gas becomes thermally unstable, it also determines how much work must be done by the jet to lift the gas to a radius where condensation can occur. Therefore, if the uplift of plasma is driving condensation, we expect to see a SFR-mass condensation efficiency relationship in the form of a SFR-$t_{cool}$ relationship. Such a relationship would appear similar to an SFR-$t_{cool}/t_{ff}$ relationship in a sample such as CLASH with a narrow range of cluster masses, since all of the clusters in CLASH have similar free-fall times in their cores.

We thus investigated the possibility that the underlying relationship we observe in Figure 5 is primarily between SFR and $t_{cool}$. The range of $t_{ff}$ at $0.025R_{500}$ in CLASH clusters with star-forming BCGs is smaller than the range of $t_{cool}$ -- while $t_{ff}$ varies by a factor of $\sim$ 1.5 across the sample, $t_{cool}$ varies by a factor of $\sim$ 3. Hence, the effect of $t_{ff}$ on the relationship with SFR is modest, and our ability to distinguish between an underlying SFR-$t_{cool}/t_{ff}$ relationship vs. a SFR-$t_{cool}$ relationship is limited. Several lines of reasoning provide evidence for a non-negligible contribution from $t_{ff}$, although we cannot rule out the interpretation that the underlying relationship we observe is solely between SFR and $t_{cool}$.

\begin{figure}
\begin{center}
\begin{tabular}{c}
\includegraphics[height=9cm]{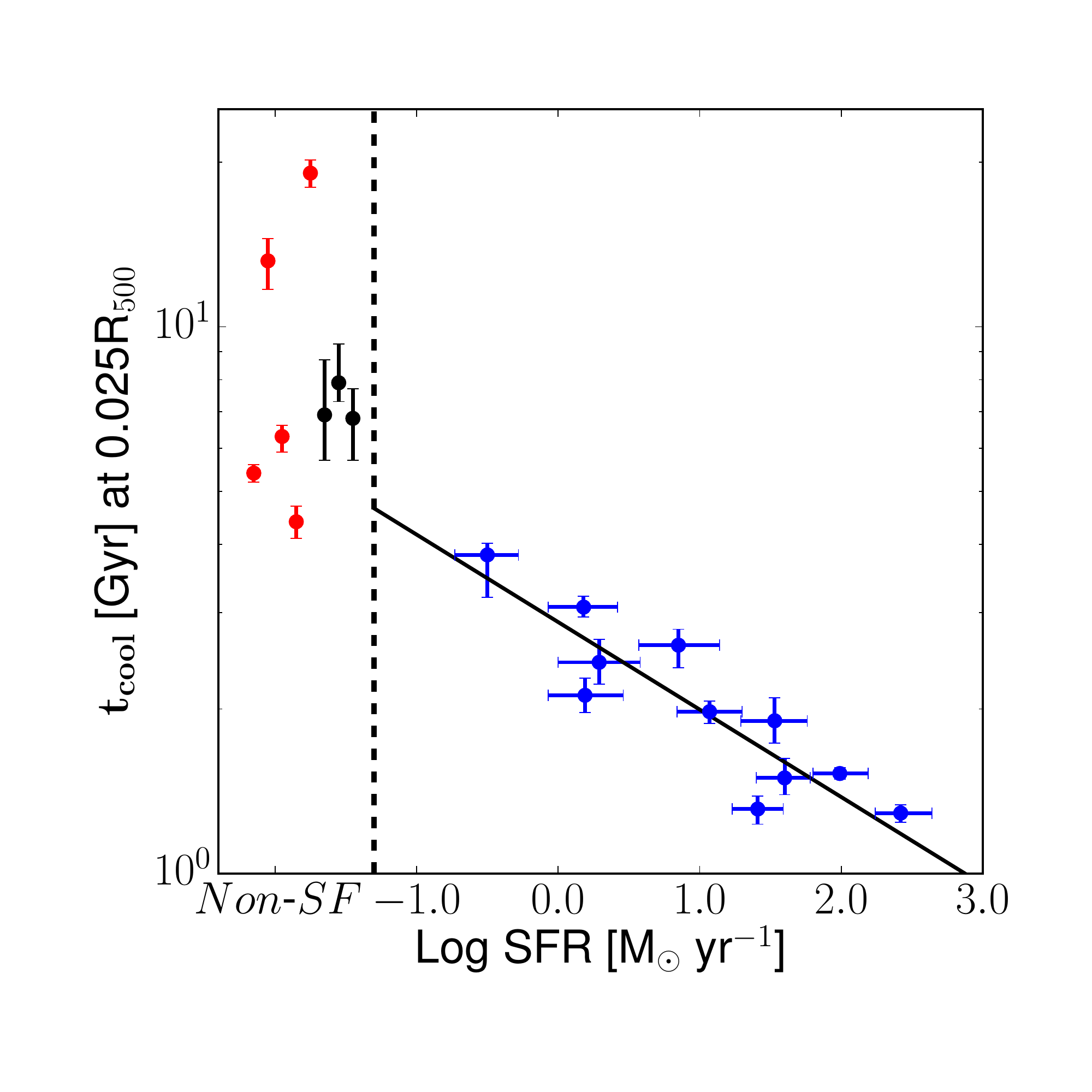}
\end{tabular}
\end{center}
\caption[]
{ \label{fig:SFR_TC} Log$_{10}$ $t_{cool}$ as a function of Log$_{10}$ SFR is shown, with color-coding and uncertainties analogous to those given in Figure~\ref{fig:SFR_Ratio}. Values of the ratio $t_{cool}$ were measured at a radius of $0.025 R_{500}$ for each cluster. Values of $t_{cool}$ at $0.025 R_{500}$ for CLASH clusters with non-starforming BCGs are shown to the left of the vertical dashed line. }
\end{figure}

\begin{figure}
\begin{center}
\begin{tabular}{c}
\includegraphics[height=9cm]{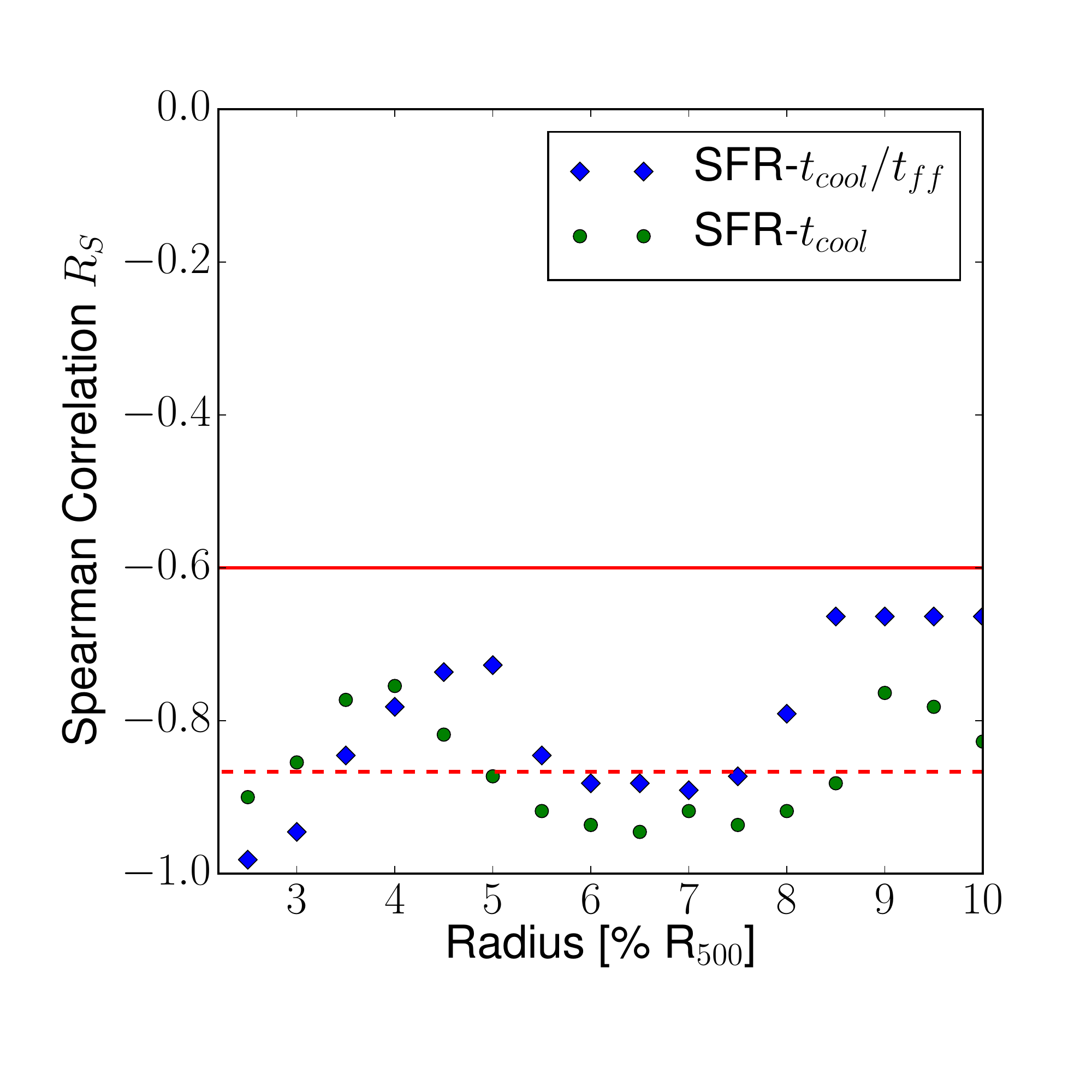}
\end{tabular}
\end{center}
\caption[]
{ \label{fig:Corr_Radius} The Spearman correlation coefficient for Log$_{10}$ SFR and Log$_{10}$ $t_{cool}/t_{ff}$ is shown as a function of the radius used to measure $t_{cool}/t_{ff}$. Correlations are measured at 5 kpc intervals, and are plotted as the blue diamonds. The solid red line denotes where the one-tailed P-value = 0.05, points below it have  P $<$ 0.05. Points below the dashed red line have P $<$ 0.0025.}
\end{figure}

SFR and $t_{cool}$ are correlated as shown in Figure \ref{fig:SFR_TC}, with a Spearman correlation coefficient of -0.90 (and a Pearson coefficient of -0.90). We calculate an intrinsic scatter between the two quantities of $0.05^{+0.03}_{-0.01}$ dex ($< 0.18$ dex at $3\sigma$). The Spearman correlation coefficient in the SFR-$t_{cool}/t_{ff}$ and SFR-$t_{cool}$ relationships as a function of sampling radius are presented in Figure \ref{fig:Corr_Radius}. We do not find a substantial difference between $t_{cool}$ and $t_{cool}/t_{ff}$ in terms of how tightly these quantities relate to the SFR. However, the distinction between $t_{cool}$ in the star-forming and non-starforming CLASH clusters is less clear than the distinction between $t_{cool}/t_{ff}$ in these two populations, and $t_{cool}$ for two clusters with non-starforming BCGs is comparable to $t_{cool}$ in the clusters with BCGs exhibiting $0.1-1$ M$_{\odot}$ yr$^{-1}$ of star formation.

\begin{figure*}
{\epsfig{file=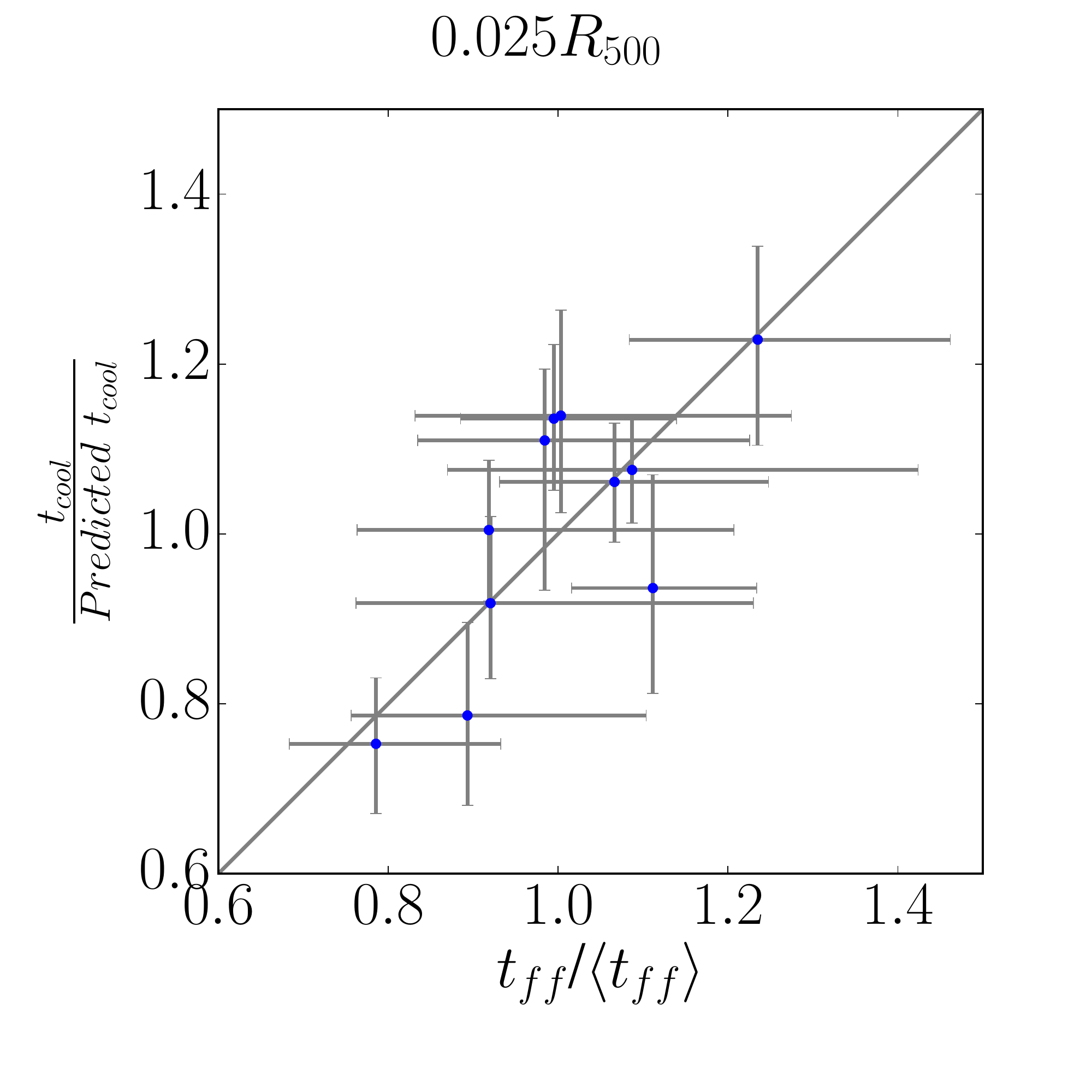,width=8.5cm,angle=0}
\label{fig:subfigure1}}
{\epsfig{file=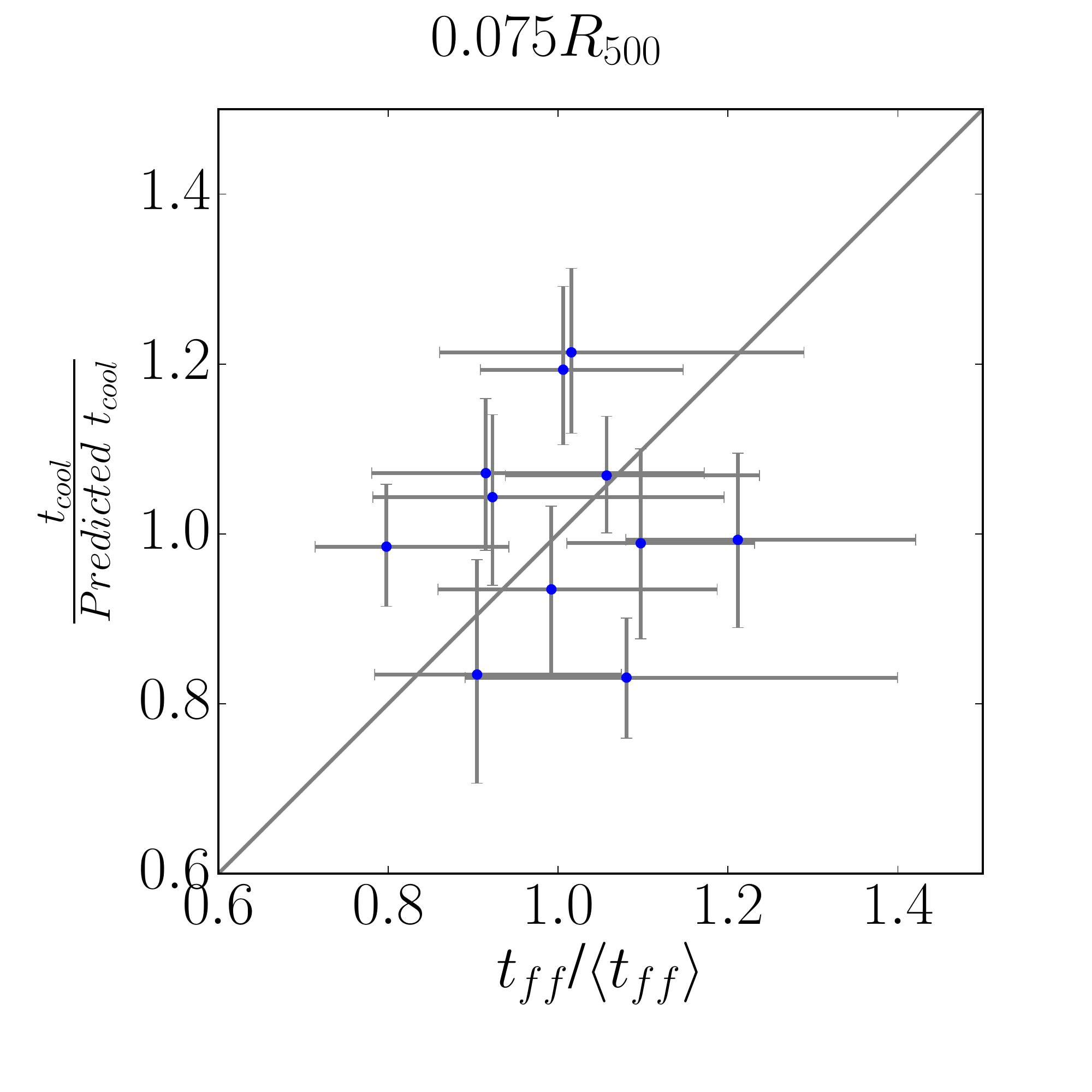,width=8.5cm,angle=0}
\label{fig:subfigure2}}
{\epsfig{file=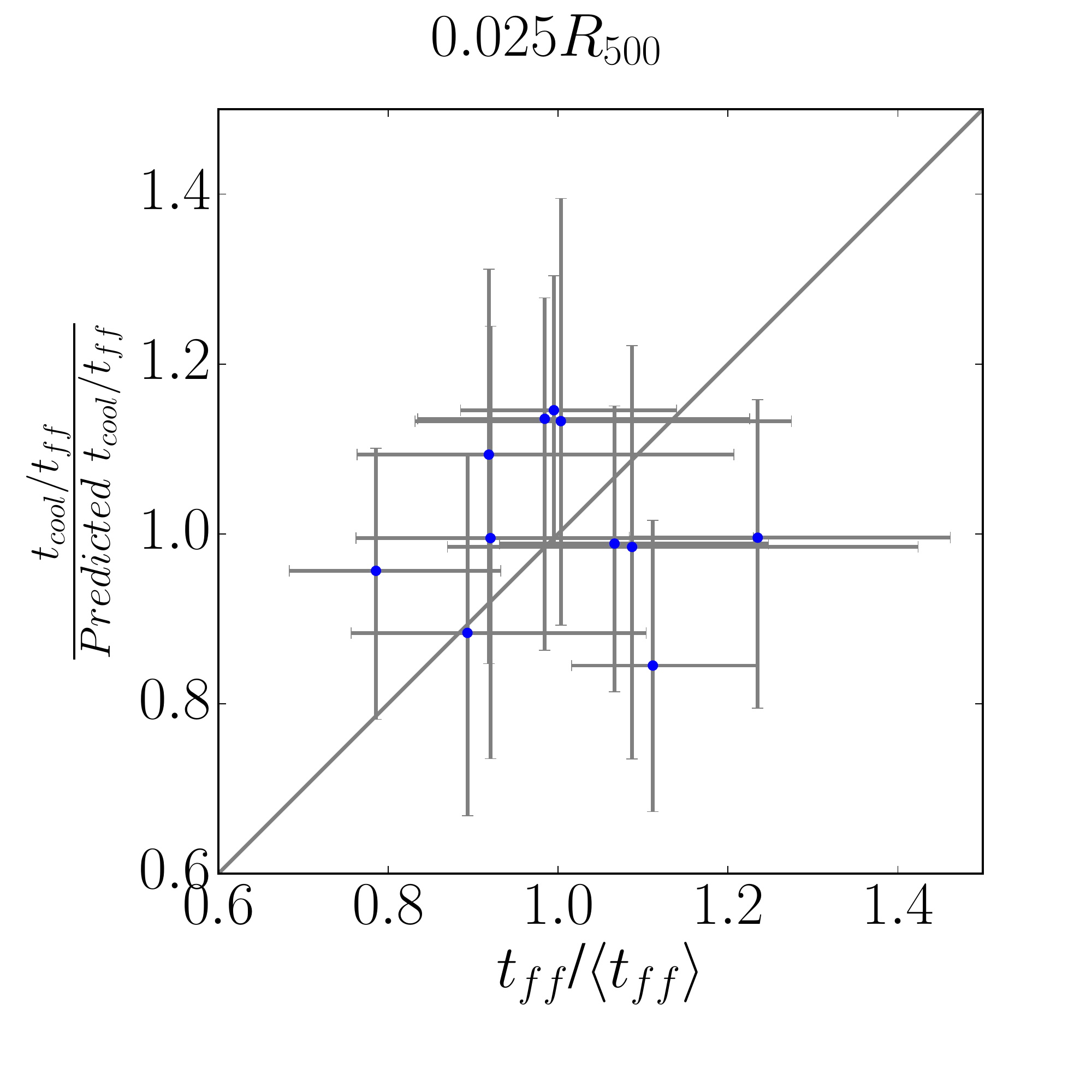 ,width=8.5cm,angle=0}
\label{fig:subfigure}}
{\epsfig{file=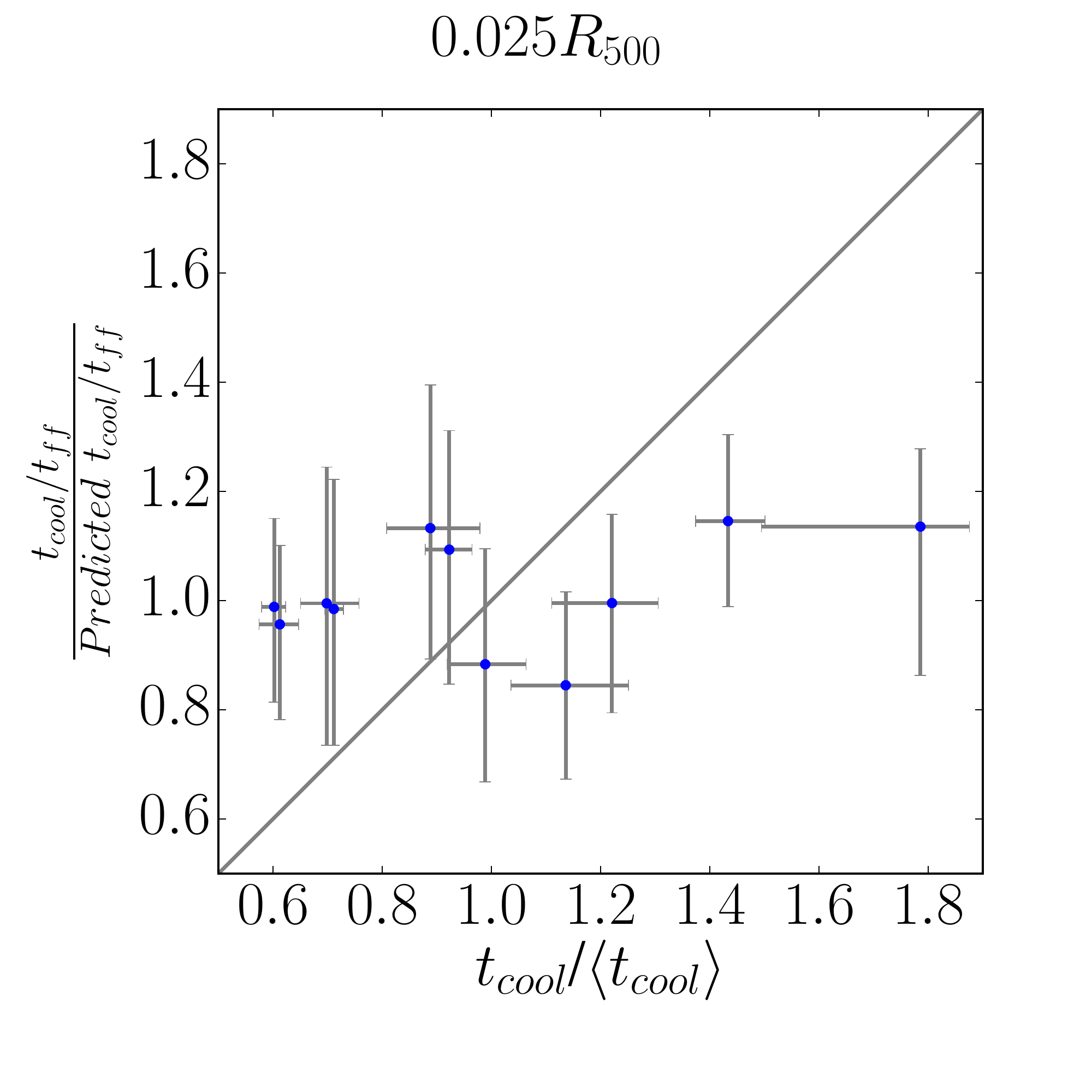 ,width=8.5cm,angle=0}
\label{fig:subfigure}}
\caption{\label{fig:Resid_Corr} Plots showing best fit residuals vs $t_{ff}$ and $t_{cool}$. The top row shows $t_{ff}$ vs. the residuals for $t_{cool}$-SFR at $0.025R_{500}$ and $0.075R_{500}$. The bottom row shows $t_{ff}$ vs. the residuals for $t_{cool}/t_{ff}$-SFR and $t_{cool}$ vs. the residuals for $t_{cool}/t_{ff}$-SFR, both at $0.025R_{500}$. The x-axis values are normalized by the mean values of these quantities for the CLASH star forming sample. In each case, the grey line denotes where the plotted quantities are equal.}
\end{figure*}

The situation becomes clearer when we examine the relationship between $t_{cool}$, $t_{ff}$, and the residuals in the fits to the CLASH SFR-$t_{cool}/t_{ff}$ and SFR-$t_{cool}$ datasets. In Figure \ref{fig:Resid_Corr}, we show $t_{ff}$ vs. the residuals when we fit a log-log relationship to the SFR-$t_{cool}$ dataset at a radius of $0.025R_{500}$ and at $0.075R_{500}$. We also show $t_{ff}$ vs. the residuals of SFR-$t_{cool}/t_{ff}$ at $0.025R_{500}$ and $t_{cool}$ vs. the residuals of SFR-$t_{cool}/t_{ff}$ at $0.025R_{500}$. We find that at $0.025R_{500}$, $t_{ff}/\langle t_{ff}\rangle$, where $\langle t_{ff}\rangle$ is the mean $t_{ff}$ for the sample, is correlated with the residuals in the fit to the SFR-$t_{cool}$ dataset, $\frac{t_{cool}}{Predicted \textrm{ }t_{cool}}$, where $Predicted \textrm{ } t_{cool}$ is $t_{cool}$ predicted by the SFR-$t_{cool}$ relationship for a given SFR. These two quantities are consistent with $\frac{t_{cool}}{Predicted \textrm{ } t_{cool}} = t_{ff}/\langle t_{ff}\rangle$, implying that dividing $t_{cool}$ by $t_{ff}$ will offset the residuals in the SFR-$t_{cool}$ relationship. The relationship between $t_{ff}$ and  $\frac{t_{cool}}{Predicted \textrm{ } t_{cool}}$ at $0.025R_{500}$ has a positive slope with $\sim 98\%$ confidence. The plot in Figure \ref{fig:Resid_Corr} does not have obvious outliers, so the scatter reduction seen at $0.025R_{500}$ by dividing $t_{cool}$ by $t_{ff}$ is not attributable to reducing the residuals in an extreme outlier. There is also no evidence that $t_{ff}$ or $t_{cool}$ are correlated with the other residuals we examined.

\subsection{BCG Activity and Low Cooling Time Gas}\label{sec-bcg-gas}
 
We find a relationship between $\Delta t_{b}$ and $t_{cool}$ that shows a possible positive correlation between these two quantities with $\Delta t_{b}$ approaching $t_{cool}$ at Gyr timescales. Since $\Delta t_{b}$ is potentially correlated with the SFR, and individual measurement uncertainties are relatively large ($\sim$0.3 -- 0.4 dex) compared to the range of $\Delta t_{b}$ values measured in the sample (1.5 dex), the observed relationship has limited power to constrain models of cluster-scale feedback evolution. However, our measurements of the starburst durations have implications for understanding how AGN-regulated feedback progresses over time.
 
Four of the 11 BCG starburst durations are reported as lower limits. These BCGs may be undergoing continuous star formation. However, given the association between longer $\Delta t_{b}$ and lower SFR, and shorter $\Delta t_{b}$ and higher SFR, we suspect that star formation in these BCGs is slowly decaying. Alternatively, BCGs undergoing feedback may initially exhibit large starbursts before settling down to a relatively steady state with anywhere between $\sim$ 1-10 M$_{\odot}$ yr$^{-1}$ of star formation. Star formation may also `flicker' on timescales that are short relative to the values of $\Delta t_{b}$ we measure.

Along with recent spectroscopic observations in the far-UV, the long burst durations we observe suggest that star formation and the thermodynamical state of the ICM in the cool-core is temporally decoupled from the gas condensation rate fueling AGN feedback, which may feature irregular spikes and dips over time. Recent observations of RXJ1532.9+3021 with Cosmic Origins Spectrograph (COS) \citep{2012Green_COS} reveal that star formation in this BCG exceeds the upper limit on gas cooling measured by \ion{N}{5} and \ion{O}{6} by a factor of $\sim$10, suggesting the initial build up of molecular gas in this system has run its course \citep{2017Donahue_RXJ1532Cooling}. Meanwhile similar spectroscopy in Abell 1795 and the Phoenix cluster reveal that gas cooling outstrips the SFR \citep{2014McDonald_A1795, 2015McDonald_PhoenixCOS}. Delayed consumption of molecular gas may allow the star formation history of BCGs to `smooth over' these intermittent spikes. Slowly decaying BCG starbursts may also trace a transition from an initial mode of rapid condensation (with $\sim$100-1000 M$_{\odot}$ yr$^{-1}$ of gas condensing) to a long-duration mode of more modest condensation (with $\sim$1-10 M$_{\odot}$ yr$^{-1}$ of condensation). If star formation lags behind cooling in making this transition, the resulting observables would be consistent both with our results and the offset between SFRs and cooling rates seen in \cite{2016Molendi_Cooling} and \cite{2017Donahue_RXJ1532Cooling}.
 
A positive correlation between $\Delta t_{b}$ and $t_{cool}$ and the existence of long-lived  ($> 1$ Gyr) starbursts are consistent with the mass condensation efficiency interpretation of the $t_{cool}/t_{ff}$-SFR scaling relationship discussed in Section~\ref{sec-tcool}. Assuming that the duration of the BCG starburst is a proxy for the duration of AGN feedback, then longer periods of star formation in a BCG imply more energy injection into the surrounding ICM, which in turn raises the cooling time of the surrounding ICM. This in turn diminishes the rate at which molecular gas condenses, and gradually shuts off both the starburst and feedback. The duration $\Delta t_{b}$ and $t_{cool}$ ought to converge in the limit of a long starburst, although feedback mechanisms could run out of fuel and shut down before the two quantities converge depending on how inefficiently mass condenses out of the ICM. Alternatively, a quasi-steady state may be reached with effectively continuous star formation in the BCG coupled with ICM plasma with $t_{cool}$ of a few Gyr being replenished by hotter gas at about the same rate it condenses. In this case, star formation and feedback may decay very slowly, or not at all. Our results do not rule this latter scenario out, leaving the following possible options:
(1) BCG activity is cyclical, (2) BCG activity approaches a slowly decaying quasi-steady state, or (3) BCG activity is cyclical but the duty cycle is comparable to the age of the cool core of the cluster.
 
Finally, it has been noted that there are few examples of BCGs with post-starburst spectra in most BCG samples \citep{2012Liu_PostStarburst, 2016Loubser_BCGSF}. Indeed, none of the SOAR or SDSS spectra of CLASH BCGs in \cite{2015Fogarty_CLASH} show post--starburst features. The lack of post-starburst BCGs is consistent with our estimates of $\sim$ Gyr duration episodes of star formation. \cite{2009Wild_PostStarburst} notes that for a galaxy with an exponentially decaying starburst to feature a post-starburst spectrum, the burst's decay timescale would have to be $\lesssim 0.1$ Gyr and the burst would have to account for at least 5-10$\%$ of the galaxy's stellar mass. Given the nature of long-duration star formation in our observations and models of feedback-regulated cooling, we would not expect to see a substantial population of post-starburst BCGs.

\subsection{Characteristics of Large Starbursts in BCGs}\label{sec-largeburst}

Large BCG starbursts (with SFRs $\gtrsim 100$ M$_{\odot}$ yr$^{-1}$) may differ from their more modest counterparts in several ways. Firstly, the SFRs and dust masses in the CLASH BCGs are consistent with using \cite{2014Hjorth_SFRDust} to describe the relationship between BCG star formation and dust mass. Specifically, MACS1931.8-2653 and the Phoenix cluster match the flattening slope and eventual turnover in the \cite{2014Hjorth_SFRDust} relation closely, raising the possibility that these BCGs harbor starbursts similar to the starbursts in the submillimeter-detected population of galaxies studied by \cite{2014Hjorth_SFRDust}. The Hjorth et al. sample extends the study of dust and star formation conducted in \cite{2010daCunha_DustMass} to cover starbursts to galaxies forming stars at $\gtrsim 1000$ M$_{\odot}$ yr$^{-1}$. Since the largest BCG starbursts in the CLASH sample also began forming stars more recently we hypothesize that, like these massive field galaxy starbursts, BCGs with large star formation rates are either forming or building up their dust reservoirs.
  
Secondly, both MACS1931.8-2653 and RXJ1532.9 +3021 boast prodigious SFRs and distinctive X-ray cavities. Furthermore, MACS1931.8-2653 exhibits an X-ray loud AGN, a feature which is not obvious in the other CLASH clusters. These features are noteworthy because the starbursts in RXJ1532.9+3021 and MACS1931.8-2653 are the youngest in our sample at log$_{10}$ $\Delta t_{b}$ = $-0.39^{+0.38}_{-0.41}$ and $-1.01^{+0.35}_{-0.34}$ Gyr, respectively. Our findings suggest that stronger X-ray features may be associated with younger, larger BCG starbursts, such as the extreme example of BCG star formation present in the Phoenix Cluster \citep{2012McDonald_Phoenix, 2013McDonald_Phoenix, 2014McDonald_PhoenixGas}.
 
\section{Conclusions}\label{sec-conclude}
 
We use multi-wavelength observations from \textit{HST, Spitzer} and \textit{Herschel} to derive SFRs, starburst durations and dust masses from fitting the SEDs of the 11 CLASH BCGs with extended UV and nebular line emission features. The SFRs and dust masses span nearly three orders of magnitude, with SFRs ranging from $\sim$0.3 to $\sim$ 250 M$_{\odot}$ yr$^{-1}$, and dust masses ranging from $\sim 10^{6}$ to $\sim 10^{9}$ M$_{\odot}$. BCG starbursts are $\lesssim$ 100 Myr to several Gyr old.

We find compelling evidence for a direct link between the thermodynamic state of the ICM and BCG star formation. Specifically, we observe a tight SFR-$t_{cool}/t_{ff}$ relationship and a relationship with an intrinsic scatter $\sigma_{i} < 0.15$ dex and a slope of $-0.15 \pm 0.03$. These results strongly suggest that thermally unstable ICM plasma with a low cooling time is the source of material that forms the reservoir of cool gas fueling star formation in the CLASH BCGs and that BCG star formation and feedback either exhausts the supply of this material on Gyr timescales or settles into a state with relatively modest ($\sim 1-10$ M$_{\odot}$ yr$^{-1}$) continuous star formation.

Even if the trend we observe between SFR and $t_{cool}/t_{ff}$ is due to an underlying trend between SFR and $t_{cool}$ we  would still find the AGN-driven condensation and precipitation model to provide a compelling explanation.
However, since we find a $\sim 2\sigma$ detection of a correlation between $t_{ff}$ and the residuals from the best-fit line SFR-$t_{cool}$ relationship, we suspect that $t_{ff}$ plays a role in the physics governing the relationship between SFR and the ICM. With larger datasets spanning a larger range of $t_{ff}$ that include both measurements of $t_{cool}$ and $t_{ff}$, it will be possible to constrain the role of $t_{ff}$ in cluster core dynamics with greater confidence.

While our results are not a direct observation of feedback-induced condensation, the condensation model provides predictions that are consistent with our observations and provides a framework for understanding why the star formation rate in BCGs would scale with $t_{cool}/t_{ff}$. If $t_{cool}/t_{ff}$ scales with the critical density for ICM perturbation collapse, $t_{cool}/t_{ff}$ measures the efficiency of condensation, and our findings can be interpreted as a relationship between SFR and the condensation efficiency of the ICM.
 
Our study also raises several questions about the life-cycle of cooling and feedback. We present evidence that star formation episodes in BCGs with larger SFRs are younger relative to BCGs with more modest SFRs, suggesting that BCG starbursts decay over time. We also present evidence that BCG star formation can persist over $\gtrsim$ Gyr timescales. However, it is not clear whether the starbursts in BCGs are cyclical, or slowly decaying single events that may eventually settle into a low level of persistent star formation. It is also possible we are observing the superposition of several shorter-lived events.

The dust and star formation in BCGs is consistent with the SFR-M$_{d}$ relationship described in \cite{2014Hjorth_SFRDust}. This consistency holds even when we include the SFR and dust mass measured for the Phoenix cluster BCG, which forms stars at a rate of $\sim 500$ M$_{\odot}$ yr$^{-1}$. Our results lead us to hypothesize that, while uncommon in BCGs, large starbursts like those in MACS1931.8-2635 and the Phoenix cluster may have properties in common with young, violent starbursts in field galaxies.
 
Our work shows a direct link between the amount of star formation occurring in a BCG and the thermodynamical state of the surrounding ICM.  The quality of the data we used in combination with the sample selected allowed us to study specifically the interaction between BCG and the ICM in a uniformly-selected sample of cool-core clusters. These clusters and their BCGs have properties that are consistent with a process of  cooling and feedback and the results presented herein bring us closer to a complete understanding of feedback in galaxy clusters.  A clear extension of this study would be to examine how the BCG-ICM relationships evolve with redshift and cluster mass by analyzing additional deep, multi-wavelength galaxy cluster surveys.

\section*{Acknowledgments}

We acknowledge insightful discussions with Mark Voit and Anton Koekemoer that helped enhance our interpretation of the observations. We thank Michael McDonald for providing the infrared SED of the Phoenix Cluster BCG. We also thank the anonymous referee for insightful comments that significantly improved the quality of this paper. This research is supported, in part, by NASA grants HSTGO-12065.01-A and HSTGO-13367. The CLASH Multi-Cycle Treasury Program is based on observations made with the NASA/ESA \textit{Hubble Space Telescope} and which is operated by the Space Telescope Science Institute. This work is also based, in part, on observations made with the \textit{Spitzer Space Telescope} and the \textit{Herschel Space Observatory}. The \textit{Spitzer Space Telescope}  is operated by the Jet Propulsion Laboratory, California Institute of Technology under a contract with NASA. \textit{Herschel} is an ESA space observatory with science instruments provided by European-led Principal Investigator consortia and with important participation from NASA. 

\bibliography{CLASH_SED_Fitting}


\newpage 
\appendix

In this appendix, we explore the impact of the choice of stellar initial mass function and of the dust models.  The attenuation law can significantly affect both the shape of the UV-through-optical of our synthetic SEDs and the normalization of the dust emission component of the SEDs. Hence, understanding if there are significant changes to the derived stellar population properties is essential in assessing the robustness of our work. We also examine the posterior probability distributions for SFR, $\Delta t_{b}$ and $M_{d}$ obtained with {\tt iSEDfit}.

We specifically investigate the sensitivity of our results when we adopt a modified \cite{2000Calzetti_Extinction} law (Table \ref{table:Calz_Params}) or a \cite{2000Witt_Extinction} clumpy SMC-like dust in a shell geometry (Table \ref{table:Witt_Params}).
Tables of SED fitting results analogous to Table \ref{table:SED_Dust_Params} are presented assuming a modified Calzetti attenuation law (Table \ref{table:Calz_Params}) and clumpy SMC-like attenuation (Table \ref{table:Witt_Params}). 
The values of SFR, $\Delta t_{b}$ and $M_{d}$ we report vary little between both attenuation models and the attenuation model assumed in the main paper. The largest difference is in $\Delta t_{b}$ between the \cite{2000Witt_Extinction} model and either model based on the \cite{2000Calzetti_Extinction} law-- burst durations fit assuming the former dust law are systematically 0.1-0.3 dex shorter. However, the differences between these sets of results are statistically insignificant, and do not qualitatively affect the interpretation of our data. 

Adopting a \cite{2003Chabrier_IMF} IMF results in a systematic shift downward in the inferred SFRs, consistent with the \cite{2003Chabrier_IMF} IMF being about 0.25 dex lighter than the \cite{1955Salpeter_IMF} IMF. However, like variations cause by assuming different dust attenuation models, these shifts are not statistically significant, and do not qualitatively change the trends between stellar and ICM parameters we observe. The systematic shift in SFR is due to the fact that the mass-to-light ratio changes when changing from a \cite{1955Salpeter_IMF} to a \cite{2003Chabrier_IMF}.

The marginal probability distributions for parameters of interest to our study (SFR, $\Delta t_{b}$ and $M_{d}$) are depicted in Figure \ref{fig:Posteriors}, and are seen to have a well-defined mode for each distribution, although there is evidence for bimodality in some of the distributions. We show an example of marginal posteriors SFR, $\Delta t_{b}$, $M_{d}$, $A_{V}$, burst decay percentage, galaxy age, $\tau$, and metallicity for Abell 383, in order to demonstrate which parameters are and are not well constrained by our fits. For five of the BCGs, the marginal distributions for $\Delta t_{b}$ are cut off by the age of the Universe at the redshift of the galaxy cluster. In \ref{fig:DTB_Posteriors}, we show the marginal posterior distributions for $\Delta t_{b}$ when the SED fit parameter {\sc fburst} is constrained to the range [0.025, 0.4]

In summary, reasonable variations in the assumed attenuation law or in the assumed IMF do not significantly change any of the key results or conclusions in this study.

\begin{table*}[h!]  
\small  
\caption{BCG Stellar and Dust Parameters assuming a Chabrier IMF and  Calzetti attenuation law}
\label{table:Calz_Params}  
\vspace{1mm}  
\centering  
{  
\begin{tabular}{lccccccc}  
\hline
\hline
 & log$_{10}$ SFR & log$_{10}$ $\Delta t_{b}$ & log$_{10}$ $M_{d}$ & A$_{\textrm{V}}$ & $R_{S}$$^{a}$ & $R_{S}$ & $R_{S}$ \\ 
 &                    &    & & & SFR$\times \Delta t_{b}$ & SFR$\times$ $M_{d}$ & M$_{\textrm{d}}\times\Delta t_{b}$ \\
BCG & (M$_{\odot}$ yr$^{-1}$) & (Gyr) & (M$_{\odot}$) & (mag) & & & \\ 
\hline
\\
Abell 383 & $-0.02^{+0.23}_{-0.24}$ & $ > -0.47$ & $7.39^{+0.78}_{-0.82}$ & $0.48^{+0.22}_{-0.24}$ & {-0.11} & {0.30} & {0.04} \\ \\
 
MACS0329.7-0211 & $1.39^{+0.19}_{-0.22}$ & $> -0.68$ & $8.39^{+0.66}_{-0.64}$ & $0.53^{+0.18}_{-0.19}$ & {0.07} & {0.01} & {-0.07} \\ \\
 
MACS0429.6-0253 & $1.32^{+0.23}_{-0.22}$ & $> -0.73$ & $8.49^{+0.67}_{-0.65}$ & $0.74^{+0.24}_{-0.22}$ & {0.10} & {0.17} & {0.09} \\ \\
 
MACS1115.9+0219 & $0.65^{+0.27}_{-0.26}$ & $> -0.75$ & $7.54^{+0.96}_{-0.88}$ & $0.43^{+0.29}_{-0.29}$ & {-0.08} & {0.45} & {-0.04} \\ \\
 
MACS1423.8+2404 & $1.21^{+0.19}_{-0.2}$ & $> -0.71$ & $8.42^{+0.74}_{-0.75}$ & $0.41^{+0.17}_{-0.18}$ & {-0.10} & {0.18} & {0.04} \\ \\
 
MACS1720.3+3536 & $0.01^{+0.26}_{-0.23}$ & $0.13^{+0.4}_{-0.41}$ & $7.61^{+0.74}_{-0.68}$ & $0.62^{+0.26}_{-0.26}$ & {-0.19} & {0.17} & {-0.0} \\ \\
 
MACS1931.8-2653 & $2.21^{+0.19}_{-0.19}$ & $-0.93^{+0.49}_{-0.51}$ & $8.86^{+0.38}_{-0.39}$ & $0.88^{+0.19}_{-0.21}$ & {-0.43} & {-0.27} & {0.11} \\ \\
 
MS2137-2353 & $0.05^{+0.27}_{-0.29}$ & $0.15^{+0.4}_{-0.44}$ & $7.43^{+0.8}_{-0.86}$ & $0.44^{+0.32}_{-0.3}$ & {-0.27} & {0.36} & {-0.15} \\ \\
 
RXJ1347.5-1145 & $0.87^{+0.22}_{-0.22}$ & $> -0.66$ & $7.74^{+0.91}_{-0.81}$ & $0.31^{+0.23}_{-0.22}$ & {-0.11} & {0.45} & {0.04} \\ \\
 
RXJ1532.9+3021 & $1.8^{+0.2}_{-0.18}$ & $-0.2^{+0.51}_{-0.55}$ & $8.79^{+0.47}_{-0.48}$ & $0.91^{+0.21}_{-0.21}$ & {-0.34} & {-0.10} & {0.07} \\ \\
 
RXJ2129.7+0005 & $-0.7^{+0.24}_{-0.21}$ & $> -0.01$ & $6.71^{+0.81}_{-0.73}$ & $0.28^{+0.14}_{-0.14}$ & {-0.30} & {0.27} & {-0.19} \\ \\
\hline
\end{tabular}  
\begin{flushleft}
$^{a}$ Spearman correlation coefficients for the sample of pairs of parameters obtained by sampling the posterior probability distribution of models. \\
$^{b}$ For log$_{10}$ $\Delta t_{b}$ posterior probability histograms that peak near the upper bound of the parameter space, we report the 1$\sigma$ confidence interval as a lower limit on log$_{10}$ $\Delta t_{b}$. \\ 
\end{flushleft}  
}  
\end{table*}

\begin{table*}[h!]  
\small  
\caption{BCG Stellar and Dust Parameters assuming a modified Calzetti attenuation law}
\label{table:Calz_Params}  
\vspace{1mm}  
\centering  
{  
\begin{tabular}{lccccccc}  
\hline
\hline
 & log$_{10}$ SFR & log$_{10}$ $\Delta t_{b}$ & log$_{10}$ $M_{d}$ & A$_{\textrm{V}}$ & $R_{S}$$^{a}$ & $R_{S}$ & $R_{S}$ \\ 
 &                    &    & & & SFR$\times \Delta t_{b}$ & SFR$\times$ $M_{d}$ & M$_{\textrm{d}}\times\Delta t_{b}$ \\
BCG & (M$_{\odot}$ yr$^{-1}$) & (Gyr) & (M$_{\odot}$) &(mag) & & & \\   
\hline
\\
Abell 383 & $0.18^{+0.24}_{-0.26}$ & $> -0.54^{b}$ & $7.41^{+0.83}_{-0.78}$ & $0.47^{+0.25}_{-0.26}$ & {-0.13} & {0.35} & {0.05} \\ \\
 
MACS0329.7-0211 & $1.6^{+0.19}_{-0.21}$ & {$> -0.31$} & $8.39^{+0.72}_{-0.69}$ & $0.53^{+0.19}_{-0.18}$ & {-0.11} & {0.08} & {0.04} \\ \\
 
MACS0429.6-0253 & $1.54^{+0.24}_{-0.25}$ & $> -0.80$ & $8.5^{+0.75}_{-0.75}$ & $0.76^{+0.24}_{-0.24}$ & {0.02} & {0.26} & {-0.01} \\ \\
 
MACS1115.9+0219 & $0.85^{+0.28}_{-0.28}$ & {$> -0.38$} & $7.61^{+0.87}_{-0.82}$ & $0.43^{+0.28}_{-0.28}$ & {-0.0} & {0.42} & {0.03} \\ \\

MACS1423.8+2404 & $1.4^{+0.21}_{-0.23}$ & $> -0.76$ & $8.37^{+0.77}_{-0.78}$ & $0.42^{+0.19}_{-0.2}$ & {0.02} & {0.21} & {-0.02} \\ \\
 
MACS1720.3+3536 & $0.19^{+0.24}_{-0.24}$ & $> -0.37$ & $7.64^{+0.68}_{-0.74}$ & $0.59^{+0.23}_{-0.24}$ & {-0.17} & {0.27} & {-0.04} \\ \\
 
MACS1931.8-2653 & $2.41^{+0.21}_{-0.19}$ & $-0.95^{+0.52}_{-0.58}$ & $8.89^{+0.37}_{-0.37}$ & $0.85^{+0.2}_{-0.2}$ & {-0.53} & {-0.25} & {0.15} \\ \\
 
MS2137-2353 & $0.27^{+0.3}_{-0.27}$ & $0.14^{+0.44}_{-0.52}$ & $7.43^{+0.75}_{-0.78}$ & $0.44^{+0.33}_{-0.34}$ & {-0.21} & {0.36} & {0.0} \\ \\
 
RXJ1347.5-1145 & $1.07^{+0.23}_{-0.21}$ & $-0.11^{+0.59}_{-0.61}$ & $7.71^{+0.92}_{-0.87}$ & $0.31^{+0.23}_{-0.22}$ & {0.05} & {0.37} & {0.09} \\ \\ 
 
RXJ1532.9+3021 & $1.99^{+0.2}_{-0.19}$ & $-0.26^{+0.57}_{-0.55}$ & $8.77^{+0.51}_{-0.49}$ & $0.88^{+0.2}_{-0.19}$ & {-0.43} & {-0.12} & {0.03} \\ \\
 
RXJ2129.7+0005 & $-0.53^{+0.24}_{-0.24}$ & $> 0.06$ & $6.65^{+0.83}_{-0.73}$ & $0.26^{+0.18}_{-0.17}$ & {-0.14} & {0.04} & {0.18} \\ \\
\hline
\end{tabular}  
\begin{flushleft}
$^{a}$ Spearman correlation coefficients for the sample of pairs of parameters obtained by sampling the posterior probability distribution of models. \\
$^{b}$ For log$_{10}$ $\Delta t_{b}$ posterior probability histograms that peak near the upper bound of the parameter space, we report the 1$\sigma$ confidence interval as a lower limit on log$_{10}$ $\Delta t_{b}$. \\ 
\end{flushleft}  
}  
\end{table*}

\begin{table*}[h!]  
\small  
\caption{BCG Stellar and Dust Parameters assuming Witt clumpy SMC-like dust in a shell geometry}
\label{table:Witt_Params}  
\vspace{1mm}  
\centering  
{  
\begin{tabular}{lccccccc}  
\hline
\hline
 & log$_{10}$ SFR & log$_{10}$ $\Delta t_{b}$ & log$_{10}$ $M_{d}$ & A$_{\textrm{V}}$ & $R_{S}$$^{a}$ & $R_{S}$ & $R_{S}$ \\ 
 &                    &    & & & SFR$\times \Delta t_{b}$ & SFR$\times$ $M_{d}$ & M$_{\textrm{d}}\times\Delta t_{b}$ \\
BCG & (M$_{\odot}$ yr$^{-1}$) & (Gyr) & (M$_{\odot}$) &(mag) & & & \\
\hline  
\\
Abell 383 & $0.36^{+0.25}_{-0.24}$ & {$> -0.39$} & $7.6^{+0.75}_{-0.78}$ & $0.9^{+0.61}_{-0.56}$ & {-0.23} & {0.27} & {-0.06} \\ \\
 
MACS0329.7-0211 & $1.61^{+0.19}_{-0.19}$ & $-0.31^{+0.74}_{-0.83}$ & $8.41^{+0.7}_{-0.69}$ & $0.58^{+0.39}_{-0.42}$ & {-0.19} & {-0.08} & {-0.01} \\ \\ 
 
MACS0429.6-0253 & $1.57^{+0.21}_{-0.2}$ & $> -1.24^{b}$ & $8.54^{+0.65}_{-0.64}$ & $1.05^{+0.56}_{-0.57}$ & {-0.10} & {0.17} & {0.01} \\ \\
 
MACS1115.9+0219 & $0.98^{+0.26}_{-0.26}$ & $> -1.06$ & $7.84^{+0.82}_{-0.81}$ & $0.67^{+0.57}_{-0.52}$ & {-0.15} & {0.27} & {-0.04} \\ \\
 
MACS1423.8+2404 & $1.51^{+0.19}_{-0.19}$ & $-0.27^{+0.72}_{-0.77}$ & $8.49^{+0.73}_{-0.75}$ & $0.56^{+0.4}_{-0.42}$ & {-0.15} & {0.05} & {-0.01} \\ \\
 
MACS1720.3+3536 & $0.43^{+0.26}_{-0.26}$ & $-0.1^{+0.57}_{-0.62}$ & $7.66^{+0.7}_{-0.71}$ & $1.12^{+0.6}_{-0.62}$ & {-0.19} & {0.11} & {-0.01} \\ \\
 
MACS1931.8-2653 & $2.36^{+0.18}_{-0.18}$ & $-1.12^{+0.54}_{-0.56}$ & $8.88^{+0.37}_{-0.37}$ & $1.09^{+0.51}_{-0.52}$ & {-0.33} & {-0.23} & {0.03} \\ \\ 
 
MS2137-2353 & $0.55^{+0.31}_{-0.31}$ & $-0.07^{+0.58}_{-0.61}$ & $7.66^{+0.79}_{-0.82}$ & $1.01^{+0.68}_{-0.68}$ & {-0.21} & {0.31} & {-0.08} \\ \\ 
 
RXJ1347.5-1145 & $1.21^{+0.19}_{-0.21}$ & $-0.26^{+0.68}_{-0.76}$ & $8.08^{+0.78}_{-0.78}$ & $0.43^{+0.32}_{-0.33}$ & {-0.09} & {0.23} & {-0.04} \\ \\ 
 
RXJ1532.9+3021 & $1.98^{+0.18}_{-0.18}$ & $-0.49^{+0.67}_{-0.81}$ & $8.79^{+0.5}_{-0.5}$ & $1.18^{+0.46}_{-0.5}$ & {-0.36} & {-0.13} & {0.04} \\ \\
 
RXJ2129.7+0005 & $-0.24^{+0.27}_{-0.24}$ & $> - 0.32$ & $6.82^{+0.75}_{-0.71}$ & $0.83^{+0.67}_{-0.62}$ & {-0.31} & {0.07} & {0.04} \\ \\
\hline
\end{tabular}  
\begin{flushleft}
$^{a}$ Spearman correlation coefficients for the sample of pairs of parameters obtained by sampling the posterior probability distribution of models. \\
$^{b}$ For log$_{10}$ $\Delta t_{b}$ posterior probability histograms that peak near the upper bound of the parameter space, we report the 1$\sigma$ confidence interval as a lower limit on log$_{10}$ $\Delta t_{b}$. \\ 
\end{flushleft}  
}  
\end{table*}

\begin{figure*}[h]
{\epsfig{file=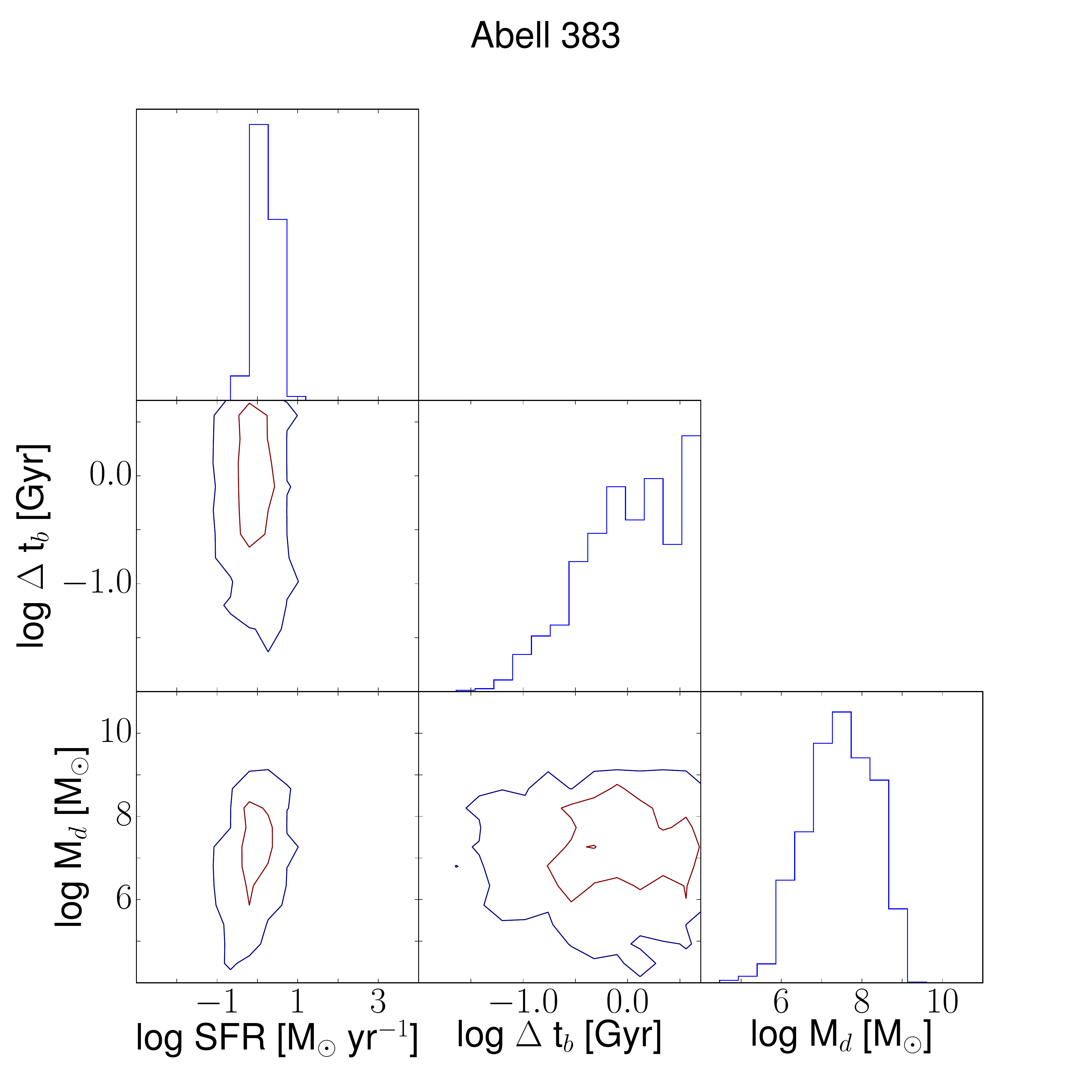,width=8.5cm,angle=0}
\label{fig:subfigure1}}
{\epsfig{file=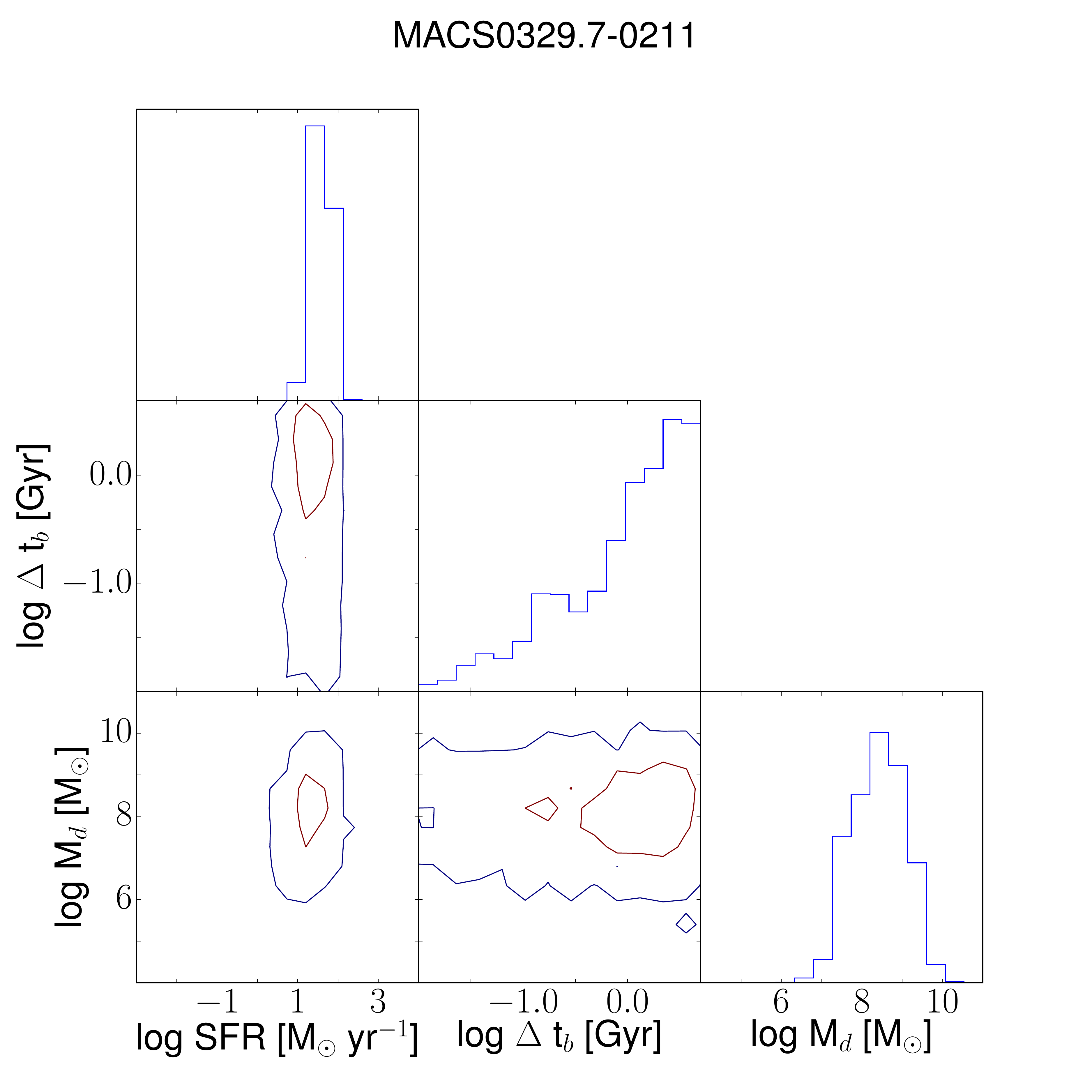,width=8.5cm,angle=0}
\label{fig:subfigure2}}
{\epsfig{file=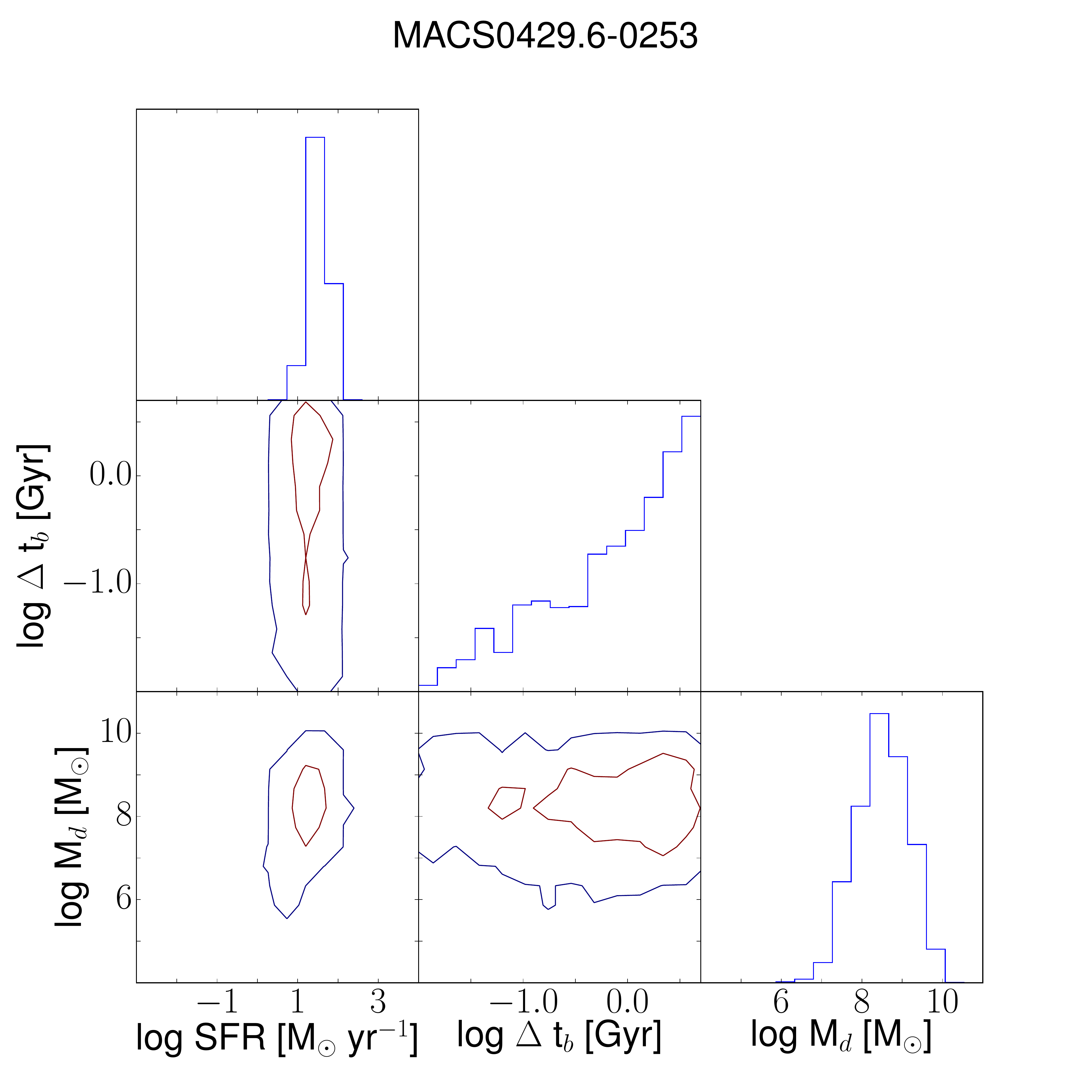,width=8.5cm,angle=0}
\label{fig:subfigure}}
{\epsfig{file=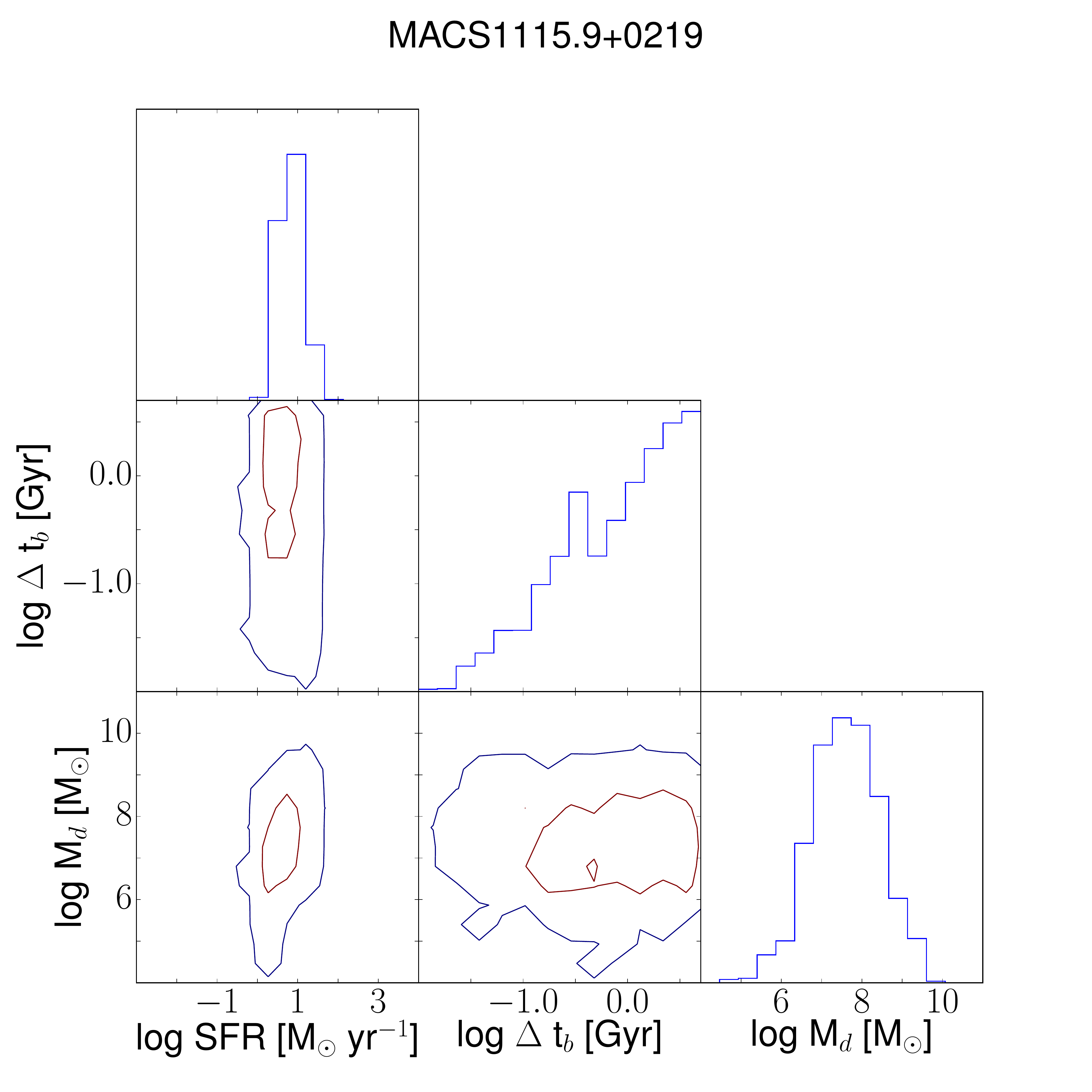,width=8.5cm,angle=0}
\label{fig:subfigure}}
\caption{\label{fig:Posteriors} Marginal posterior probability distributions for the SFR, burst duration $\Delta t_{b}$, and dust mass $M_{d}$ are shown for each BCG. Marginal distributions for individual parameters are obtained from the distribution of each parameter for the  model SEDs sampled using {\tt iSEDfit}, and are depicted by the histograms in the diagonal sub-plots of each figure. Two-dimensional slices of the posterior probability distribution (for SFR$\times \Delta t_{b}$, SFR$\times M_{d}$, and $M_{d}\times\Delta t_{b}$), are obtained from the distribution of pairs of parameters for each model SED, and are depicted by the contours drawn on the two-dimensional histograms in the off-diagonal sub-plots. The contours represent the 68.3$\%$ and 99.7$\%$ credible intervals, which correspond to the 1$\sigma$ and 3$\sigma$ contours for Gaussian distributions.}
\end{figure*}
\begin{figure*}
\ContinuedFloat
{\epsfig{file=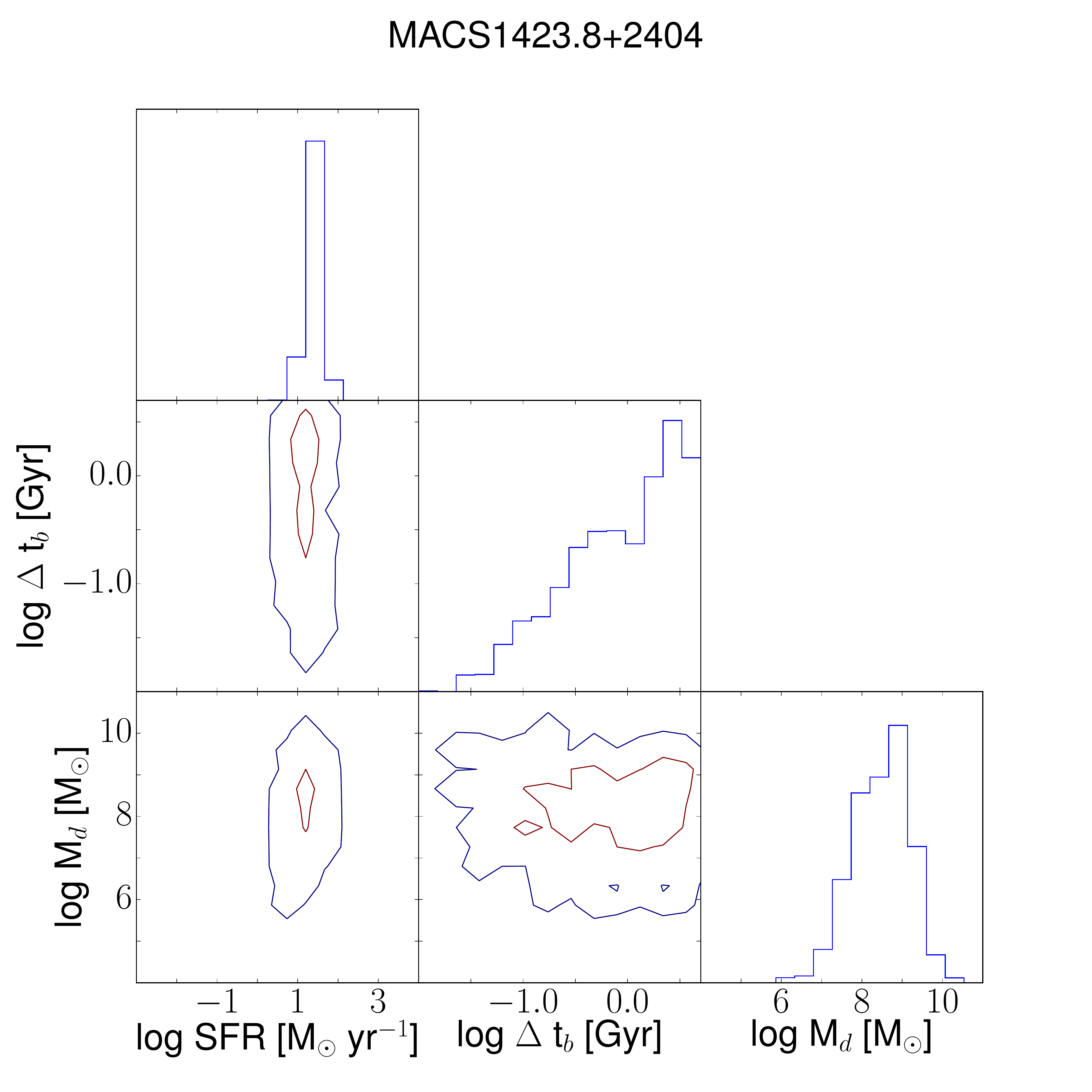,width=8.5cm,angle=0}
\label{fig:subfigure}}
{\epsfig{file=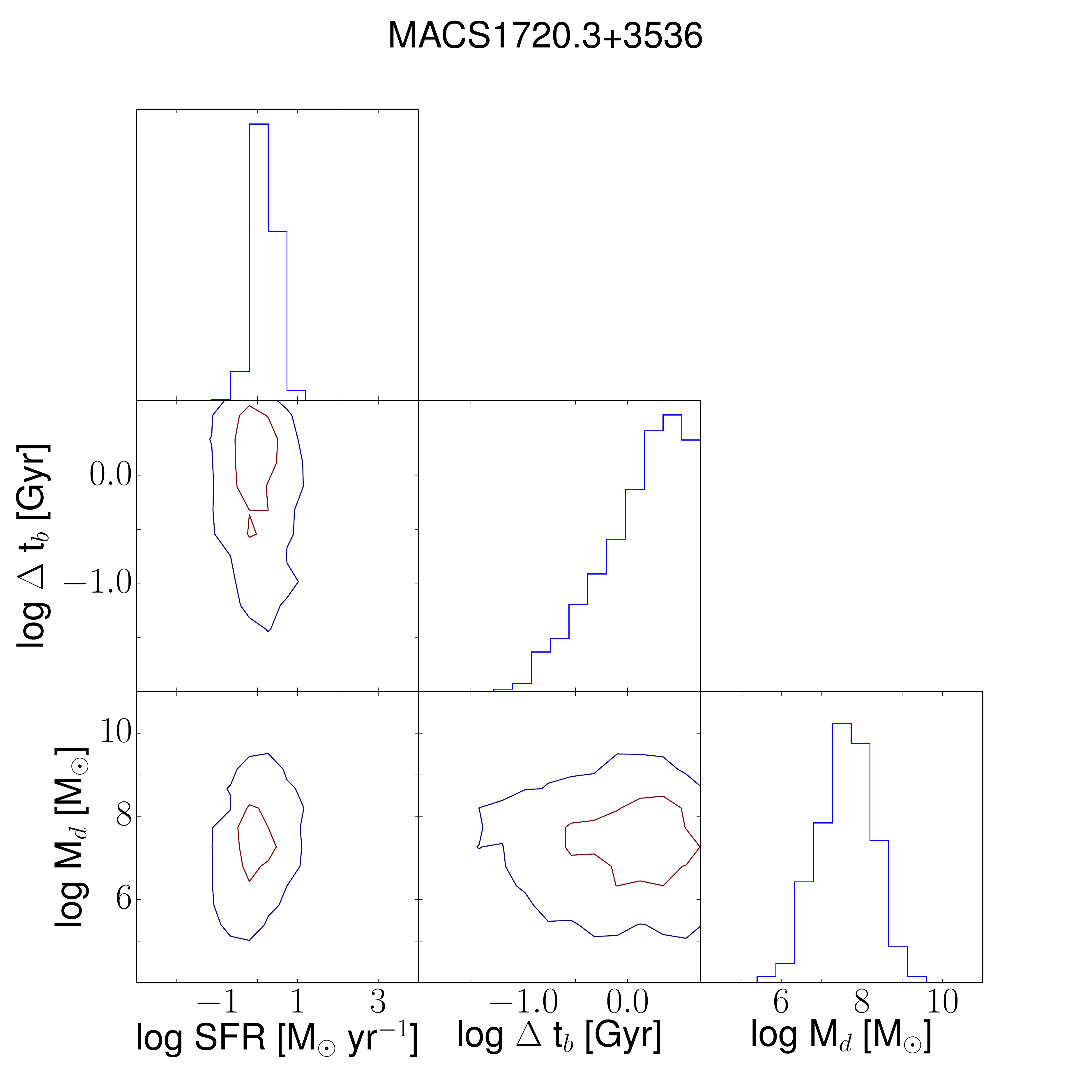,width=8.5cm,angle=0}
\label{fig:subfigure}}
{\epsfig{file=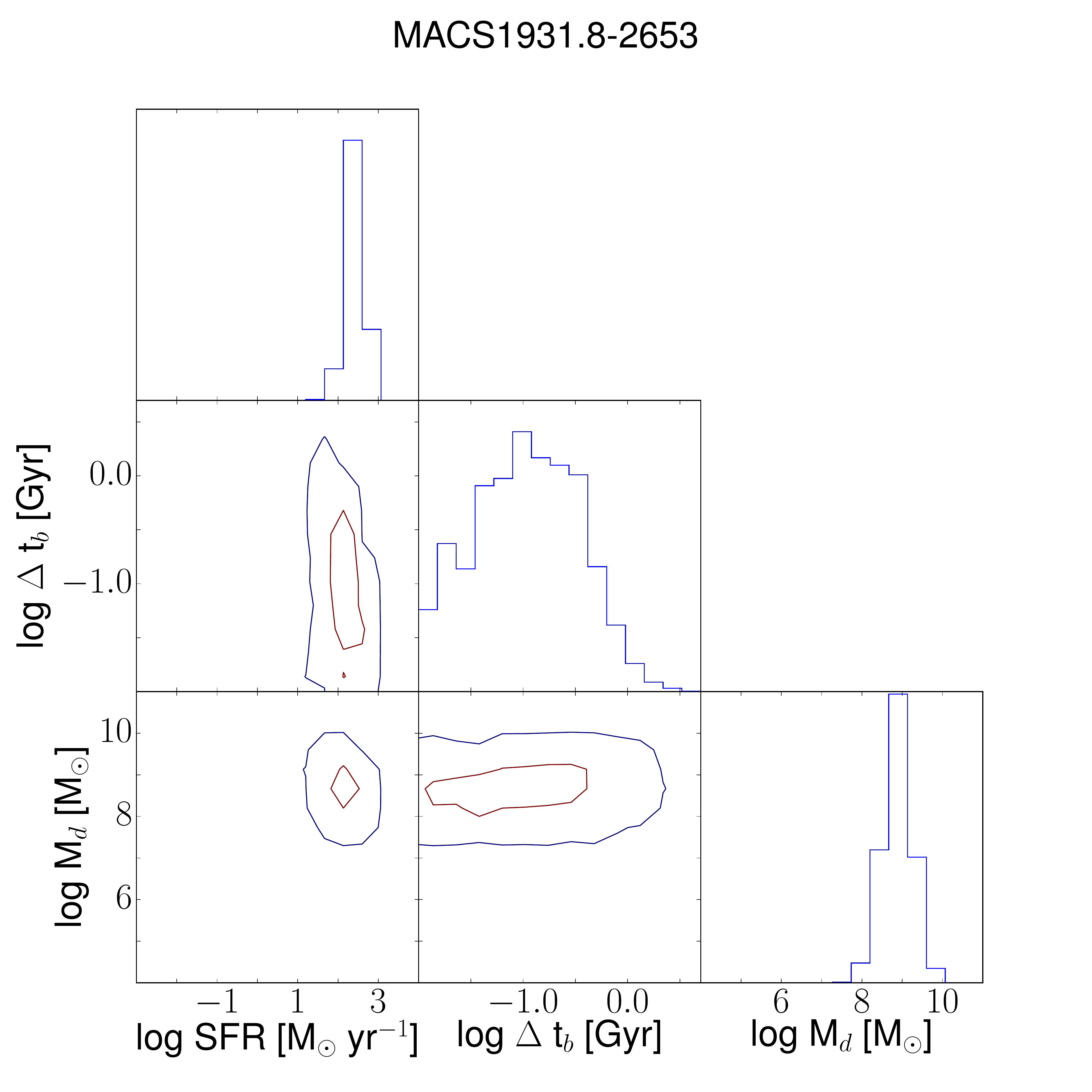,width=8.5cm,angle=0}
\label{fig:subfigure}}
{\epsfig{file=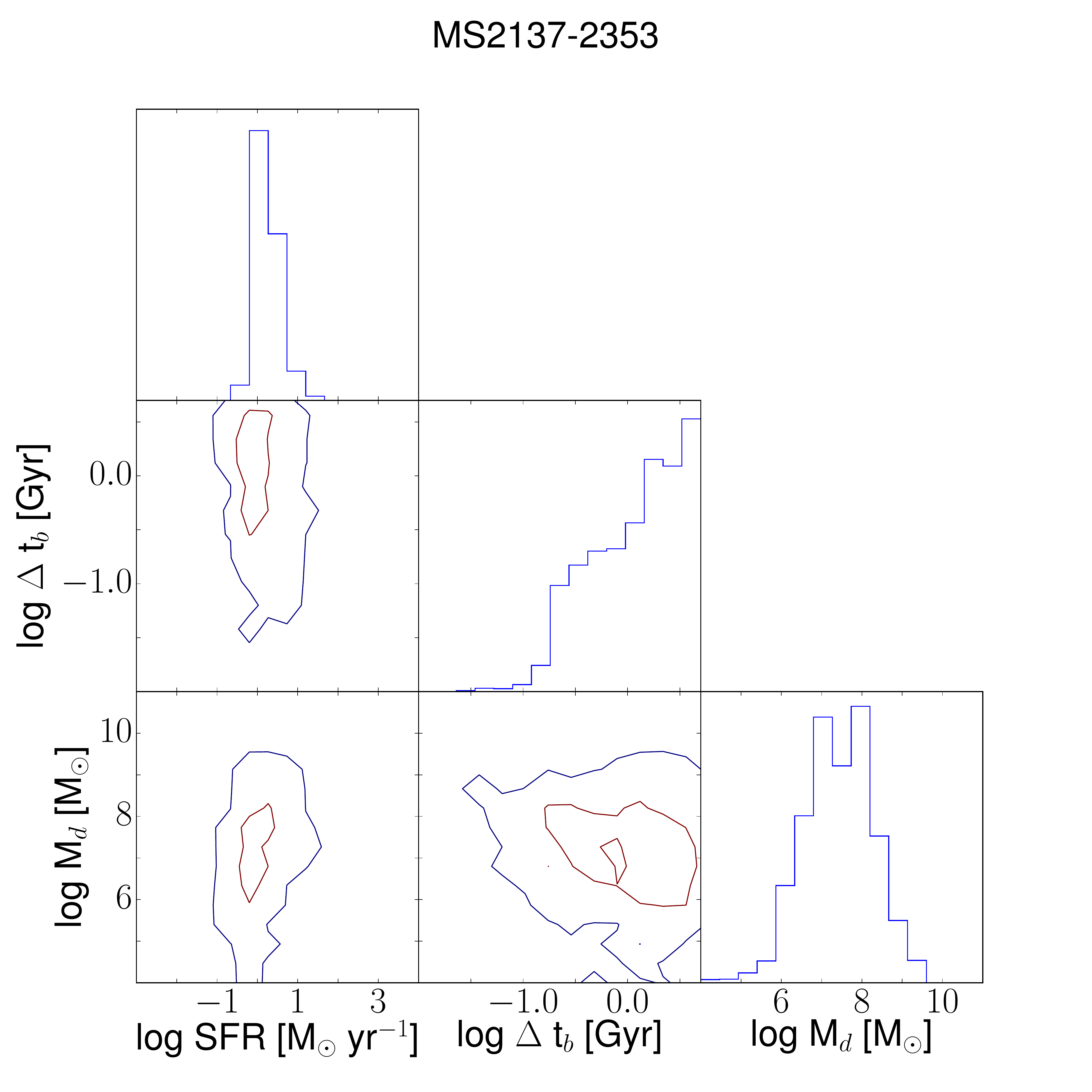,width=8.5cm,angle=0}
\label{fig:subfigure}}
\caption{\textit{Continued}}
\end{figure*}
\begin{figure*}
\ContinuedFloat
{\epsfig{file=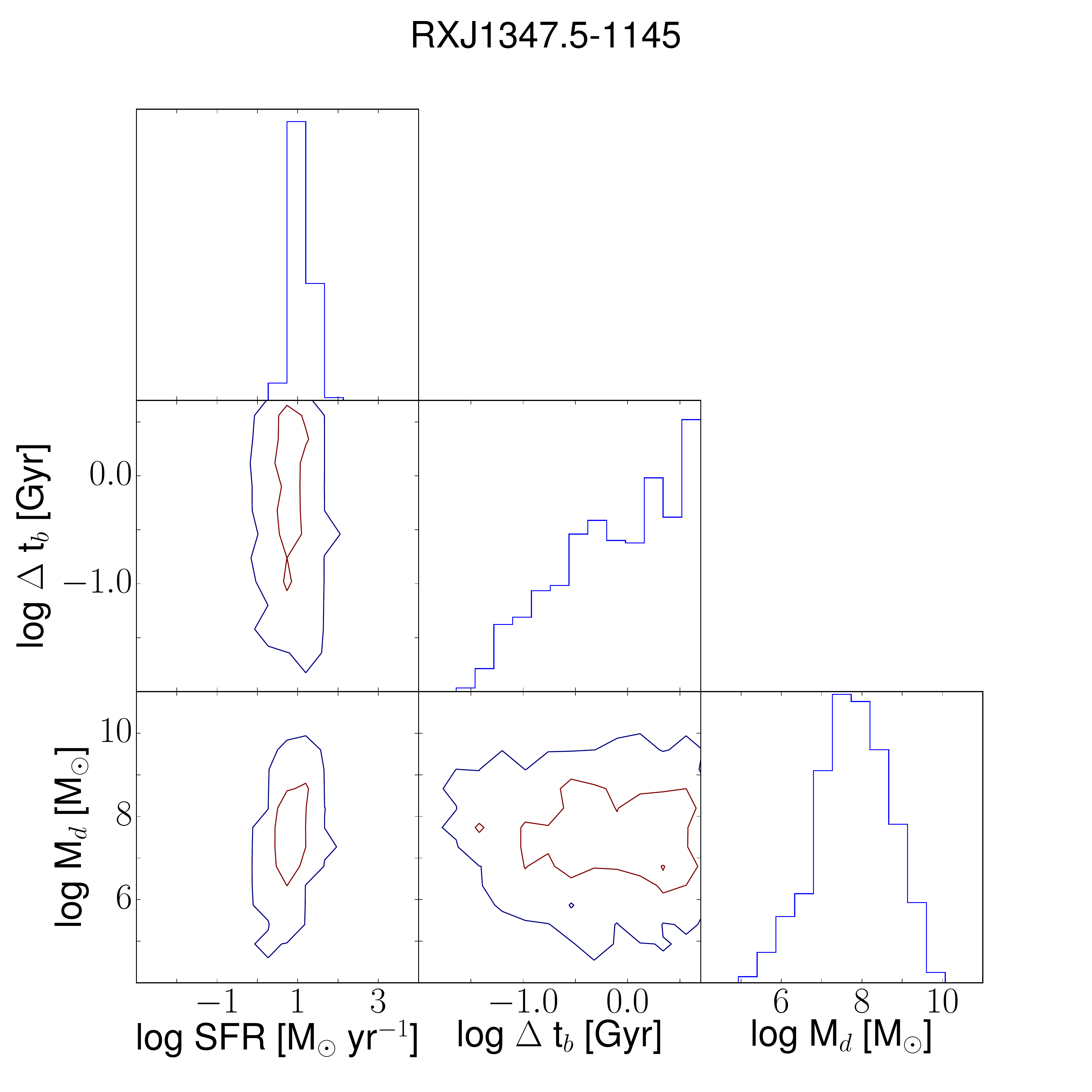,width=8.5cm,angle=0}
\label{fig:subfigure}}
{\epsfig{file=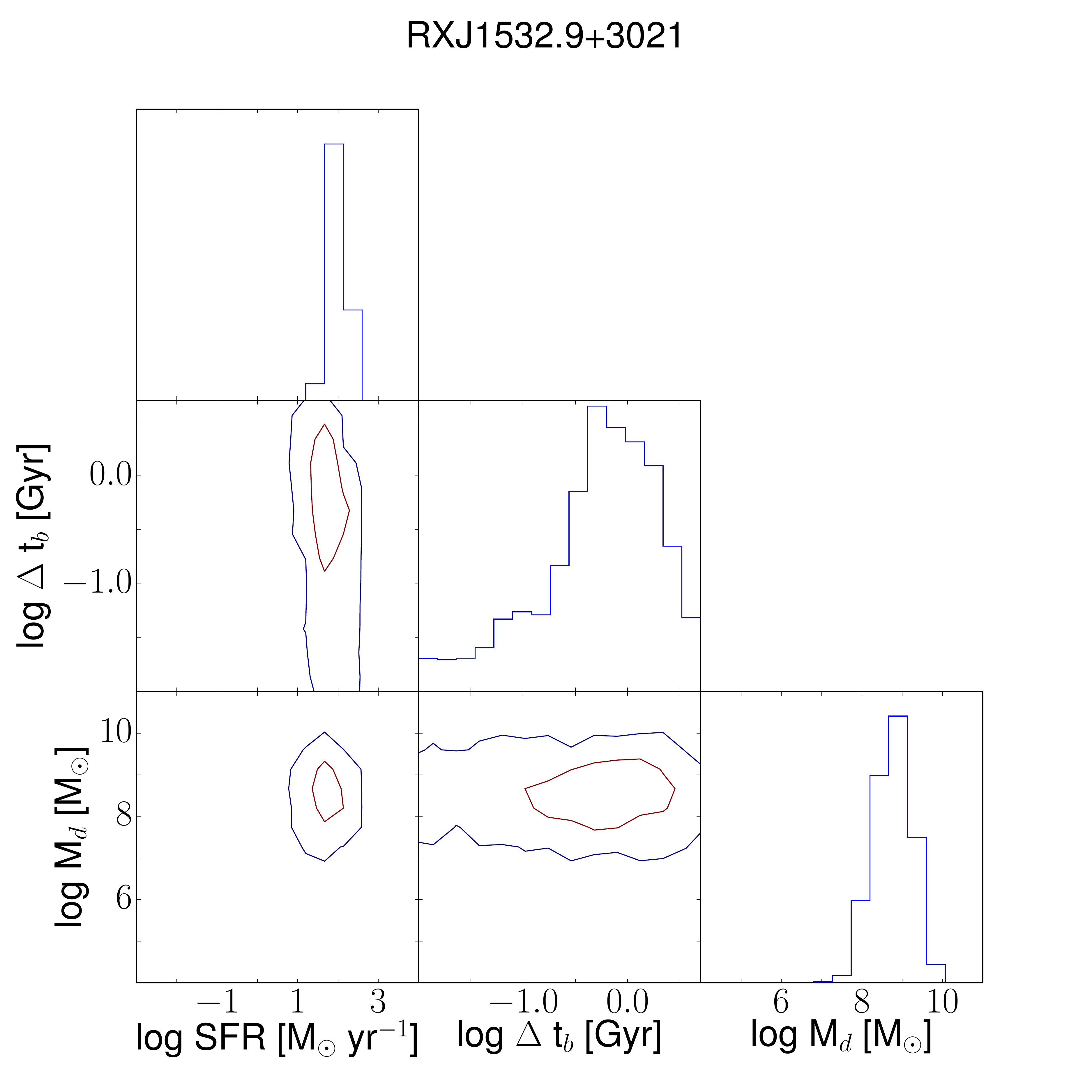,width=8.5cm,angle=0}
\label{fig:subfigure}}
{\epsfig{file=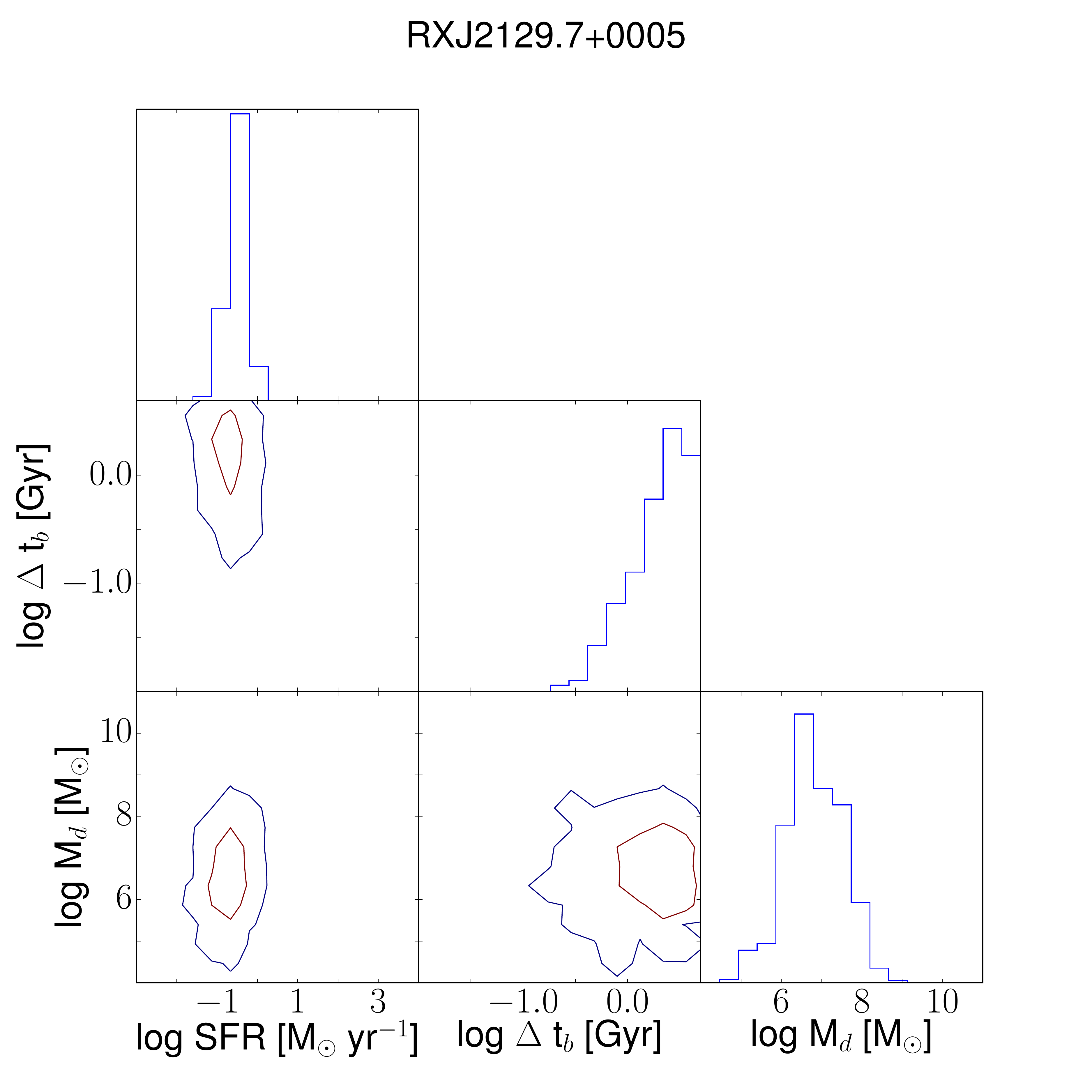,width=8.5cm,angle=0}
\label{fig:subfigure}}
\caption{\textit{Continued}}
\label{fig:Triptychs}
\end{figure*}

\begin{figure*}
\begin{center}
\begin{tabular}{c}
\includegraphics[height=16cm]{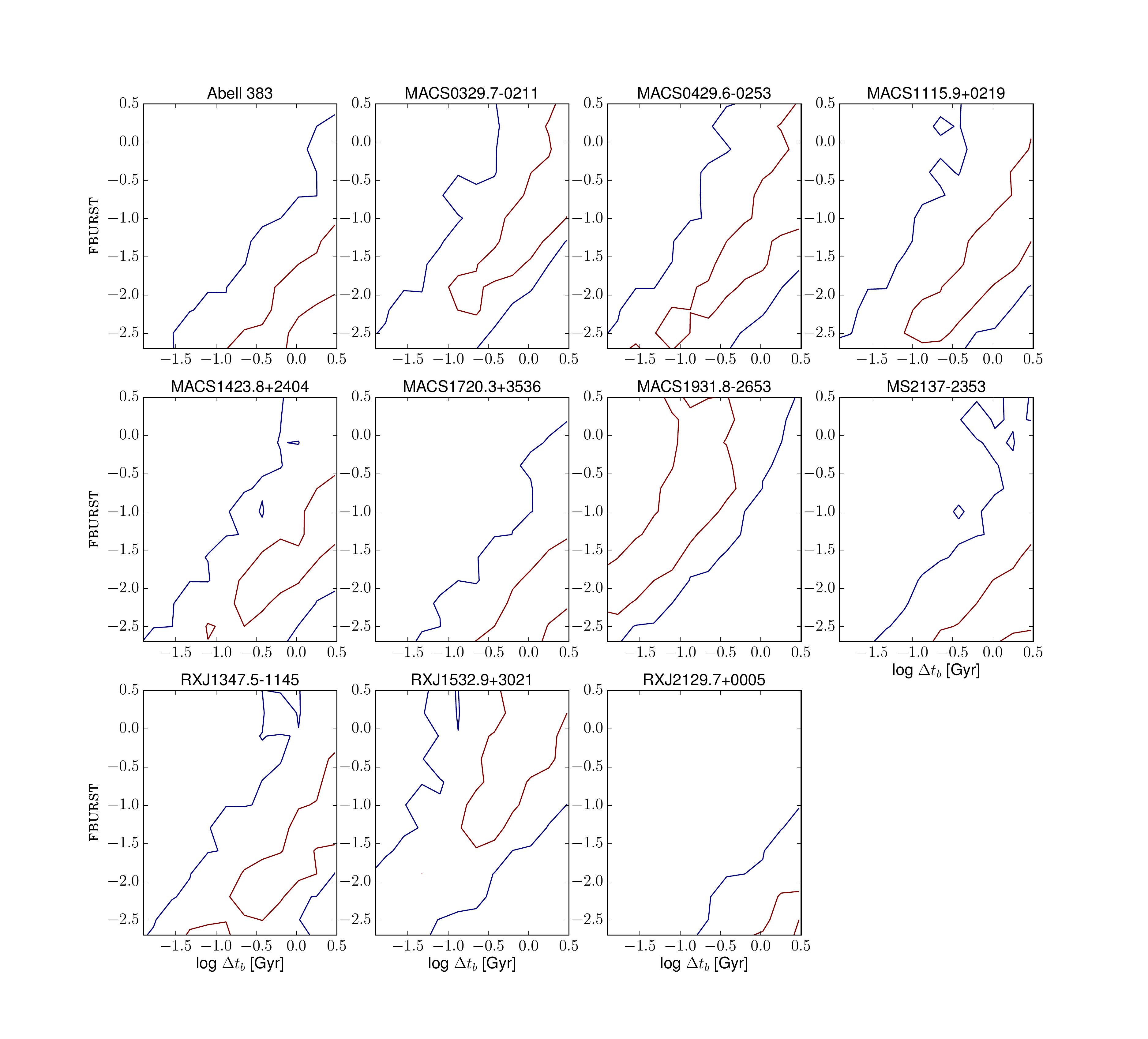}
\end{tabular}
\end{center}
\caption[]
{ \label{fig:DTB_FB} Marginal posterior probability distributions are shown for the burst duration, $\Delta t_{b}$, in log Gyr, and {\sc fburst}, also in log units.}
\end{figure*}

\begin{figure*}
\begin{center}
\begin{tabular}{c}
\includegraphics[height=14cm]{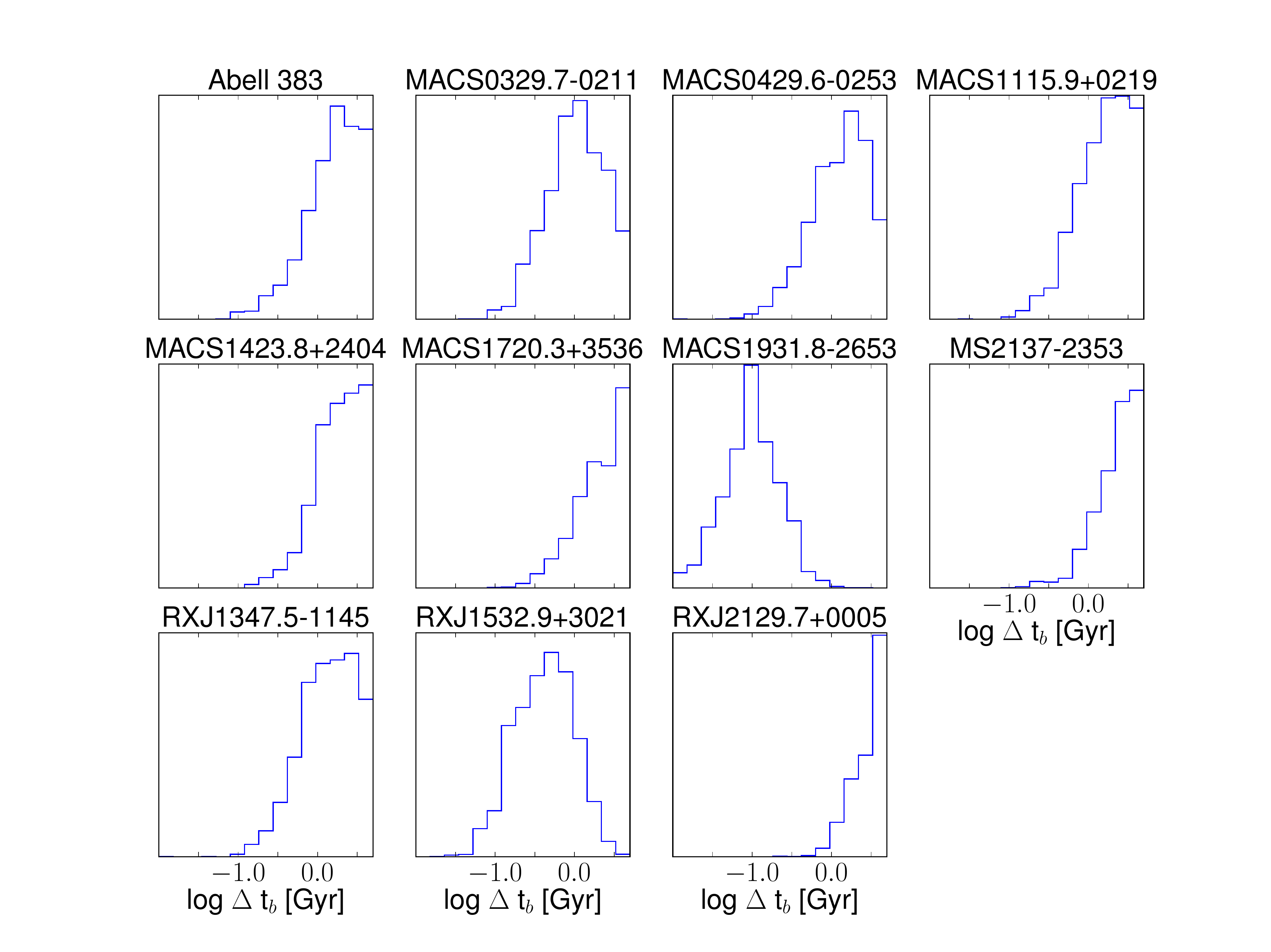}
\end{tabular}
\end{center}
\caption[]
{ \label{fig:DTB_Posteriors} Marginal posterior probability distributions are shown for the burst duration, $\Delta t_{b}$, in log Gyr, when {\sc fburst} is restricted as described in Section~\ref{sec-results}. Results were obtained by generating a new model grid for each BCG with the new prior on {\sc fburst} and re-fitting the data.}
\end{figure*}

\begin{figure*}
\begin{center}
\begin{tabular}{c}
\includegraphics[height=20cm]{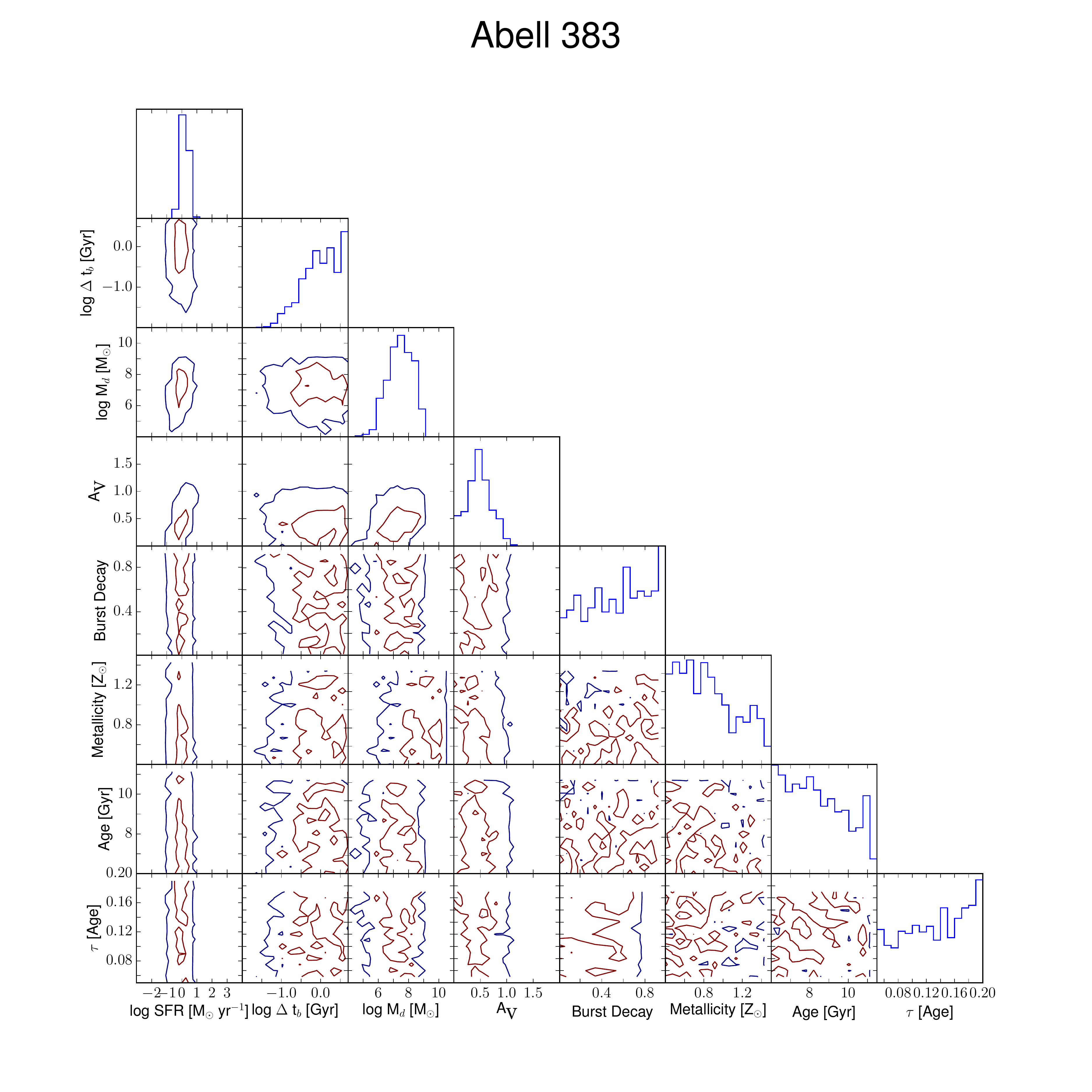}
\end{tabular}
\end{center}
\caption[]
{ \label{fig:Giant_Marginal} Marginal posterior probability distributions for the SFR, burst duration $\Delta t_{b}$, and dust mass $M_{d}$, dust attenuation $A_{V}$, burst decay percentage, metallicity, galaxy age, and $\tau$ shown for Abell 383. Plots are analogous to the marginal distributions presented in Figure \ref{fig:Posteriors}.}
\end{figure*}

\begin{figure*}[h]
{\epsfig{file=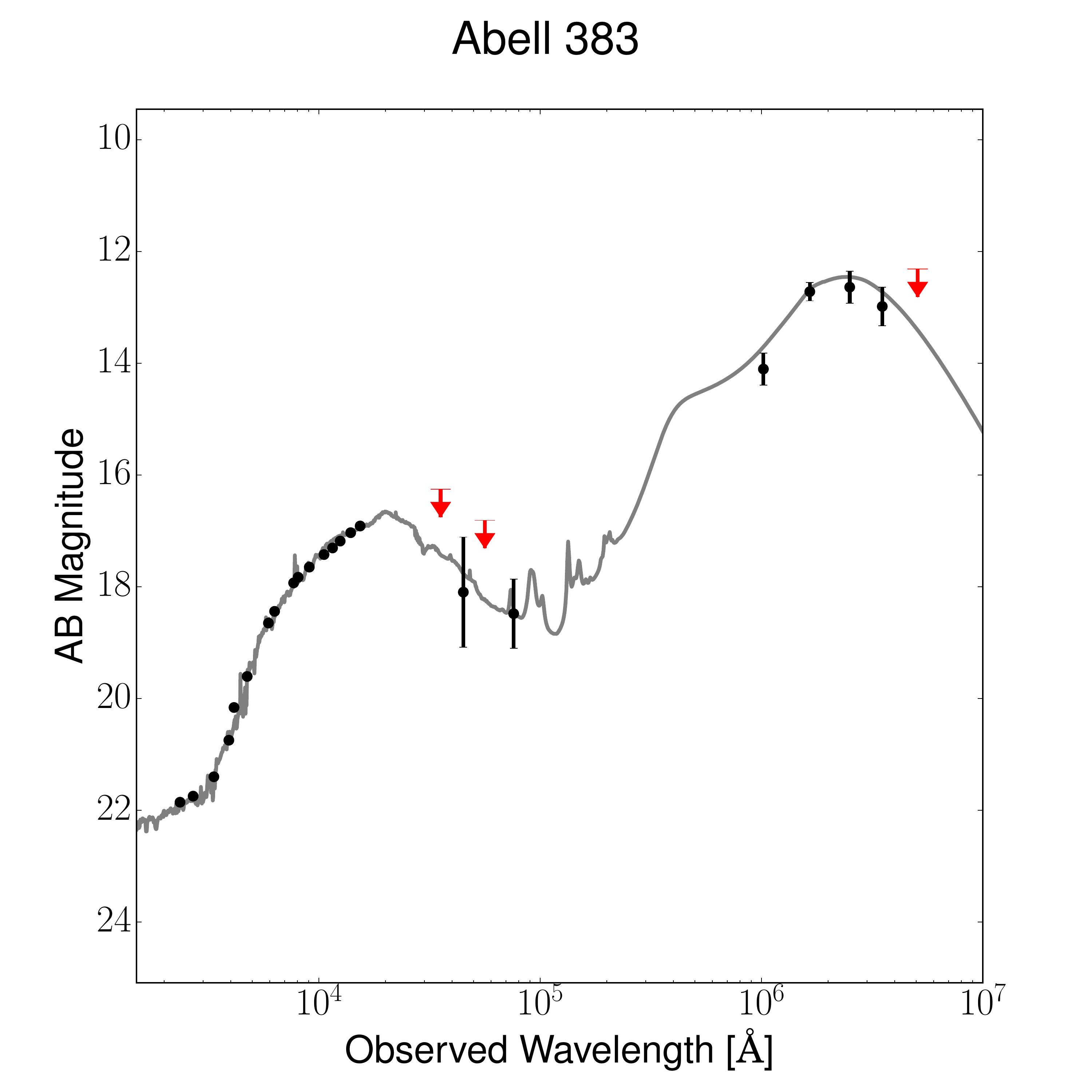,width=8.5cm,angle=0}
\label{fig:subfigure1}}
{\epsfig{file=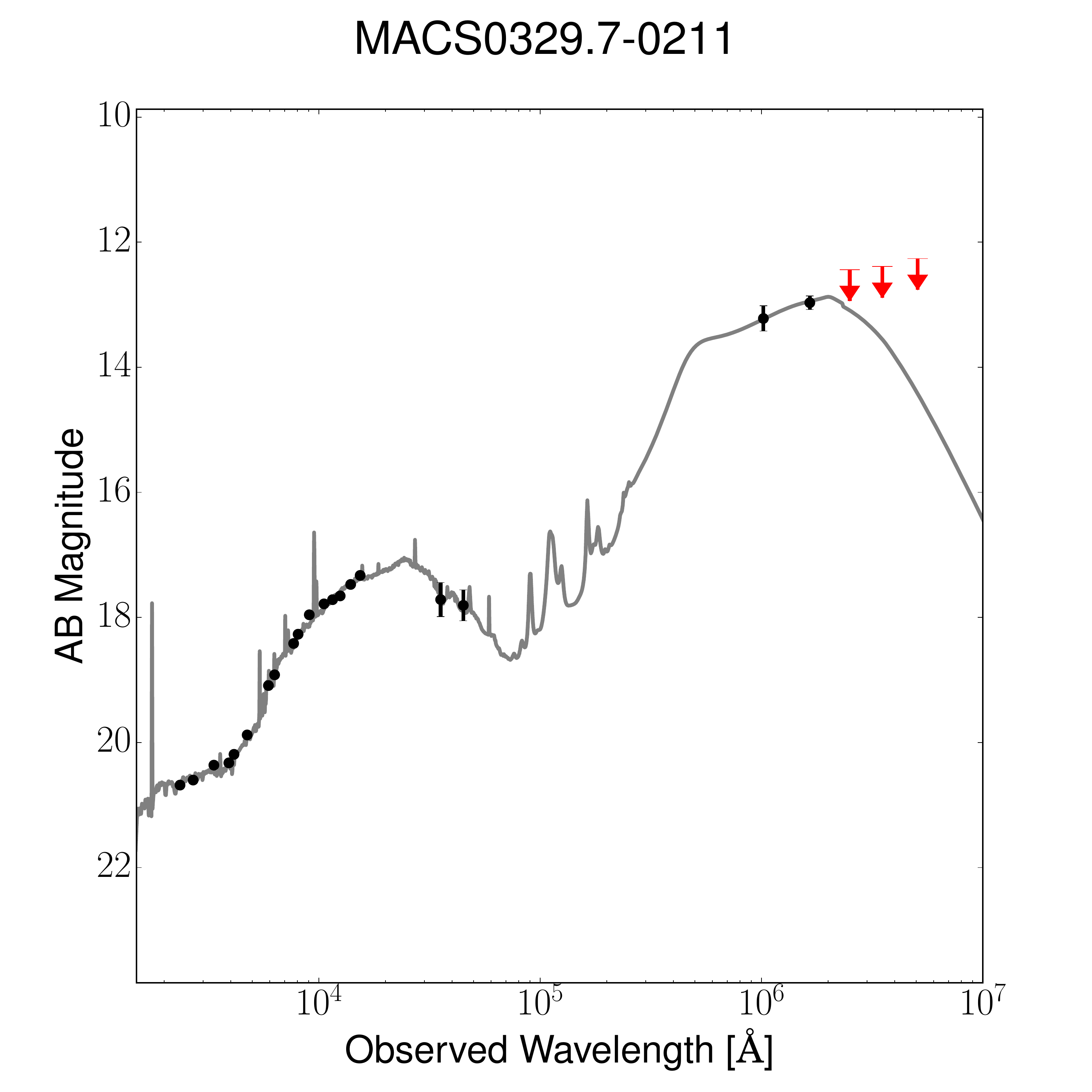,width=8.5cm,angle=0}
\label{fig:subfigure2}}
{\epsfig{file=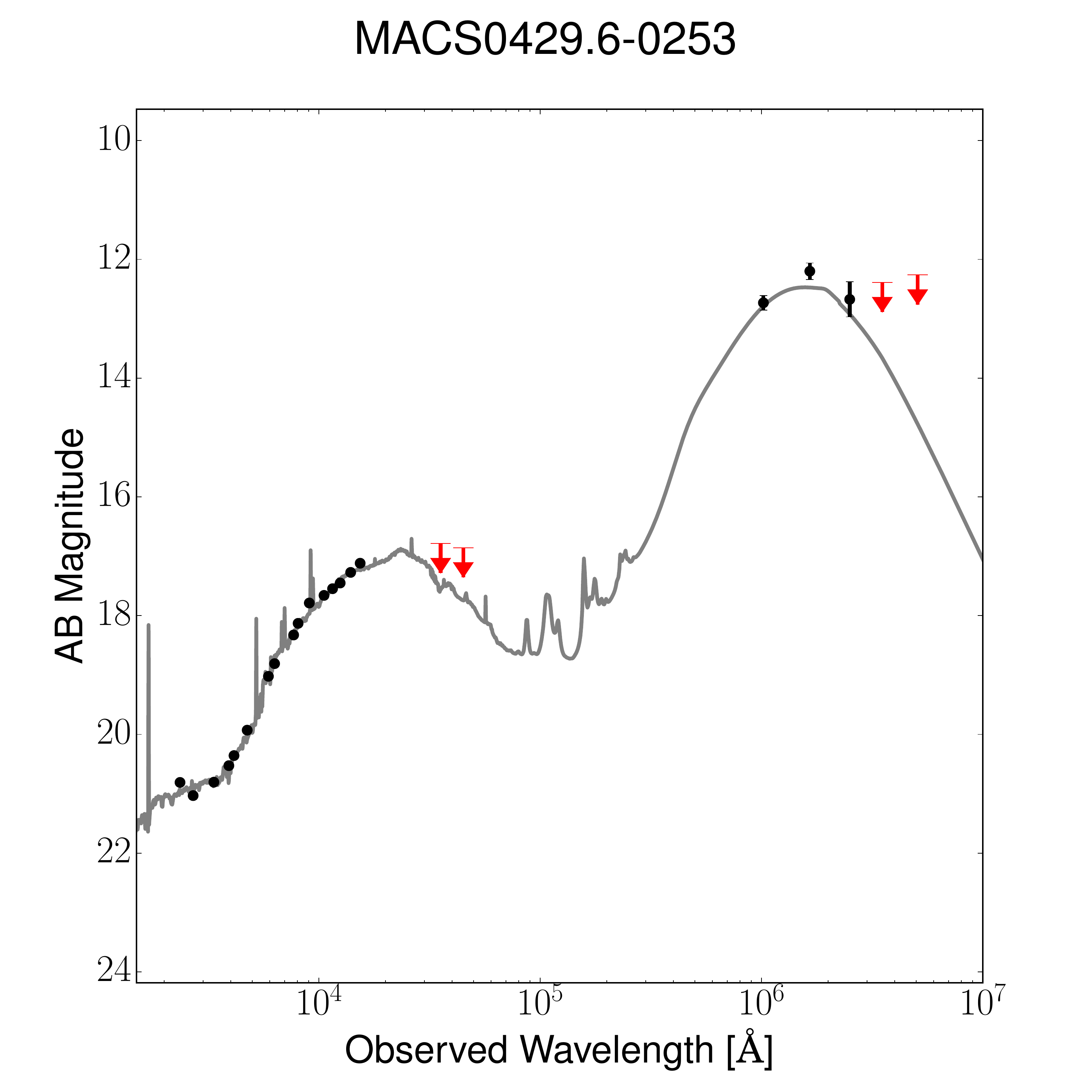,width=8.5cm,angle=0}
\label{fig:subfigure}}
{\epsfig{file=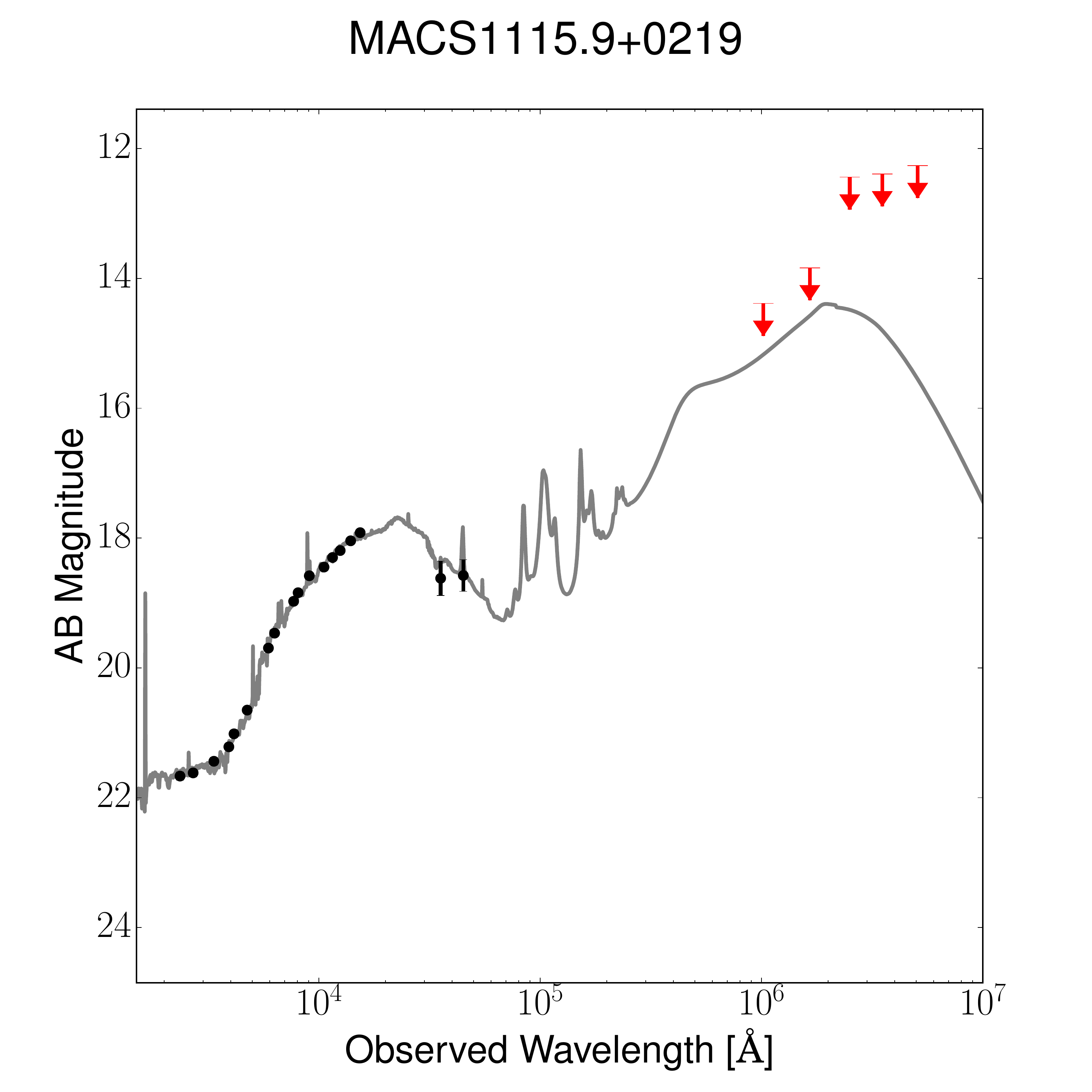,width=8.5cm,angle=0}
\label{fig:subfigure}}
\caption{\label{fig:Best_Fits} Best fit SEDs for each CLASH star forming BCG. SED photometry data points with 1$\sigma$ error bars are shown as black points or as 3$\sigma$ upper limits denoted by red arrows. Grey lines depict the `best fit' synthetic spectra, where `best fit' is defined to be the synthetic spectrum producing the smallest reduced $\chi^{2}$ in the {\tt iSEDfit} Monte Carlo grid.}
\end{figure*}
\begin{figure*}
\ContinuedFloat
{\epsfig{file=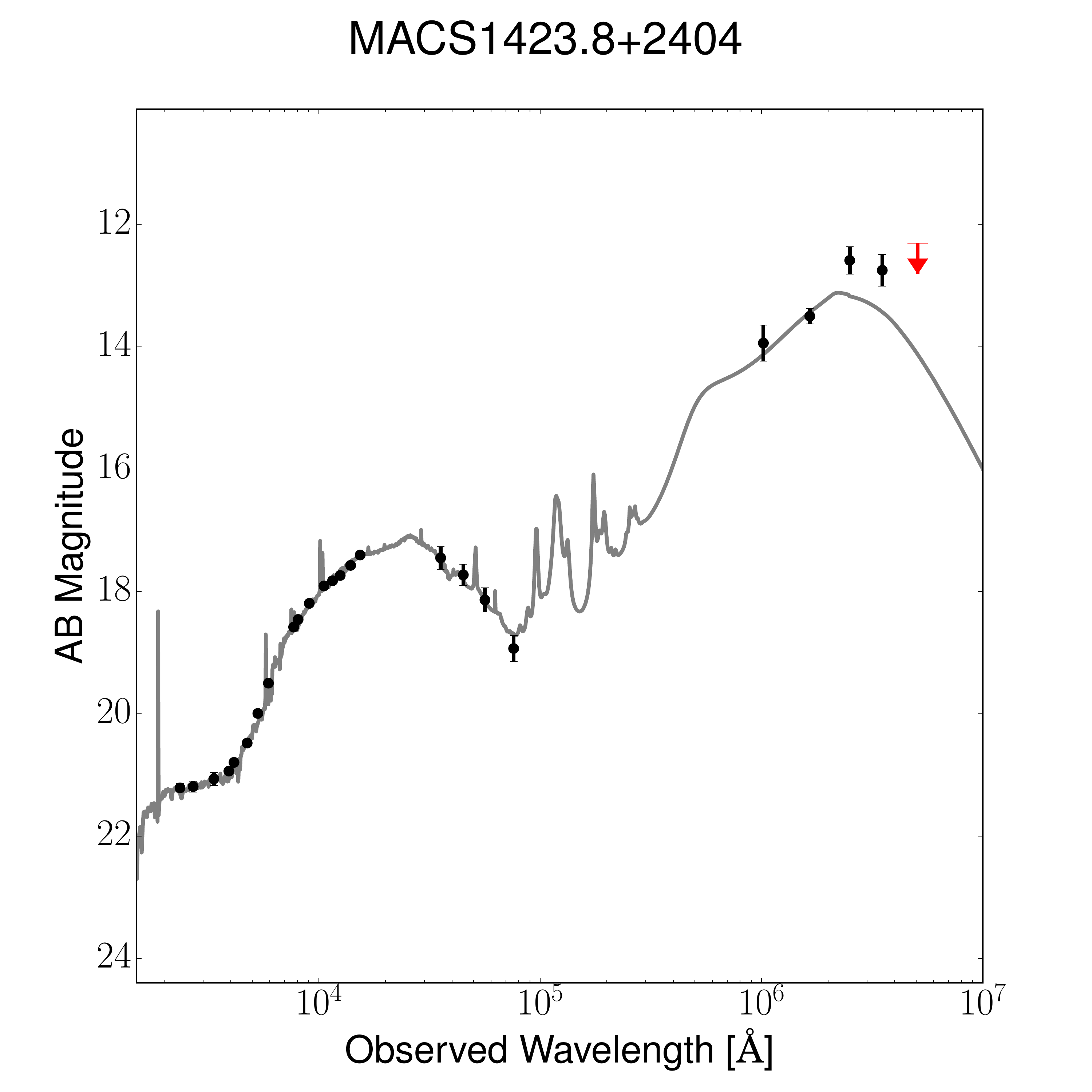,width=8.5cm,angle=0}
\label{fig:subfigure}}
{\epsfig{file=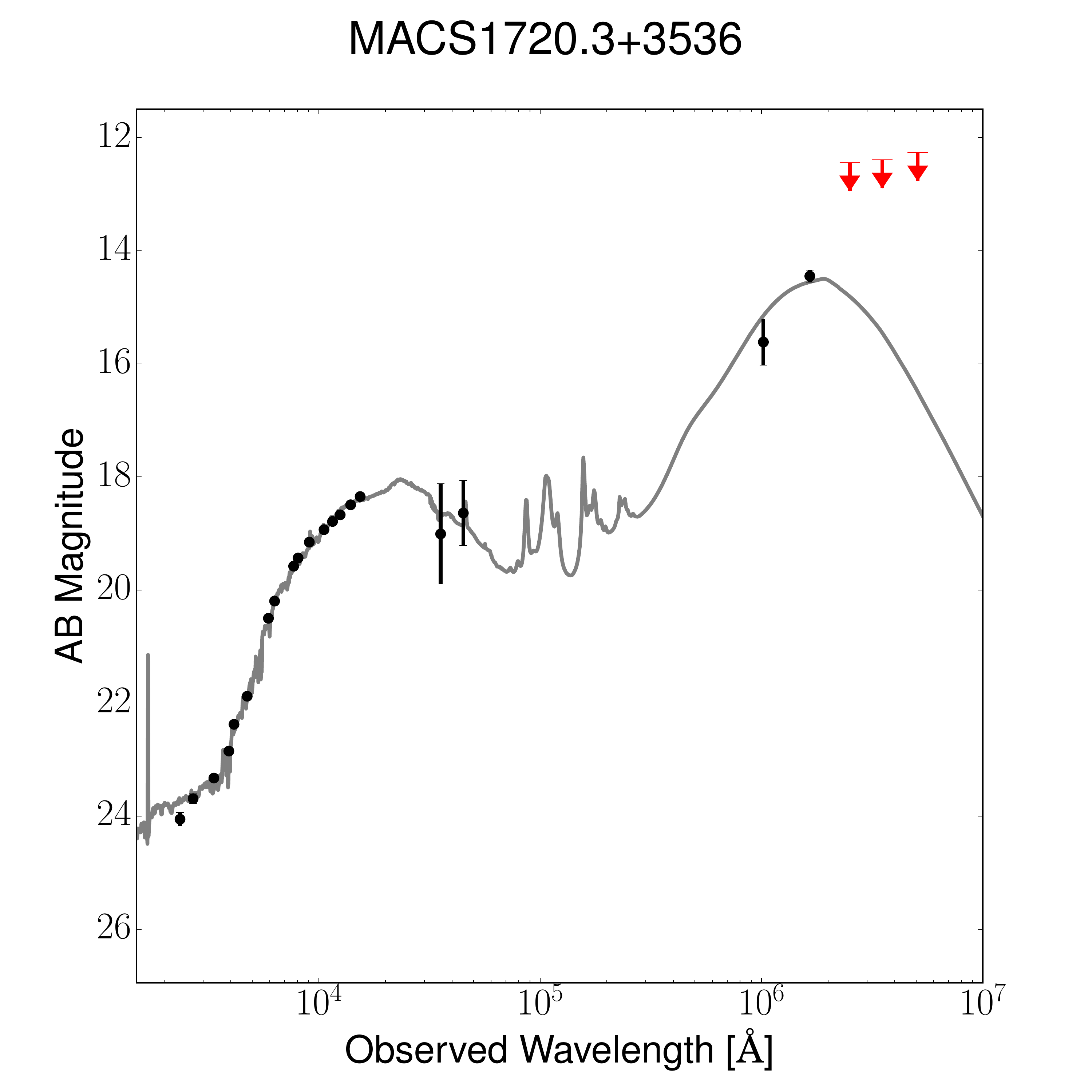,width=8.5cm,angle=0}
\label{fig:subfigure}}
{\epsfig{file=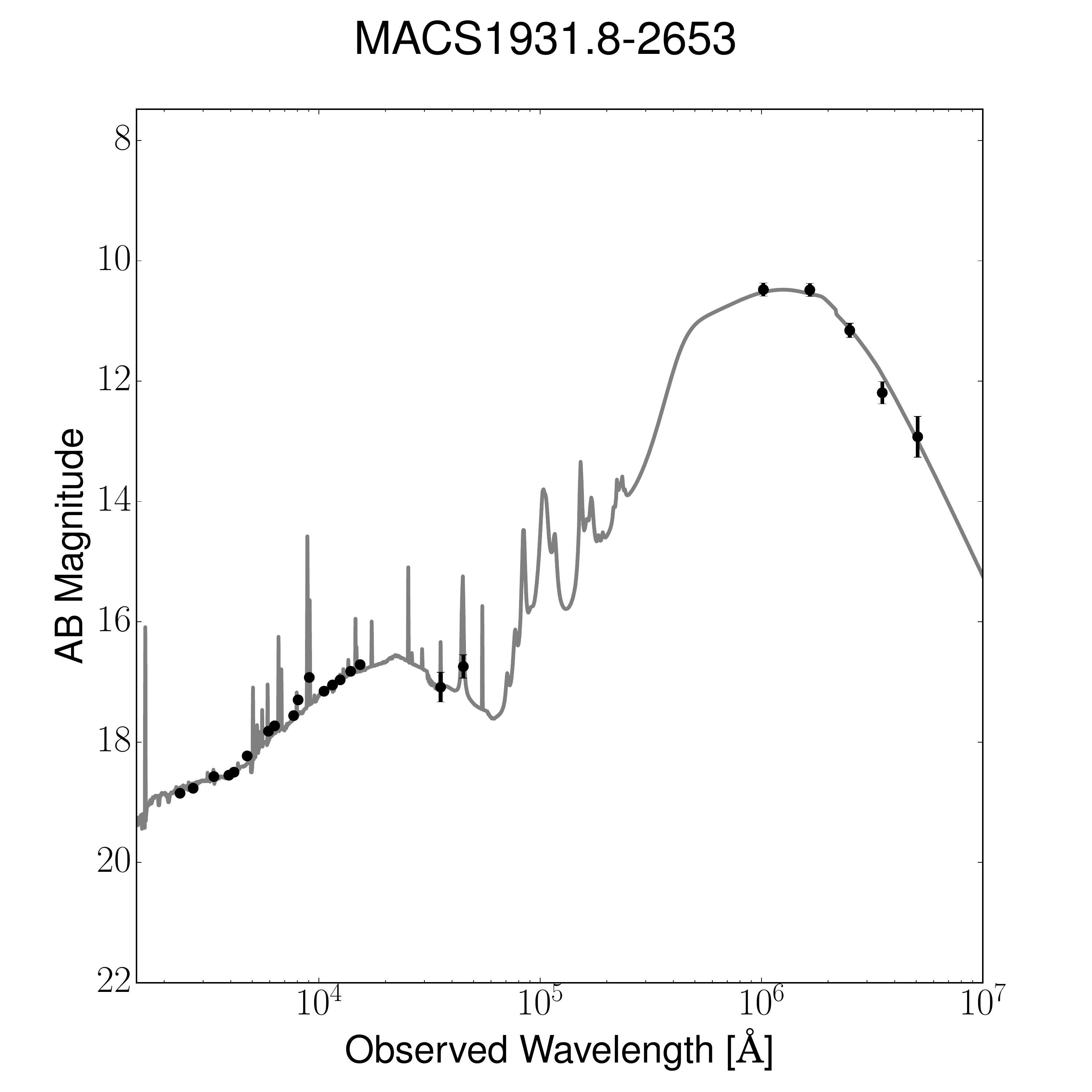,width=8.5cm,angle=0}
\label{fig:subfigure}}
{\epsfig{file=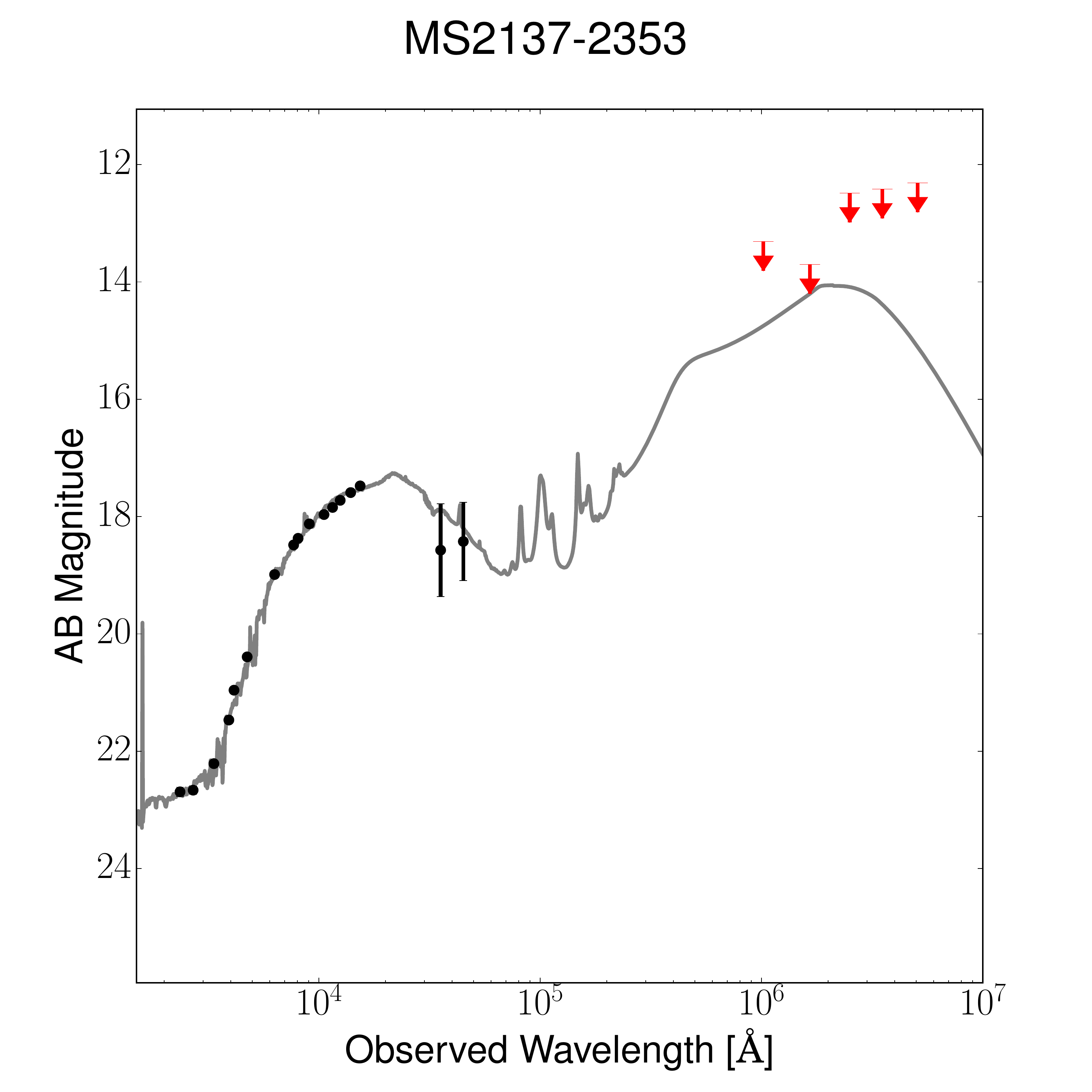,width=8.5cm,angle=0}
\label{fig:subfigure}}
\caption{\textit{Continued}}
\end{figure*}
\begin{figure*}
\ContinuedFloat
{\epsfig{file=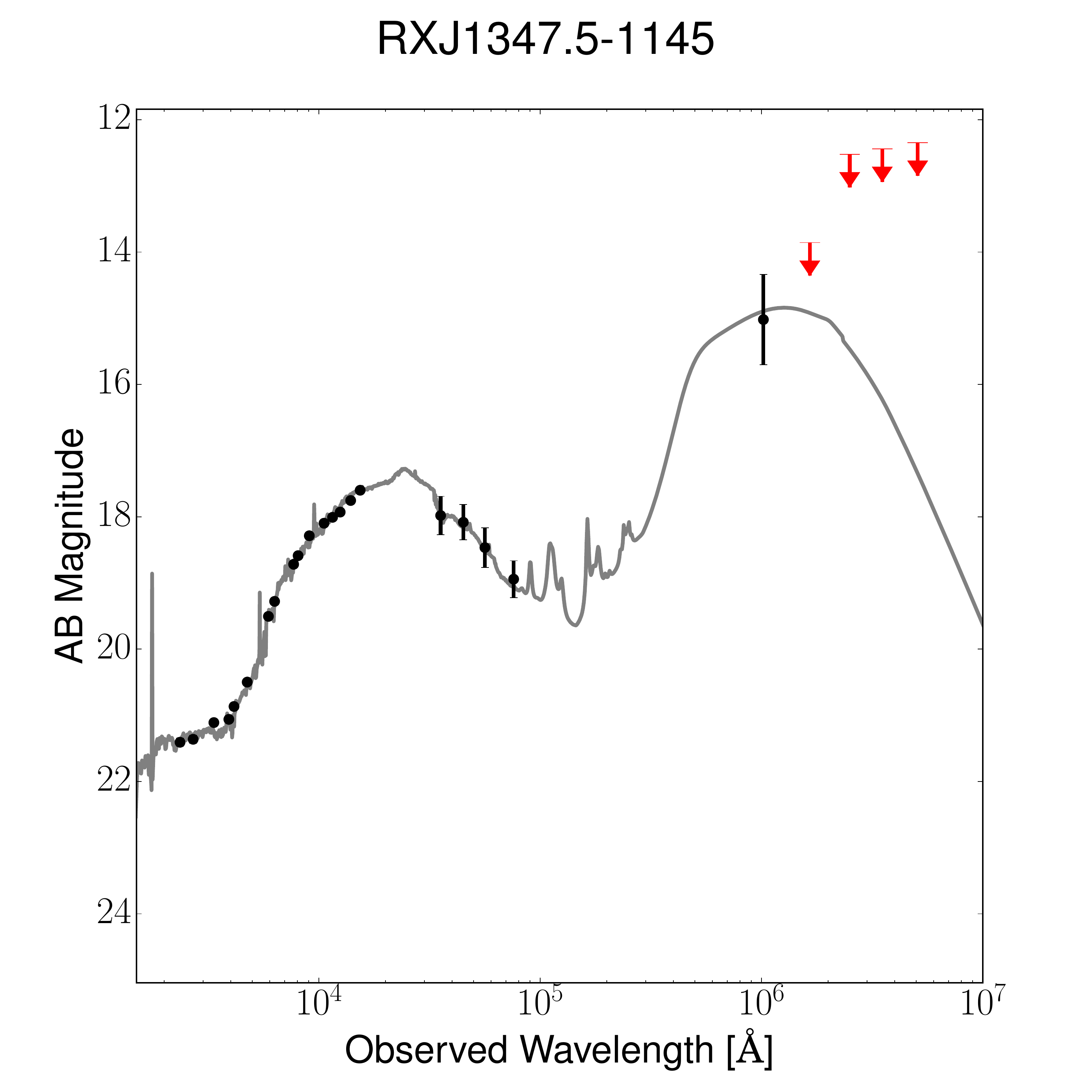,width=8.5cm,angle=0}
\label{fig:subfigure}}
{\epsfig{file=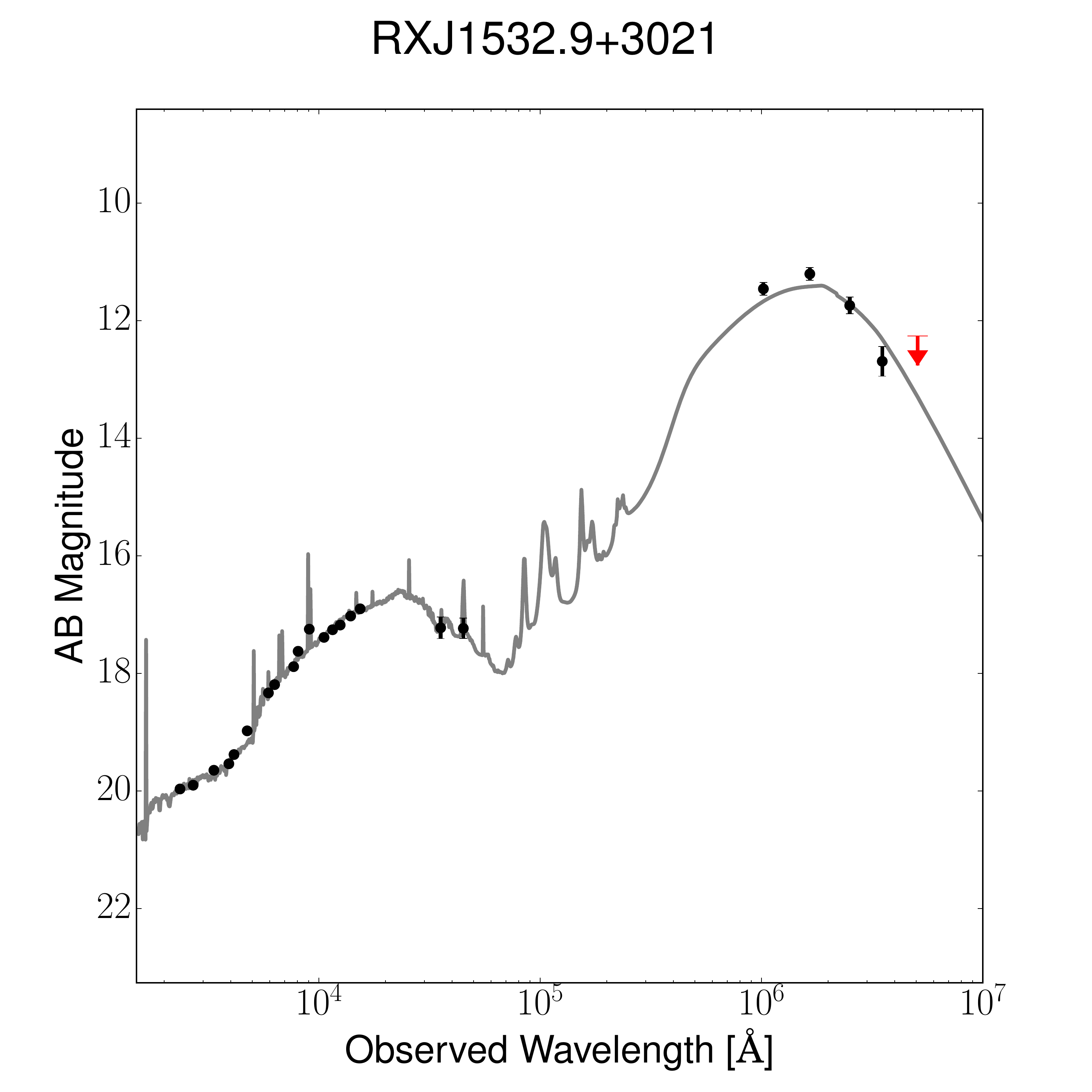,width=8.5cm,angle=0}
\label{fig:subfigure}}
{\epsfig{file=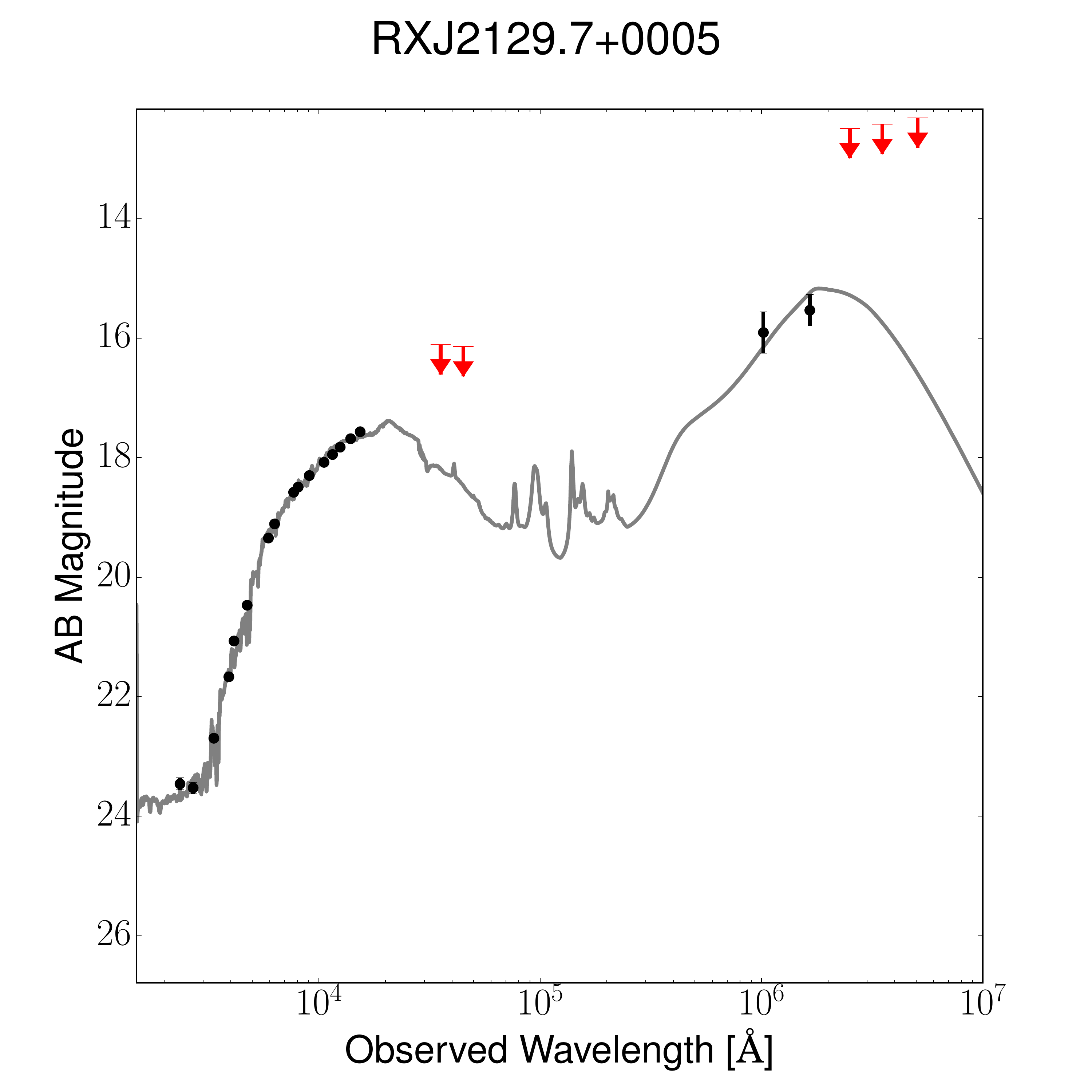,width=8.5cm,angle=0}
\label{fig:subfigure}}
\caption{\textit{Continued}}
\label{fig:Triptychs}
\end{figure*}

\end{document}